\documentclass[12pt]{article}

\usepackage{psfig}
\usepackage{amsfonts}
\usepackage{latexsym}
\newtheorem{lemma}{Lemma}

\def\FF{\hbox to 8.33887pt{\rm I\hskip-1.8pt F}}
\def\NN{\hbox to 9.3111pt{\rm I\hskip-1.8pt N}}
\def\PP{\hbox to 8.61664pt{\rm I\hskip-1.8pt P}}
\def\QQ{\rlap {\raise 0.4ex \hbox{$\scriptscriptstyle |$}}
{\hskip -4.5pt Q}}
\def\RR{\hbox to 9.1722pt{\rm I\hskip-1.8pt R}}
\def\ZZ{\hbox to 8.2222pt{\rm Z\hskip-4pt \rm Z}}

\renewcommand{\thesection}{\Roman{section}}

\newcommand{\resetequ}{\setcounter{equation}{0}}
\newcommand{\fr}{{\cal F}}           %%%%% F call
\newcommand{\tree}{{\cal T}}           %%%%% T call
\newcommand{\be}{\begin{equation}}
\newcommand{\ee}{\end{equation}}
\newcommand{\bqa}{\begin{eqnarray}}
\newcommand{\eqa}{\end{eqnarray}}
\newcommand{\ba}{\begin{array}}
\newcommand{\ea}{\end{array}}
\newcommand{\p}[1]{{\partial\over \partial{#1}}}
\newcommand{\no}{\nonumber}

\newcommand{\lp}{\left (}
\newcommand{\rp}{\right )}
\newcommand{\qed}{\hfill\vrule height .6em width .6em depth 0pt
\goodbreak}
\newcommand{\al}{\alpha}
\newcommand{\bt}{\beta}

\newcommand{\de}{\delta}
\newcommand{\ep}{\epsilon}
\newcommand{\vep}{\varepsilon}

\newcommand{\la}{\lambda}

\newcommand{\si}{\sigma}
\newcommand{\up}{\upsilon}

\newcommand{\Om}{\Omega}
\newcommand{\Si}{\Sigma}
\newcommand{\La}{\Lambda}

\newcommand{\De}{\Delta}

\newcommand{\Up}{\Upsilon}
\newcommand{\bpsi}{\bar{\psi}}

\begin{document}

\title{Interacting Fermi liquid in three dimensions at finite temperature: 
Part I: Convergent Contributions}

%\centerline{\Large \bf Interacting Fermi liquid}
%\centerline{\Large \bf in three dimensions at finite temperature}  
%\centerline{\Large \bf Part I: Convergent Contributions}
%\vskip 2cm

\author{M. Disertori$^a$, J. Magnen$^b$ and V. Rivasseau$^b$}
%\centerline{M. Disertori$^a$, J. Magnen$^b$ and V. Rivasseau$^b$}

\maketitle

\centerline{a) Institute for Advanced Study}
\centerline{Einstein Drive, Princeton}
\centerline{NJ 08540,  USA}

\centerline{b) Centre de Physique Th{\'e}orique, CNRS UMR 7644}
\centerline{Ecole Polytechnique}
\centerline{91128 Palaiseau Cedex, FRANCE}

\vskip 1cm
\medskip
\noindent{\bf Abstract}

In this paper we complete the first step,
namely the uniform bound on completely convergent contributions,
towards proving that a three dimensional interacting system of Fermions 
is a Fermi liquid in the sense of Salmhofer.
The analysis relies on a direct space decomposition of the propagator,
on a bosonic multiscale cluster expansion
and on the Hadamard inequality, rather than on a Fermionic expansion and an 
angular analysis in momentum space, as was used in the recent proof by two of
us of Salmhofer's criterion in two dimensions.
 
\section{Introduction}

Conducting electrons in a metal at low temperature are well described
by Fermi liquid theory. However we know that the Fermi liquid theory is
not valid down to zero temperature. Indeed below the BCS critical temperature
the dressed electrons or holes which are the excitations of the Fermi liquid
bound into Cooper pairs and the metal becomes superconducting.
Even when the dominant electron interaction is repulsive, the 
Kohn-Luttinger instabilities prevent the Fermi liquid theory to be 
generically valid down to zero temperature.

Hence Fermi liquid theory 
(e.g. for the simplest case of a jellium model with a spherical Fermi surface)
is only an effective theory above some non-perturbative transition 
temperature, and it is not obvious to precise its mathematical definition.  
Recently Salmhofer proposed such a mathematical definition 
[S]. It consists
in proving that (under a suitable
renormalization condition on the two-point function), perturbation 
theory is analytic in a domain $ |\lambda \log T| \le K$, where $\lambda$ 
is the coupling constant and $T$ is the temperature, and that 
uniform bounds hold in that domain for the self-energy and its first and
second  derivatives. This criterion in particular
excludes Luttinger liquid behavior, which has been proved to hold in
one dimension [BGPS-BM], and for which 
second momentum-space derivatives of the self-energy
are unbounded in that domain.

Recently two of us proved Salmhofer's criterion for the two dimensional
jellium model [DR1-2]. However the proof relies in a key way on the
special momentum conservation rules in two dimensions. In three dimensions
general vertices are not necessarily planar in momentum space. This has
drastic constructive consequences (although perturbative
power counting is similar in 2 and 3 dimensions). In particular it seems
to prevent, up to now, 
any constructive analysis based on angular decomposition
in momentum space. The only existing constructive result for three 
dimensional Fermions relies on the use of a bosonic method 
(cluster expansion) 
together with the Hadamard inequality [MR]. It proves
that the radius of convergence of the theory in a single momentum
slice of the renormalization group analysis around the Fermi surface 
is at least a constant independent of the slice. 

In this paper, we build upon the analysis of [MR], extending it to
many slices. We use a multiscale bosonic cluster expansion
based on a direct space decomposition
of the propagator, which is not the usual momentum decomposition 
around the Fermi sphere. We bound uniformly the sum of
all convergent polymers in the Salmhofer domain $| \lambda \log T| \le K$.
Hence our result is the three dimensional analog of [FMRT] and [DR1]. 
Because of its technical nature, this result is stated precisely 
only in section III.6, after the definition of the multiscale 
cluster expansion.

Using a Mayer expansion 
we plan in a future paper (which would be the three dimensional analog of 
[DR2]) to perform renormalization of the two point subgraphs and to study 
boundedness of the self energy and of its first and second momentum space
derivatives. That would complete the proof of 
Fermi liquid behavior in three dimensions.

Remark however that the optimal analyticity radius 
of the Fermi liquid series should be given by 
$| \la \ln T|= K_{BCS}$ where $ K_{BCS}$ is a numerical 
constant given by the coefficient 
of a so called  ``wrong-way'' bubble graph [FT2].
In this paper we prove analyticity in a domain  $\la |\ln T|\leq  K$
but our constant is not the expected optimal one,  $K_{BCS}$, not
only because of some lazy bounds, but also because of a fundamental
difficulty linked to the use of the Hadamard inequality.
Actually the kind of Hadamard bound relevant for a model
of fermions with two spin states is 
$\sum_n \frac{\la ^n}{n!} \det^2 A_n  \le \frac{n^n}{n!} a^n$, where $A_n$  
is an $n\times n$ matrix whose coefficients are all bounded by $a$. Hence 
(using Stirling's formula),the
radius of convergence in $\la$ of that series is only shown
to be at least $1/ea$ by this bound,
whether $1/a$ would be expected from perturbation theory. 

For this reason it seems to us that 
the analyticity radius obtained by any method based on
Hadamard bound is smaller than the optimal
radius by a factor at least $1/e$, and we do not know how to cure
this defect.

\section{Model}

We consider the simple model of isotropic jellium in three spatial dimensions
with a local four point interaction.
We use the formalism of non-relativistic field theory
at imaginary time of [FT1-2-BG] to describe the interacting 
fermions at finite temperature. Our model is therefore similar to the 
Gross-Neveu model, but with a different, non relativistic propagator.

\subsection{Free propagator}
Using the Matsubara formalism, the propagator at temperature $T$,
$C(x_0,\vec{x})$, is antiperiodic in the variable
$x_0$ with antiperiod ${1\over T}$. This means that the
Fourier transform defined by
\be
 \hat{C}(k)= \frac{1}{2}\int_{-{1\over T}}^{1\over T} dx_0
\int d^3x \; e^{-ikx}\; C(x)
\ee
is not zero only  for   discrete
values (called the Matsubara frequencies) :
\be
k_0 =   \frac{2n+1}{\beta} \pi \ , \quad n \in \ZZ \ , 
\label{discretized}
\ee
where $\beta=1/T$ (we take $ /\!\!\!{\rm h} =k =1$). Remark that only
odd frequencies appear, because of  antiperiodicity.

Our convention is that a four dimensional vector is denoted by
$x = (x_0, \vec{x})$
where $ \vec{x}$ is the three dimensional spatial component.
The scalar product is
defined as $kx := - k_0 x_0 + \vec{k}.\vec{x}$. 
By some slight abuse of notations we may write either
$C(x-\bar{x})$ or $C(x,\bar{x})$, where the first 
point corresponds to the field
and the second one to the antifield (using translation invariance of the
corresponding kernel).

Actually  $\hat{C}(k)$  is obtained
from the real time propagator by changing $k_0$ in $i k_0$ and is equal to:
\be
\hat{C}_{ab} (k) = \de_{ab} \frac{1}{ik_0-e(\vec{k})},
\quad \quad e(\vec{k})= \frac{\vec{k}^2}{2m}-\mu \ ,
\label{prop}
\ee
where $a,b \in \{\uparrow,\downarrow\}$ are the
spin indices. The vector $\vec k$ is three-dimensional. Since our theory
has three spatial dimensions and one time dimension,
there are really four dimensions.
The parameters $m$ and $\mu$ correspond to the effective mass and to
the chemical potential (which fixes the Fermi energy).
To simplify notation we put $2m= \mu=1$, so that, if $\rho=|\vec k|$,
$e(\vec{k})= e(\rho) = \rho^2-1$.
Hence,
\be
C_{ab}(x) =\frac{1}{(2\pi)^3\beta}\; \sum_{k_0} \; \int d^3k\; e^{ikx}\;
\hat{C}_{ab}(k) \ .
\label{tfprop}\ee

The notation $\sum_{k_0}$ means really the discrete sum over the integer
$n$ in (\ref{discretized}).
When $T \to 0$ (which means $\beta\to \infty$) $k_0$
becomes a continuous variable, the corresponding discrete sum becomes an
integral, and the corresponding propagator
 $C_{0}(x)$ becomes singular
on the Fermi surface
defined by $k_0=0$ and $|\vec{k}|=1$.
In the following to simplify notations we will write:
\be
\int d^4k \; \equiv \; {1\over \beta} \sum_{k_0} \int d^3k
\quad , \quad 
\int d^4x \; \equiv \; {1\over 2}
\int_{-\beta}^{\beta}dx_0 \int d^3x \ . \label{convention}
\ee

\subsection{Ultraviolet cutoff}

It is convenient to add a continuous ultraviolet cut-off
(at a fixed scale $\La_{u}$) to the propagator
(\ref{prop}) for two reasons: first because it makes its Fourier transformed
kernel in position space  well defined, and second because a non relativistic
theory does not make sense anyway at high energies. To preserve physical (or
Osterwalder-Schrader) positivity one should introduce this ultraviolet cutoff
only on spatial frequencies [FT2]. However for convenience
we introduce this cutoff both on spatial and on Matsubara frequencies
as in [FMRT]; indeed
the Matsubara cutoff could be lifted with little additional work.
The propagator (\ref{prop})
equipped with this cut-off is called
$C^{u}$ and   is defined as:
\be
\hat{C}^{u}(k) :=  \hat{C}(k)\left .
\left [u(r)\right]\right |_{r=k_0^2+e^2(\vec{k})}
\ee
where the compact support function 
$0\leq u(r)\in{\cal C}_{0}^\infty({\rm R})$
satisfies: $u(r)=1$ for $r\leq 1$,  $u(r)=0$ for $r>10$.
 
\subsection{Position space}

In the following we will use the propagator in position space. The
key point for further analysis is to write it as
\be
C^{u}(\vec{x},t) = \frac{1}{1+|\vec{x}|}\;  \frac{1}{1+f(t)+|\vec{x}|} \; 
F(\vec{x},t) \label{cdecomp}
\ee
where $f(t)$ is defined by
\be
f(t) := \left | {\sin \lp 2\pi Tt\rp \over 2\pi T}\right |
= \vep(t) {\sin \lp 2\pi Tt\rp \over 2\pi T} \quad 
t\in \left [-{1\over T}, {1\over T} \right ]
\label{ftem}\ee
and 
$\vep(t)$ is the sign of $\sin \lp 2\pi Tt\rp$.

This is useful since the remaining function $F$ has 
a spatial decay scaled  with $T$, and no global scaling factor
in $T$, as proved in the following lemma.
\begin{lemma}
For any $p\ge 1$, there exists $K_p$ such that
the function $F(\vec{x},t)$ defined by (\ref{cdecomp}) satisfies
\be
|F(\vec{x},t)| \leq {K_p \over \lp 1+ T|\vec{x}|\rp^p} \quad \quad 
\forall p\geq 1.
\label{cdecdec}
\ee
\end{lemma}
\paragraph{Proof.}
In radial coordinates the propagator is written as
\be
 C^{u}(\vec{x},t) = 
 \frac{T}{(2\pi)^3}\sum_{k_0} \int_0^{2\pi}\hspace{-0.3cm} d\phi
\int_0^\pi \hspace{-0.3cm}d\theta\ \sin \theta\ \int_0^\infty \hspace{-0.3cm}
d\rho \; \rho^2 \;
 \frac{e^{i\rho|\vec{x}|\cos\theta-ik_0 t}}{ik_0-e(\rho)} 
u\left [k_0^2+e^2(\rho) \right ]\ .
\ee
Now we write the integral over $\theta$  as
\be
\int_0^\pi d\theta\; \sin \theta\; e^{i\rho|\vec{x}|\cos\theta} = \int_{-1}^1 dv\;
e^{i\rho|\vec{x}| v}  
\ee
and  applying twice the identity 
\be
e^{i\rho|\vec{x}| v} \; =\;  \frac{1}{(1+|\vec{x}|)}
\lp 1-\frac{i}{\rho}\frac{d}{dv} \rp \; e^{i\rho|\vec{x}| v}
\ee
we obtain
\bqa
\lefteqn{
\int_{-1}^1 \hspace{-0.3cm} 
dv\; e^{i\rho|\vec{x}| v} =
\frac{1}{1+|\vec{x}|}  \int_{-1}^1 \hspace{-0.3cm} dv\; 
\left [ 1-\frac{i}{\rho}\frac{d}{dv} \right ]
e^{i\rho|\vec{x}| v}} \\ 
&&
%\hspace{-0.8cm}
= \frac{1}{1+|\vec{x}|} \left [  \int_{-1}^1\hspace{-0.3cm} 
dv\; e^{i\rho|\vec{x}| v} 
+ \frac{1}{i\rho} \lp e^{i\rho|\vec{x}|} - e^{-i\rho|\vec{x}|} 
\rp\right ]\no\\
&&
%\hspace{-0.8cm}
=\left [ \frac{1}{(1+|\vec{x}|)^2} \int_{-1}^1\hspace{-0.3cm} 
dv\; e^{i\rho|\vec{x}| v} 
\right ]
 +  \left \{
\left [ \frac{(2+ |\vec{x}|)}{(1+|\vec{x}|)^2} \right ]
\frac{1}{i\rho} \lp  e^{i\rho|\vec{x}|} - e^{-i\rho|\vec{x}|} \rp
\right \}\no
\eqa
We decompose further, introducing for the first term 
$1= \chi (|\vec{x}| \le 1) + 
\chi (|\vec{x}| >1)$, where $\chi$ is the characteristic function of the event
indicated, and perform the $v$ integration for the second term only.
In this way the function $F$ can be written as a sum of two terms
$F=F_1+F_2$ where
\bqa
&&\hspace{-1.5cm}
F_1 = \chi (|\vec{x}| \le 1) 
\frac{(1+f(t)+|\vec{x}|) }{(1+|\vec{x}|)}  
\frac{T}{(2\pi)^2}
\no \\&&
\sum_{k_0} 
\int_{-1}^1 \hspace{-0.3cm} dv \int_0^\infty \hspace{-0.3cm} d\rho \; \rho^2\;
e^{i\rho|\vec{x}|v-ik_0 t}{u[k_0^2+e^2(\rho)] \over ik_0-e(\rho)}\; 
\eqa
\bqa
&&\hspace{-1.5cm}
F_2 = \left [  \frac{(2+|\vec{x}|+ \chi (|\vec{x}| > 1)/|\vec{x}|) 
(1+f(t)+|\vec{x}|)}{(1+|\vec{x}|)} 
\right ] 
 \frac{T}{(2\pi)^2}
\no \\&&
\sum_{k_0}\sum_{\si =\pm 1}
\int_0^\infty\hspace{-0.3cm}  d\rho \  
{\si \rho\over i}\ e^{i\si\rho|\vec{x}|-ik_0 t}
\frac{
 u[k_0^2+e^2(\rho)]}{ik_0-e(\rho)}
\eqa
%where $\si\in \{1,-1\}$. 
Now we apply on $F_1$ 
and on $F_2$ the identity
\be
\left [1+f(t)-\frac{i}{2} \vep(t) a_i |\vec{x}|\right] 
e^{i a_i \rho |\vec{x}|  -ik_0 t} =
\left [1+ \vep(t)\lp i\frac{\De}{\De k_0} -\frac{1}{2} 
\frac{d}{d\rho} \rp\right ]
 e^{i a_i \rho |\vec{x}| -ik_0 t} 
\label{tident}\ee
where we defined   $a_1=:v$  for $F_1$ and $a_2=:\si$ for $F_2$, and
where the discretized derivative $\frac{\De}{\De k_0}$
on a function $F(k_0)$ is defined by
\be
\frac{\De}{\De k_0}  F(k_0) = \frac{1}{4\pi T} \left [
F(k_0+2\pi T) - F(k_0-2\pi T)\right ] .
\ee
Hence integrating by parts the $F_i$'s are written as
\bqa
&&\hspace{-2cm}
F_1(\vec{x},t) =   \frac{T}{(2\pi)^2}\sum_{k_0} \; 
\int_{-1}^1 \hspace{-0.3cm} dv f_1 (\vec{x},t,v)
\int_0^\infty \hspace{-0.3cm} d\rho e^{i\rho|\vec{x}|v-ik_0t}
G_1 (k_0, \rho)\; 
\\&&
 G_1(k_0, \rho) 
=  [1+\vep(t) \De ] 
\left [ { \rho^2 \; u\lp k_0^2+e^2(\rho) \rp\over ik_0-e(\rho)}\; 
\right ]\no
\eqa
\bqa
&&\hspace{-2cm}F_2(\vec{x},t) = \frac{T}{(2\pi)^2}
\sum_{k_0}\sum_\si f_2(\vec{x},t,\si)  
\int_0^\infty \hspace{-0.3cm} d\rho \;{\si\over i} 
\;  G_2(k_0, \rho)\;  e^{i\si \rho|\vec{x}|-ik_0 t}\\
&&  
G_2(k_0, \rho) = \left [1+\vep(t) \De
\right ]
\left [ {\rho\; u\lp k_0^2+e^2(\rho) \rp\over ik_0-e(\rho)}\; 
\right ]\no
\eqa
where we have defined 
\be
\label{rcl1}
f_1(\vec{x},t,v) = \chi (|\vec{x}| \le 1) 
\frac{(1+f(t)+|\vec{x}|) }{(1+|\vec{x}|) 
( 1+f(t)-\frac{i}{2}\vep(t) v |\vec{x}| )} 
\ee
\be
f_2(\vec{x},t,\si) = {2+ |\vec{x}| + \chi (|\vec{x}| > 1)/|\vec{x}|
\over 1+|\vec{x}|}
{1+f(t)+ |\vec{x}| \over ( 1+f(t)-\frac{i}{2}\vep(t) |\vec{x}|\si)}
\label{rcl2}\ee
\be
\De = \lp \frac{1}{2} \frac{d}{d\rho} -i\frac{\De}{\De k_0} \rp 
\ee
We remark that these functions are uniformly bounded in modulus ($f_1$
is bounded by 1 and $f_2$ by 6).
The signs and coefficients 
in $\De$ have been optimized in order to
obtain a positive factor $1+f(t)$ and to minimize the action of
$\De$ on $(ik_0-e(\rho))^{-1}$. After a tedious 
but trivial computation, we find
\bqa
&&\hspace{-0.6cm}G_i=:[1+ \ep (t)
\De]\left [ \rho^{b_i}{ u[k_0^2+e^2(\rho)]
\over ik_0-e(\rho)} \;
\right ] = \left\{\rho^{b_i} + {\ep (t) b_i \rho^{b_i-1} \over 2}  
 \right\}{u[k_0^2+e^2(\rho)]  
\over ik_0-e(\rho)}
\no\\ 
&&\hspace{-0.5cm}
+ \rho^{b_i} \ep (t)\left \{ 
 u'[k_0^2+e^2(\rho)] \left [  {2\rho(\rho^2-1)\over 
ik_0-e(\rho)}    
-  { i k_0 (ik_0-e(\rho)) \over \left [ i k_0-e(\rho)\right
]^2 + 4\pi^2 T^2 } \right ] \right .
%+ { (ik_0-e(\rho)) O(T)  \over 
%\left [ i k_0-e(\rho)\right ]^2 + 4\pi^2 T^2  \right ]  } 
\no\\ 
&&\hspace{-0.5cm}     + u[k_0^2+e^2(\rho)] 
{(\rho-1) [ i k_0-e(\rho) ]^2  + 4\pi^2 T^2 \rho \over 
[ i k_0-e(\rho) ]^2 \;
\lp \left [ i k_0-e(\rho)\right ]^2 + 4\pi^2 T^2\rp } 
\no\\ 
&&\hspace{-0.5cm}  \left .   +  { O (T) \over 
\left [ i k_0-e(\rho)\right ]^2 + 4\pi^2 T^2 } 
 \right \}
\eqa
where  $b_1=2$ for $G_1$ and $b_2=1$ for $G_2$.

Using these explicit expressions 
it now easy to check that $F_1$ and $F_2$ are uniformly
bounded by some constant $K$ (independent of $T$ as $T \to 0$).
To complete the proof of Lemma 1, there remains to check
that these functions $F_1$ and $F_2$ also
decay like any power as $T|\vec x| \to \infty$.
For $F_{1}$ there is obviously nothing to check remarking 
the function $\chi(|\vec{x}| \le 1)$ in (\ref{rcl1}).
Hence we have only to prove 
\be
\lp 1 + |\vec{x}|T \rp^p \; |F_2| \; \leq \; K_p 
\ee
for some constant $K_p$ independent from $T$. Since
\be
\lp 1 + |\vec{x}| T \rp^p \; e^{i \si |\vec{x}|\rho} = 
\lp  1-T \frac{i}{\si }\frac{d}{d\rho}\rp^p e^{i \si |\vec{x}|\rho} 
\ee
we have
\be
|\lp 1 + |\vec{x}|T \rp^p F_2(\vec{x},t)| 
\leq K_1\; \sup_{\si = \pm 1}
\sum_{k_0}  \int_0^\infty d\rho
\left |\lp  1+T \frac{i}{\si }\frac{d}{d\rho}\rp^p 
\; G_2(k_0, \rho)
\right |  \label{boubound}
\ee
where we bounded  the factors $|f_i|$ are bounded by  constants. 
Now, performing the change of variable $w=\rho^2-1$, using the
fact that the $u$ function has compact support, and the
fact that the sum over $k_0$ is bounded away from 0
since by (\ref{discretized}) $|k_0 | \ge T$, 
it is a trivial power counting
exercise to check that (\ref{boubound}) is actually bounded by a constant.
\qed

Remark that it is not possible to improve significantly Lemma 1. 
Actually if we try in (\ref{cdecomp}) to obtain e.g.
more factors such as $(1+f(t)+|x|)$, 
identity (\ref{tident}) should be applied
several times and the action of two or more $\De$ operators
on the free propagator (\ref{prop}) would generate terms that
diverge when $T\rightarrow 0$. Similarly, if the factor
$(1+|x|)$ appears more than one time, some corresponding factors 
$f_i$ would not remain bounded when $|x|\rightarrow \infty$. 

\begin{figure}
\centerline{\psfig{figure=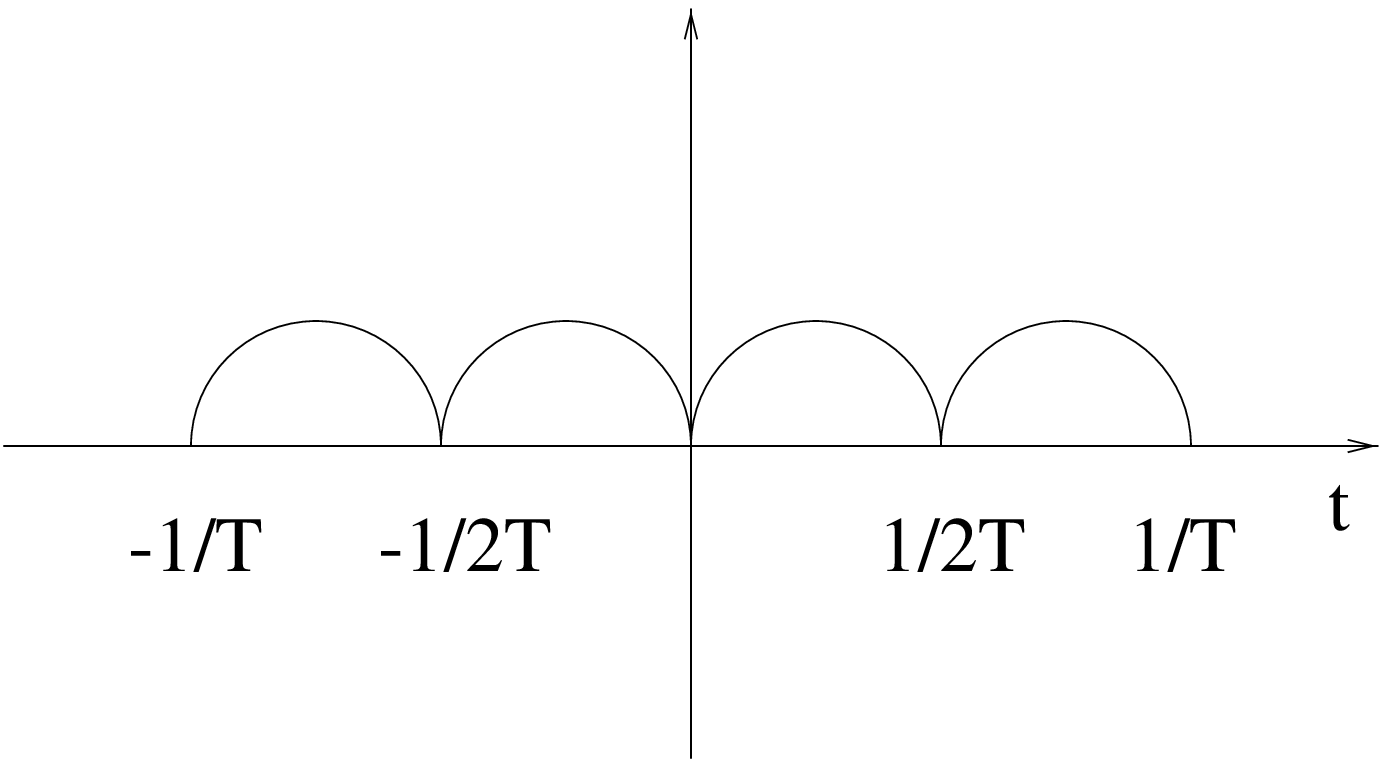,width=8cm}}
\caption{The function $\vep(t) \sin(2\pi T t)$}
\label{had}
\end{figure}

In the following we will use the spatial decay of the propagator
to integrate and  the following lemma will be useful
\begin{lemma}
Let the interval $\biggl[ -{1\over T}, {1\over T}\biggl[ $
be divided into eight sub-intervals 
\be
I_j=: \biggl[-{1\over T} + {j-1\over 4T}, -{1\over T} + 
{j\over 4T} \biggl[ \quad
1\leq j \le 8 
\ee
Then
\be 
\frac{1}{1+f(t)+|\vec{x}|} \leq  \frac{1}{1+{2\over \pi}|t-t_j|+|\vec{x}|} 
\ee
where $t_j=-{1\over T} + {j-1\over 4T}$ for $j$ even number and 
$t_j=-{1\over T} + {j\over 4T}$ for $j$ odd number.
\end{lemma}
\paragraph{Proof.}
Remember that $f(t)=\vep(t) {\sin 2\pi T t\over 2\pi T}$ is 
positive and periodic with period $1/2T$ (see Fig.\ref{had}).

In each interval $I_j$ with $j$ even, the function 
$\vep(t) \sin 2\pi T t$ is higher or equal to the line $4T(t-t_j)$ 
while for $j$ odd it is higher that the decreasing line $-4T(t-t_j)$.
The proof follows\footnote{Splitting $C=\sum_{j=1}^{8}C_{j}$ 
according to which interval we
are in, and taking $t_{j}$ as the new origin, we
could in fact obviously restrict ourselves to proving the main result
of this paper for $j=5$, where $t\ge 0$ and $f(t)$ is simply $t$.}. 
\qed

\subsection{Slice decomposition}

To introduce multiscale analysis we can work directly in position
space. We then write the propagator as 
\be
C^{u}(\vec{x},t) = \sum_{j=0}^{j_M+1}  C^{j}(\vec{x},t) \ ; \quad 
C^{j}(\vec{x},t) = C^{u}(\vec{x},t)\  \chi_{\Om_j}(\vec{x},t) 
\ee
where $\chi_{\Om}(\vec{x},t)$ is the characteristic function of
the subset $\Om\subset R^4$
\bqa
\chi_{\Om} (\vec{x},t) &=& 1 \quad {\rm if} \quad (\vec{x},t)\in \Om\no\\
&=& 0 \quad {\rm otherwise} 
\eqa
and the subset $\Om_j$ is defined as follows:
\be
\ba{rll}
\Om_j  =&  \{ \ (\vec{x},t) \ | \ M^{j-1} \leq 
(1+|\vec{x}|)^{\frac{3}{4}} \; 
 (1+f(t)+|\vec{x}|)^{\frac{1}{4}} <  M^{j} & \} 
\; 0\leq j\leq j_M\\
  =&  \{ \ (\vec{x},t) \ | \ M^{j_M} \leq 
(1+|\vec{x}|)^{\frac{3}{4}} \; 
 (1+f(t)+|\vec{x}|)^{\frac{1}{4}} & \} 
\; j=j_M+1 \\
\ea
\label{Oj2}\ee
where $M>0$ is a constant that will be chosen later. In  Appendix A
we discuss why the relative powers 3/4 and 1/4 for $(1+|\vec{x}|)$
and $(1+f(t)+|\vec{x}|)$ are convenient. 
$j_M$ is defined as the temperature scale 
$M^{j_M}\simeq  1/T$, more precisely
\be
j_M =1+ I\left [ \frac{\ln   \lp T^{-1} \rp }{\ln M}\right  ] 
\label{jM}\ee
where $I$ means the integer part. With these definitions
\be
\sum_{j=0}^{j_M+1}   \chi_{\Om_j}(\vec{x},t) =1\ .
\ee

This decomposition is somewhat dual  to the usual slice decomposition
in momentum space of the renormalization group.
Now, for each slice $j$ we can introduce a corresponding lattice 
decomposition.
We work at finite volume
$\La:=[-\bt,\bt]\times \La'$, where  $\La'$ is a finite volume in the 
three dimensional space. For  $j\leq j_M$ we  
partition $\La$ in cubes of side
$M^j$ in all directions, forming the lattice ${\cal D}_j$. 
For that we introduce the function
\bqa
\chi_\De(x) & =& 1 \qquad {\rm if} \quad x\in \De\no\\
& =& 0 \qquad {\rm otherwise} 
\eqa
satisfying $\sum_{\De\in {\cal D}_j}\chi_\De(x)= \chi_\La(x) $. 
For  $j=j_M+1$ we  
partition $\La$ in cubes of side
$M^{j_M}$ in all directions, forming the lattice 
${\cal D}_{j_M+1}={\cal D}_{j_M}$. 
We define the union of all partitions ${\cal D}= \cup_j {\cal D}_j$.

\paragraph{Auxiliary scales} The function $\chi_{\Om_j}$ 
actually mixes temporal and spatial coordinates. In order 
to sharpen the analysis of $\vec{x}$ and $t$, we will need
later an  auxiliary 
slice decoupling for each scale $j$:
\be
C^{j}(\vec{x},t) = \sum_{k=0}^{k_M(j)} C^{j, k}(\vec{x},t) \ ; \quad
C^{j, k}(\vec{x},t)  =
C^{j}(\vec{x},t) \ \chi_{\Om_{j, k}}(t) 
\label{ascale1}\ee
where, for any $j\leq j_M$ we defined 
\be 
\ba{rll}
\Om_{j, k} = &  \{ \ t \ | \ M^{j+k-1} \leq 
 f(t) <  M^{j+k} & \} \quad k>0\\
 =&  \{ \ t\  |\quad  0 \leq  
 f(t) < M^{j} & \}\quad  k=0\\
\ea
\label{tdec}\ee
and $k_M(j)$ is defined as
\be
k_M(j) =\min \{  j_M-j,\ 3j \}
\ee
The  bound $k\leq j_M-j$ 
is obtained observing that $f(t)\leq M^{j_M}$ in
any case by periodicity. 
The  bound $k\leq 3j$ 
is obtained observing that $(1+f(t))^{1\over 4}\leq M^{j}$.

The case $j=j_M+1$ is special. In this case we must have 
$ 0 \leq  f(t)\leq M^{j_M}$ by periodicity, therefore there is no
$k$ decomposition. Actually we say that $k=0$ and  we define
\be 
\Om_{j_M+1, 0} =   \{ \ t\  |\quad  0 \leq  
 f(t)\leq M^{j_M} \}
\ee

\paragraph{Spatial constraints} For any $j$ and $k$ fixed,
the spatial decay is constrained too. We must distinguish three cases:
\begin{itemize}

\item{} $j< j_M$ and $k>0$: then 
there is a 
 non zero contribution only for
\be
 M^{j-\frac{k}{3}-\frac{4}{3}}\  2^{-\frac{1}{3}} \  
\leq \  
(1 + |\vec{x}|) \ \leq  \ 
  M^{j-\frac{k}{3}+\frac{1}{3}}\
\label{xdec}\ee

\item{} $j\leq j_M$ and $k=0$: then 
there is a non zero contribution only for
\be
 M^{j-\frac{4}{3}}\  2^{-\frac{1}{3}} \  
\leq \  
(1 + |\vec{x}|) \ \leq  \ 
M^{j} 
\label{xdec1}\ee

\item{} $j= j_M+1$: then 
there is a non zero contribution only for
\be
M^{j_M}\  2^{-\frac{1}{3}} \  
\leq \  (1 + |\vec{x}|) 
\ee
\end{itemize}

\paragraph{Power counting and scaled decay of the propagator}
Now for each $j$ and $k$ we can estimate 
more sharply the propagator $C^{jk}$. We distinguish three 
cases:
\begin{itemize}
\item{} 
for $j< j_M$ and $k>0$ we have 

\be
\left |C^{j, k}(\vec{x},t) \right | \ 
\leq \ K_1\  M^{-2j-\frac{2}{3}k} \ M^{\frac{7}{3}}\ 2^{\frac{1}{3}}\
\chi_{j, k}\lp \vec{x},f(t) \rp\label{sdec1}\ee
where the function $\chi_{j, k}$ is defined by
\bqa
\chi_{j, k}(\vec{x},t) &=& 1 \qquad {\rm if \ } \        
|\vec{x}| \leq M^{j-\frac{k}{3}+\frac{1}{3}}\ ,\
f(t) \leq M^{j+k}\no\\
                 &=& 0 \qquad {\rm otherwise \ }
\eqa
and the function $F(\vec{x},t)$ is bounded by $K_p$. 
\item{} 
for $j\leq j_M$ and $k=0$ we have 
\be
\left |C^{j, k}(\vec{x},t) \right | \ 
\leq \ K_1\ M^{-2j} \ M^{\frac{8}{3}}\ 2^{\frac{2}{3}}\
\chi_{j, 0}\lp \vec{x},f(t) \rp\label{sdec2}\ee
where the function $\chi_{j, 0}$ is defined by
\bqa
\chi_{j, 0}(\vec{x},t) &=& 1 \qquad {\rm if \ } \        
|\vec{x}| \leq M^{j}\ ,\
f(t) \leq M^{j}\no\\
                 &=& 0 \qquad {\rm otherwise \ }
\eqa

\item{} 
for $j= j_M+1$  we have 
\be
\left |C^{j_M+1, 0}(\vec{x},t) \right | \ 
\leq \ M^{-2j_M} \ 2^{\frac{2}{3}}\
\chi_{j_M+1, 0}\lp f(t)\rp
\ \frac{K_p}{\lp 1+ M^{-j_M}|\vec{x}|\rp^p}
\label{sdec3}\ee
where the function $\chi_{j_M+1, 0}$ is defined by
\bqa
\chi_{j_M+1, 0}(t) &=& 1 \qquad {\rm if \ } \    
f(t) \leq M^{j_M}\no\\
                 &=& 0 \qquad {\rm otherwise \ }
\eqa
and the spatial decay for $|\vec{x}|$ comes from the
decay of the function $F$ in (\ref{cdecdec}).
\end{itemize}

In the following, the multiscale analysis is essentially performed
using the $j$ index. The auxiliary structure will
be introduced only in section IV. 
In that section we will also need to exchange the sums over $j$ and $k$.
The constraints on the maximal value of $k$,
$k_M(j)$, are then changed into constraints on $j$:
\be
\sum_{j=0}^{j_M+1}  \sum_{k=0}^{k_M(j)} C^{j, k} = 
\sum_{k=0}^{\frac{3j_M}{4}} 
\sum_{j\in J(k)}  C^{j, k}
\ee
where 
\bqa
J(k) &=& [\frac{k}{3},\  j_M-k] \quad {\rm for} \ k>0
\label{ascale3}\\
J(0) &=& [0,\ j_M+1] 
\eqa

\subsection{Partition function}

We introduce now the local four point interaction 
\be
I(\psi,\bpsi) = \la \int_\La d^4x \; (\bar{\psi}_\uparrow \psi_\uparrow)
(\bar{\psi}_\downarrow \psi_\downarrow) =
 \la \int_\La d^4x \; \prod_{c=1}^4 \psi_c \ ,
\ee
where $\psi_c$ is defined as: 
\be
\psi_1= \bar{\psi}_{\uparrow} \quad \psi_2= \psi_{\uparrow}\quad 
 \psi_3= \bar{\psi}_{\downarrow} \quad  \psi_4= \psi_{\downarrow} 
\label{psic}\ee
The partition function is then defined as
\bqa
Z_\La^{u} &=& \int
d\mu_{C^{u}}(\psi,\bpsi)
e^{I(\psi,\bpsi)}
= \sum_{n=0}^\infty \frac{1}{n!} \int
d\mu_{C^{u}}(\psi,\bpsi)
I(\psi,\bpsi)^n\no\\
&=& \sum_{n=0}^\infty \frac{1}{n!} \int
d\mu_{C^{u}}(\psi,\bpsi)
\prod_{v\in V} I_v(\psi,\bpsi)
\eqa
where $V$ is the set of $n$ vertices and $I_v(\psi,\bpsi)$ denotes
the local interaction at  vertex $v$. 
Now we can introduce slice decomposition over fields:
\be
\psi_c = \sum_{j=0}^{j_M+1} %\sum_{k=0}^{k_M(j)} 
\psi_c^{j} 
\ee
hence
\be
I_v(\psi,\bpsi) = \la \sum_{J_v} 
%\sum_{k^v_1,...,k^v_4}  
\int_\La d^4x_v \; 
\prod_{c=1}^4 \psi^{j^v_c}_c
\ee
where $x_v$ is the position of the vertex $v$,  
$J_v=(j^v_1,j^v_2,j^v_3,j^v_4)$ gives the  slice indices for the fields 
hooked to $v$. 
%Finally  $K_v$ gives the set of auxiliary slice indices  
%$0\leq k^v_c\leq k_M(j^v_c)$ $\forall c$.    
Now we write
\be
I(v) = \la \sum_{J_v} 
 \sum_{\De_v}
\int_{\De_{v}} d^4x_v \; 
\prod_{c=1}^4 \psi^{j^v_c}_c
\ee
where $\De_v\in {\cal D}_{0}$ and
\bqa
&&Z_\La^{u} = \sum_{n=0}^\infty \frac{\la^n}{n!} 
\sum_{J_V}
\sum_{\De_V}\label{partfunc}\\
&&\qquad \left [\prod_{v} \int_{\De_{v}} d^4x_v \right]\;\int
d\mu_{C^{u}}(\psi,\bpsi) 
\left [\prod_{v} \lp \prod_{c=1}^4 
\psi^{j_c^v }_c(x_v)\rp \right ]\ , \no
\eqa
where we denoted  any set $\{a_v\}_{v\in V}$ by
$a_V$.

The Grassmann functional integral at the  $n$-th order 
in (\ref{partfunc}) can be  written as a determinant
\be
\int
d\mu_{C^{u}}(\psi,\bpsi) 
\left [\prod_{v} \lp \prod_{c=1}^4 
\psi^{j_c^v}_c(x_v)\rp \right ] = 
 \det M(J_V,\{x_v\})
\label{det}\ee
where $M(J_V,\{x_v\})$ is a
$2n\times 2n$ matrix, whose rows correspond to  fields and whose columns 
correspond
to  antifields. Therefore, for a given vertex $v$, $\psi_1(x_v)$ 
and $\psi_3(x_v)$ correspond to
columns and $\psi_2(x_v)$ 
and $\psi_4(x_v)$ correspond to
rows. The matrix element is then
\be
M_{v c;{\bar v} {\bar c}}  =\ 
\de_{j_{c}^v,j_{{\bar c}}^{{\bar v}}} \ 
%\de_{k_{c}^v,k_{{\bar c}}^{{\bar v}}} \ 
C^{j_c^v}(x_v,x_{{\bar v}}) 
\label{matrelement}\ee
where  $c\in C=: \{2,4\}$ are field indices and ${\bar c}\in {\bar C}=: \{1,3\}$ 
are antifield indices.

\begin{figure}
\centerline{\psfig{figure=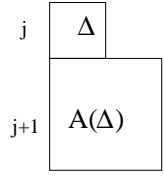,width=2cm}}
\caption{Ancestor}
\label{ancestor}
\end{figure}

\paragraph{Notations}
For each cube $\De$ we denote by $i_\De$ its slice index,
that is $\De\in{\cal D}_j$ with $j=i_\De$.
We  call 
ancestor of any cube $\De\in{\cal D}_j$, ${\cal A}(\De)$, 
the unique cube $\De'\in{\cal D}_{j+1}$ satisfying
$\De\subset \De'$ (see Fig.\ref{ancestor}). In the same way
for any set $S$ of cubes in ${\cal D}_j$, 
we call ancestor of $S$ the set  
${\cal A}(S)=\cup_{\De\in S}{\cal A}(\De)$.
We call $\De_v^j$, the unique cube $\De\in {\cal D}_j$, 
for any  $j\geq i_{\De_v}$, 
satisfying $\De_v\subset \De$ 
(for $j=i_{\De_v}$ we have 
$\De_v^j=\De_v$). (We remark that for 
the moment all $i_{\De_v }=0 \ \forall \De_v$).

\begin{figure}
\centerline{\psfig{figure=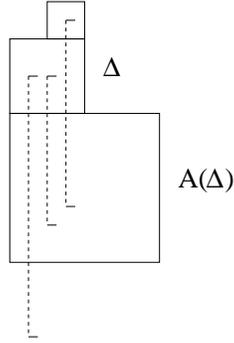,width=3cm}}
\caption{External fields for  $\De$}
\label{link}
\end{figure}

In the following we will denote by  
$h_c^v$ the half-line corresponding to the field 
$\psi_c^{j_c^v}(x_v)$. 
We say that $h_c^v$ is external field for the cube $\De$ 
if $\De_v\subseteq \De$, $i_\De < j_c^v$ and there exist at least one
field  $h_{c'}^v$ hooked to $v$ (different from $h_c^v$)  with
attribution $j_{c'}^v \leq i_\De$ (see Fig.\ref{link}). 
We call  $E(\De)$ %and ${\bar E}(\De)$ 
%respectively 
the set of external fields 
and antifields of $\De$.
In the same way we denote by 
 $E(S)= \cup_{\De\in S} E(\De) $ 
the set of external fields and antifields 
of the subset $S\subset {\cal D}_j$.

We need also to introduce some notations for the fields with smallest
index attached to a vertex $v$. We call $i_v$ 
the smallest scale of the vertex $v$,  $n_v$ 
the number of fields hooked to $v$ with band index $j=i_v$
($1\leq n_v\leq 4$)  and $\si_v$ 
the set of indices of these $n_v$ fields with $j=i_v$, which is 
necessarily non-empty.
Finally we distinguish the particular field in $\si_v$ with lowest value of 
$c$, which we call $c_v$.
\be
i_v = \inf\  \{j_{c}^v\ | \ 1\leq c \leq 4\}\  ;\  
\si_v = \{c \ |\  j_{c}^v = i_v \}\ ;\ 
n_v = |\si_v|\  ; \ c_v = \inf\  \{ c \in \si_v\}
\label{iv}\ee

We say that a vertex $v$ {\em belongs} to a cube $\De\in {\cal D}_j$ 
if $x_v\in \De$, and we denote the corresponding set of
vertices by
\be
V(\De) = \{v \ |\De_v\subseteq \De \} 
\ee
In the same way we denote by $V(S)= \cup_{\De\in S} V(\De) $
the set of  vertices  belonging to 
the subset $S\subset {\cal D}_j$.

We then say that a
 vertex $v$ is  {\em internal} for a cube $\De\in {\cal D}_j$ if
$v$ belongs to $\De$ and $i_v\leq j$. The set of 
internal vertices of $\De$ is therefore defined as
\be
I(\De) = V(\De)\; \cap \ \{v \ |\ i_v\leq j\} 
\ee
We remark that there
may be vertices in $V(\De)\backslash I(\De)$).
In the same way we denote by $I(S)= \cup_{\De\in S} I(\De) $
the set of internal vertices  for 
the subset $S\subset {\cal D}_j$. 
Remark that, if $v\in I(\De)$, then $v\in I(\De')$ for any $\De'$
such that $\De\subseteq\De'$. 

\section{Connected functions}

In order to compute physical quantities, we need to extract
connected functions. For instance ${\cal Z}$ in perturbation theory is the
sum over all vacuum graphs corresponding to the full expansion
of the determinant in (\ref{det}), and we know 
that the logarithm of ${\cal Z}$ is the same sum 
but restricted to connected graphs. 
But while in ordinary graphs the connectedness can be read directly from
the propagators joining
vertices, here we need for constructive reasons to test the connection between 
different cubes in ${\cal D}$ by a multiscale cluster expansion. 
Then the computation of $\log \; {\cal Z}$
is achieved through a Mayer expansion [R].

For this purpose we must introduce two
kinds of connections, vertical connections between cubes at adjacent levels 
$j-1$ and $j$, whose {\it scale} is defined
as $j$, and horizontal connections between cubes at the same level $j$,
whose scale is defined as $j$. (We remark that there is therefore no
vertical connection of scale 0). The difficulty is that our definition
of these connections is inductive, starting from the scale zero towards
the scale $j_M$.

We define a {\em connected polymer} $Y$ 
as a subset of cubes in ${\cal D}$, such that for any two
cubes $\De,\De'\in Y$, there exists a chain of 
cubes $\De_1,...,\De_N\in Y$ such that $\De_1=\De$, $\De_N=\De'$ and
there is a connection between 
$\De_i$ and $\De_{i-1}$ for any $i=2,...,N$.

For each scale $j$ we define connected subpolymers at scale $j$ as 
subsets of cubes belonging to $\cup_{q=0}^j{\cal D}_q$, that 
are connected through connections of scale $\le j$. 
These are the analogs of the quasi-local subgraphs in [R]. 
As for usual graphs, we call $Y^j_k$ ($k=1,...,c(j)$) the   
$c(j)$  connected polymers at scale $j$ and 
$y^j_k$ their restriction to ${\cal D}_j$. 
The set of external fields for  $Y^j_k$ then corresponds to 
the set of external fields for   $y^j_k$, which is denoted by $E(y^j_k)$.

\medskip
\noindent{\bf Connections} 

{\bf 1)} For any pair $\De,\De'$, with   
$\De,\De'\in{\cal D}_j$ and $\De\neq \De'$, we say that 
there is a horizontal connection, or $h$-connection
$(\De,\De')$ between them if there exists
a propagator $C^j(x_v,x_{v'})$ with $\De_v\subseteq \De$ and 
$\De_{v'}\subseteq \De'$ in the expansion of 
the determinant of (\ref{det}). (This definition is not inductive).

It is also convenient to introduce generalized notions:
a "generalized cube" $\tilde\De$ of scale $j$ is a subset
of cubes of scales $j$ and a generalized horizontal connection,
or $gh$-connection $(\tilde\De,\tilde\De')$ is a propagator 
$C^j(x_v,x_{v'})$ with $\De_v\subseteq \tilde\De$ and 
$\De_{v'}\subseteq \tilde\De'$ in the expansion of 
the determinant of (\ref{det}).

{\bf 2)} For each connected subpolymer at scale $j$, denoted by $Y$,
we suppose by induction that we have defined all subconnections for the
subpolymers in $Y$ of scales $\le j$.
We now say that there is a vertical connection between
each cube of $y= Y\cap {\cal D}_j$ 
and its ancestor if one of the following conditions hold:

\begin{itemize}

\item{} we can associate to $y$ a new internal vertex $v$ in $Y$ that
has never been associated previously by the inductive process 
to any other vertical connection at scale $j'\le j$. 
(We remark that several
vertical connections can be associated to a single vertex, and that 
vertical connections can form loops (see Fig.\ref{had3})). In this
case the connection is called a $v$-connection. The set
of all such $v$-connections for a fixed $y$ and 
a fixed new internal vertex $v$ is called the $v$-block
associated to the vertex $v$. 

\begin{figure}
\centerline{\psfig{figure=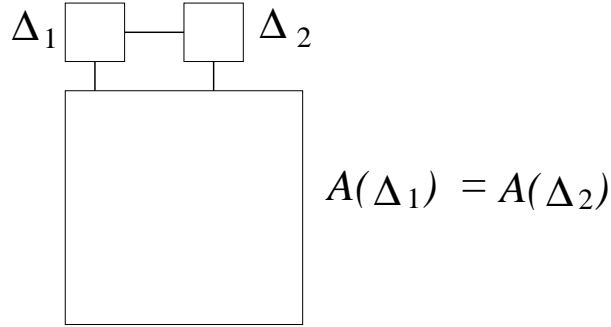,width=8cm}}
\caption{Example of vertical and horizontal connections}
\label{had3}
\end{figure}

\item{} there is no such new internal vertex $v$ for $y$, but
$|E(y)| > 0$. In this
case the connection is called an $f${\it -connection}, and
$|E(y)|$ is called the strength of the connection. The set
of all such $f$-connections for a fixed set of external lines 
is called the $f$-block associated to these external lines. 

\end{itemize}

In fact in this paper we will restrict ourselves to the
analysis and bound for connected subpolymers for which in the
second case, we always have $|E(y)| \ge 6$, since the other cases
need renormalization. 

When there is no vertical connection, i.e. no new vertex, and 
$|E(y)| =0$, we call $Y$ simply a (vacuum) polymer.

\subsection{Polymer structure}
\begin{figure}
\centerline{\psfig{figure=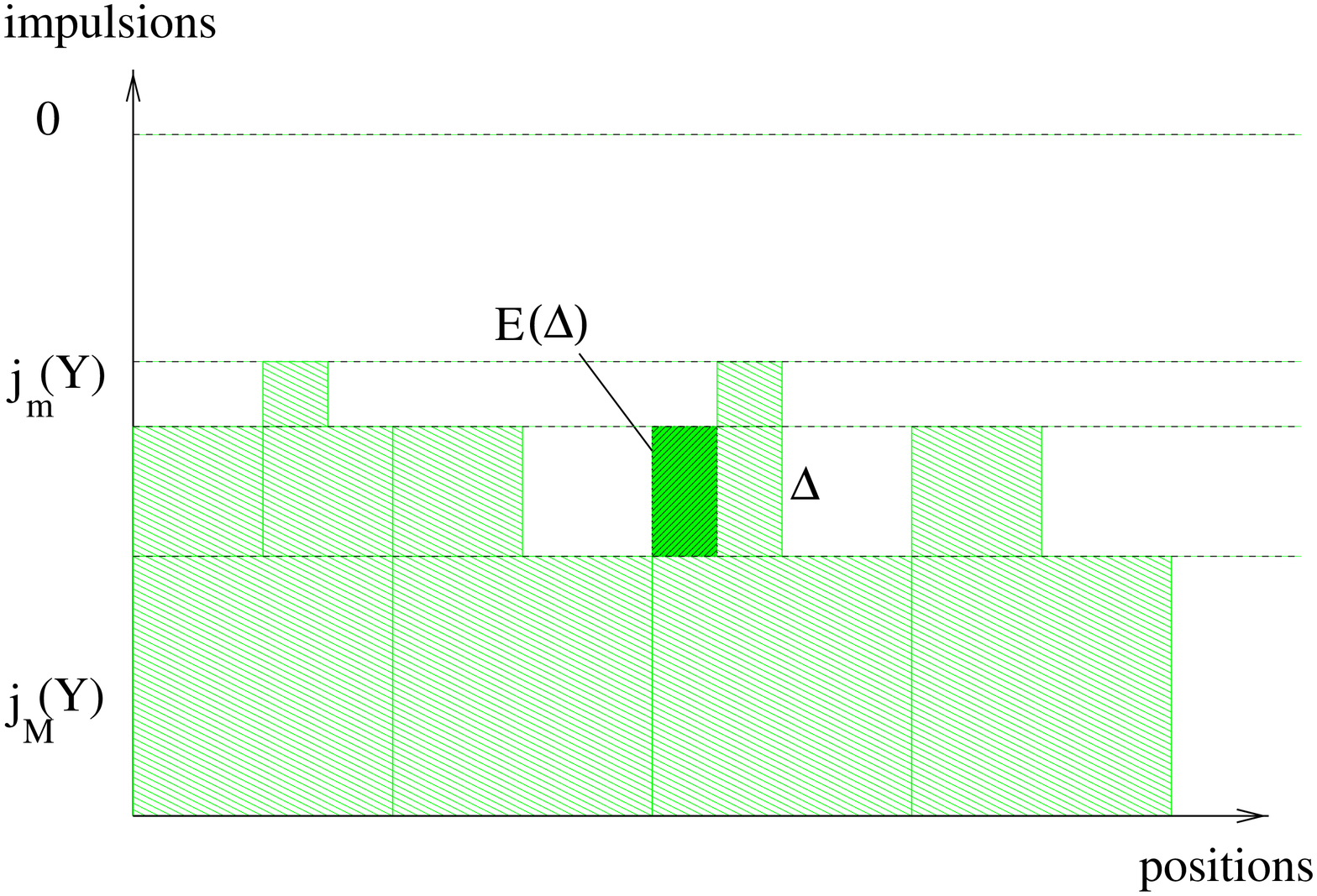,width=8cm}}
\caption{An example of polymer $Y$.}
\label{had1}
\end{figure}

With these definitions in phase space (in our usual representation,
for which index space is vertical) all polymers have a  ``solid on solid'' 
profile (see Fig.\ref{had1})\footnote{This is not the unique possible
choice. In [AR1] polymers with holes or overhangs are allowed.
Here we choose  polymers without holes for simplicity.}.

We define the highest and lowest slice index of each polymer $Y$ as
\be
\ba{rcl}
m_Y &=& \min_{\De\in Y}\  i_\De\\
M_Y &=& \max_{\De\in Y}\  i_\De.\\
\ea
\ee
For each cube $\De\in Y$, we define the ``exposed volume of $\De$''
as
\be
Ex(\De) = \cup_{\left \{ {\De'\in {\cal D} \ {\rm with}\atop  
\De={\cal A}(\De') \ {\rm and}\
\De'\not\in Y}\right \}}\  \De'.
\ee
In other words this is the part of $\De$ that 
contains no other cube of $Y$, and  is therefore at the upper
border of the polymer (see  Fig.\ref{had1}).
An element $\De\in Y$ is called a ``summit cube'' if
$Ex(\De)\neq \emptyset$, and we define the ``border of $Y$'', $B(Y)$,
as the union of all  summit cubes: 
$E(Y) = \cup_{\{\De\ |\  Ex(\De)\neq \emptyset\}}\  \De$. We remark that
$\{Ex(\De)\}_{\De\in B(Y)}$ is a partition
of the volume occupied by $Y$, and the sum over $\De_v$ for any 
$v$ in $Y$  can be written as
\be
\sum_{\De_v\in {\cal D}_0} \int_{\De_v} dx_v = 
\sum_{\De_v\in B(Y)} \int_{Ex(\De_v)} dx_v \ .
\label{locsumcub}
\ee
and we say that the vertex $v$ is {\it localized} in the summit cube
$\De_v\in B(Y) $.

\paragraph{Trees and Forests} 

The connections among cubes in a polymer are the constructive analogs 
of lines in a graph. It is useful to select among these connections a minimal
set i.e. a tree connecting the cubes of the polymer. This is the purpose
of the expansion defined below. 
But we perform this task in two steps. In the main step,
called the multiscale cluster expansion, we select vertices,
external lines and propagators 
which form $v$-blocks, $f$-blocks and $gh$-connections (still containing loops,
see  Figure \ref{had3}); then in a second, auxiliary step, called
the tree and root selection, we
eliminate some redundant connections from the $v$-blocks and $f$-blocks,
and we localize the $gh$-connections into
ordinary $h$-connections, in order to obtain an ordinary
tree connecting all cubes of the polymer; moreover we
select for any subpolymer a particular cube called the root, in a coherent 
way. 

Just like the definition
of the connections, our expansion is inductive. The 
multiscale expansion starts from the slices 
with lowest index towards the ones with higher index. 
The tree and root selection works also inductively but
in the inverse order, from the slices 
with highest index towards the ones with lower index.

In the end the particular 
connections which are selected by the expansion to form the tree 
will be called {\it links} (more precisely $v$-link, $f$-link, 
or $h$-link, if they correspond to a $v$-connection, an $f$-connection, or an 
$h$-connection).

Therefore by construction for each subpolymer $Y^j_k$, 
the set of horizontal and vertical links of scale $j'\le j$ 
forms a subtree $\tree_j$ spanning the subpolymer; and for the union
$\cup_k Y^j_k$ of subpolymers at scale $j$ it forms a forest $\fr_j$
(i.e. a set of disjoint trees). 

The forest $\fr_j$  at scale $j$ is built from
the forest  $\fr_{j-1}$ at scale $j-1$, by adding a set of 
$v$-links or $f$-links of scale $j-1$ and a set of $h$-links of scale
$j$. Therefore $\fr_0\subset \fr_1 \subset...\subset \fr_{j_M+1}:=\fr$
(such a growing sequence of forests is technically called a ``jungle''[AR2]).

\subsection{Multiscale Cluster Expansion}

In this first step we build connected polymers by choosing $v$-blocks,
$f$-blocks and $gh$-links which ensure the connectedness
of the polymer. This is done through Taylor expansions with integral
remainders, inductively from scale 0 to  scale $j_M$.

We build the connected subpolymers at scale
$j+1$, knowing already the connected subpolymers at scales $j'<j+1$.
We perform first the vertical expansion, then the horizontal one,
except for the first slice, for which we start with
the horizontal one.

\subsubsection{Vertical expansion}

For each connected subpolymer $Y^j_k$, we define
$I_j(Y^j_k)$ as the subset of internal vertices  
that have been selected until the step $j$.  
We can also define the union  of
all vertices already selected at scale $j$:
$I_j(\fr_j) = \cup_k I_j(y^j_k)$.  
We extract first the 
$v$-blocks, then the $f$-blocks of scale $j+1$
(all other connections at scale $j'\leq j$ being already fixed).

\paragraph{$v$-blocks}

First we test the existence of a $v$-block associated to a vertex.
We want therefore to know whether
$I(y^j_k)\backslash I_j(y^j_k)\neq \emptyset$
for each  $y^{j}_k$, namely whether there is 
at least one internal vertex $v$ 
that has not already been selected, 

For this purpose we  introduce into (\ref{partfunc})
the identity 
\be
1= \prod_{v\in V\backslash I_j(\fr_j)}
\left [\prod_{c=1}^4 \lp \up^{j}(j^v_c) +  \up^{>j}(j_c^v)\rp \right ] 
\label{vfact}\ee
where we defined 
\be
\ba{lll}
\up^{j}(j_c^v) &= & 1 \quad {\rm \ if\ } \  j_c^v\leq j \\
             &= & 0 \quad {\rm otherwise}\\
\ea
\ee
and $\up^{>j}(j_c^v)= 1 -  \up^{j}(j_c^v)$.
Remark that $v$ is internal
vertex for $\De^j_v$ if there is at least one field hooked to
$v$ with $j_c^v\leq j$. Therefore, to select one  new internal vertex for  
$y_j^k$ we define the function
\be
F(w'_{y_k^j}) = \prod_{v\in V(y)\backslash I_j(y)}\prod_{c=1}^4 
\left [\lp    w'_{y_k^j} \up^{j}(j_c^v) +  \up^{>j}(j_c^v)\rp\right ] 
\ee
The identity (\ref{vfact}) corresponds to $F(w'_{y_k^j}=1)$. Now we apply
the first order Taylor formula:
\be
F(1) = F(0) + \int_0^1 dw'_{y_k^j}  \ F'(w'_{y_k^j})
\ee
where 
\be
F(0) = \prod_{v\in V(y_k^j)\backslash I_j(y_k^j)}    
\left [ \prod_{c=1}^4  
 \up^{>j}(J_v)\right ]
\ee
means there is no new internal vertex for $y_j^k$
(hence $I_j(y_j^k)=I(y_j^k)$),  and we must go to the next paragraph to
test for the existence of external fields ($f$-blocks). 
On the other hand, the integral remainder 
\bqa
F'(w'_{y_k^j}) &=& \sum_{v\in V(y_k^j)\backslash I_j(y_k^j)}  \sum_{\al_v=1}^4 
 \up^{j}(j_c^v) 
\ \int_0^1 dw'_{y_k^j} 
\prod_{c'\neq \al_v}  \lp w'_{y_k^j} \up^{j}(j_{c'}^v) +  
\up^{>j}(j_{c'}^v)\rp  \no\\ 
&&\prod_{v'\in V(y_k^j)\backslash I_j(y_k^j) \atop v'\neq v} 
\left [ \prod_{c=1}^4 \lp w'_{y_k^j}\ 
\up^{j}(j^{v'}_c) +  \up^{>j}(j^{v'}_c)\rp\right ] . \label{alphav}
\eqa
extracts one new  internal vertex for $y_j^k$, choosing the field with
$c=\al_v$ to have $j_c^v\leq j$. To simplify this expression 
we  define
\be
\Up_j(v,c) = \lp w'_{y_v^j}\ 
\up^{j}(j^{v}_c) +  \up^{>j}(j^{v}_c)\rp
\ee
Hence the remainder term is written
\be
F'(w'_{y_k^j}) = \hspace{-0.3cm}\sum_{v\in V(y_k^j)\backslash I_j(y_k^j)}  
\sum_{\al_v=1}^4 
 \up^{j}(j_c^v) 
\ \int_0^1 dw'_{y_k^j} 
\prod_{c'\neq \al_v}  \Up_j(v,c')\hspace{-0.3cm}  
\prod_{v'\in V(y_k^j)\backslash I_j(y_k^j) \atop v'\neq v} 
\prod_{c=1}^4 \Up_j(v',c)  .
\ee
When this remainder term is selected, we have built the $v$-block
corresponding to $y_k^j$ and to the vertex $v$.

This analysis is performed for each connected component $y_j^k$ 
before going on. 

\paragraph{$f$-blocks}

If $w'_{y_k^j}=0$, that is 
$I(y_k^j)\backslash I_j(y_k^j) = \emptyset$,
there is no $v$-block connecting $y_k^j$ to its ancestor, 
therefore we must test for the existence of external fields
($f$-block). 

Fo each $v\in I(y_k^j)$ (actually in this case 
$I_j(y_k^j)=I(y_k^j)$) we can write the sum over field attributions
as follows
\be
\sum_{J_v} = \sum_{n_v,\si_v} \sum_{i_v\in I_v} \sum_{J'_v} 
\ee 
where we recall that $i_v= \min \{j^v_c\ | \ c=1,...,4\}$, $\si_v$ gives
the indices of the fields with $j_c^v=i_v$ and $n_v = |\si_v|$
(\ref{iv}). The attribution $i_v$ can belong 
only to the interval $I_v= [0, l_v]$ where $l_v$ is the scale 
where the vertex $v$ has been associated to a vertical block.
Remark that $l_v\leq j-1$ because
this vertex has been extracted as internal vertex for some
$y_{k'}^{j'}$ with $j'<j$.
Finally $J'_v$ gives the  band indices for the $4-n_v$ fields that
do not belong to the band $i_v$: $j^v_c>i_v$, $\forall c\not\in\si_v$.
Remark that if the field $c=\al_v$ does not
belong to $\si_v$ then it satisfies the constraint
$i_v<\al_v\leq l_v\leq  j-1$. The interpolating function  $F$ is now 
\be
F(w''_{y^j_k}) = 
\prod_{v\in I(y_k^j)} \prod_{c\not\in \si_v \atop j_c^v>j}\ 
w''_{y^j_k}
\ee
We want to extract external lines
until we have convergent power counting. Since in this theory
two and four point functions a priori require renormalization
[FT1-2], we push the Taylor formula in $w''$ to sixth order:
\be
F(w''=1) = \sum_{p=0}^5 F^{(p)}(w''=0) + \int_0^1 dw'' \ F^{(6)}(w'')
\label{taylor6}
\ee
where all terms with $p$ odd are zero by parity and  the  
term  $F^{(p)}(w''=0)$
for $p=0,2,4$  corresponds to the case of 0, 2 and 4 external
fields. Finally  the integral remainder corresponds to the case of  6
external legs or more.
When a field is derived by the Taylor formula at scale $j$, 
hence is chosen as 
external field, its band attribution is constrained to
the set $j_c^v>j$. The highest band is constrained to
$i_v\leq j$, but this was already true because external
fields  only  hook to vertices
that have been extracted at some level $j'\leq j$ 
(therefore $i_v\leq j-1$).

Remark that the same field may be chosen as external field at different
scales.

When any term in (\ref{taylor6}) is selected except the one with $p=0$ 
we build the $f$-block corresponding to $y_k^j$ and to the corresponding 
set of selected external lines, and we say that this $f$-block has 
a corresponding {\it strength} of $p=2, 4,$ or 6\footnote{In part 
II of this study we plan to perform 
renormalization of the two point function and to simply bound logarithmic
divergences such as those of the 4-point function using the smallness
of the coupling constant like in [DR2]. For that purpose we 
need to complicate slightly this 
definition, and to introduce holes in the vertical direction of our polymers
when  $f$-blocks have {\it strength} 2 or 4. These complications
are not necessary here so we postpone them to this future publication.}.

This analysis is again performed for each connected component $y_j^k$ 
before going on. 

\subsubsection{Horizontal expansion}

The extraction of the vertical blocks has fixed a certain set of generalized 
cubes at scale $j+1$, called ${\tilde{\cal D}_{j+1}}$.
The elements of ${\tilde{\cal D}_{j+1}}$ are the connected components at scale
$j+1$, taking into accounts all previous connections,
that is the connections of scale $j'\le j$
and the vertical connections of the $v$ and $f$-blocks
of scale $j+1$ that have just been built.

In order to complete the construction of the connected 
subpolymers at scale $j+1$, we must test horizontal connections 
between these generalized cubes, that is $gh$-connections. 
Extracting these $gh$-connections
actually corresponds to extracting forests made of
such $gh$-connections at scale $j+1$
over these generalized cubes. We denote such a forest by $\fr^h_{j+1}$.
This is done using a so called forest formula.

\paragraph{Forest formula}

To simplify notation we work at scale
$j$ instead of $j+1$.
Forest formulas are  Taylor expansions with integral
remainders which test connections (here the $gh$-connections at
scale $j$)
between $n\geq 1$ points (here the generalized cubes at
scale $j$)
and stop as soon as the final connected components are built.
The result is a sum over forests, a forest being a set of disjoint trees.

We use the {\em unordered Brydges-Kennedy Taylor formula}, which states [AR2]
that for any smooth function $H$ of the $n(n-1)/2$ variables
$u_{l}$,  $l \in P_n = \{(i,j)| i,j\in \{1,..,n\}, i\neq j\}$,
\be
H |_{h_{l}=1} = \sum_{u-\fr} \prod_{q=1 }^{k}
\lp \int_0^1 dw_q \ 
\rp
\lp \prod_{q=1 }^{k}\p{h_{l_{q}}}  H \rp ( h^{\fr}_{l}({w_{q}}), l \in P_n)
\label{bloblo}
\ee
where $u-\fr$ is any unordered forest, made of  $0\le k\le n-1$
lines $l_{1},...,l_{k}$ over the $n$ points. To each line $l_{q}$  $q=1,...,k$
of $\fr$ is associated the parameter $w_{q}$, and to each pair $l=(i,j)$
is associated the weakening factor
$ h^{\fr}_{l}({w_{q}})$. These factors replace the variables
$u_{l}$ as arguments of the derived function $\prod_{q=1 }^{k}\p{h_{l_{q}}} H$
in (\ref{bloblo}).
These weakening factors $ h^{\fr}_{l}({w})$ are themselves functions
of the parameters $w_{q}$, $q=1,...,k$ through the formulas
\bqa
h^{\fr}_{i,i}(w)&=&1\no\\
h^{\fr}_{i,j}(w)&=&\inf_{l_{q}\in P^{\fr}_{i,j}}w_{q}, \quad\quad
\hbox{if $i$ and $j$ are
connected by $\fr$}\no\\&&
\hbox{where $P^{\fr}_{i,j}$ is the unique path in the forest
$\fr$ connecting $i$ to $j$}\no\\
h^{\fr}_{i,j}(w)&=&0 \quad \quad\hbox{if $i$ and $j$ are not
connected by $\fr$}.
\label{w-factor}\eqa
\medskip

In our case, the $H$ function is the determinant,
$P_n $ is the set of pairs of generalized cubes
at scale $j$ 
\be
P_n = \{({\tilde \De},{\tilde \De}') \ | \ {\tilde \De},{\tilde \De}'\in 
 {\tilde {\cal D}}_j   \}
\ee 
We apply the forest formula
(\ref{bloblo}) at scale $j$ and we denote the corresponding
forest by $\fr^h_j$. 
Therefore the interpolation parameter $h^{\fr^h_j}_{{\tilde \De}{\tilde\De'}}$ 
is inserted besides the matrix  element defined in  (\ref{matrelement}):
\be
M_{v c;{\bar v} {\bar  c}}  = 
\de_{j_{c}^v,j_{\bar c}^{\bar v}} \
\left [ C^{j}_{\De^j_v,\De^j_{\bar v}}(x_v,x_{\bar v})\right]_{j=j_c^v} 
\ee
where we defined 
\be
C^{j}_{\De^j_v,
\De^j_{\bar v}}(x_v,x_{\bar v}) 
=:
\chi_{\De^j_v}(x_v)  \ 
C^{j}(x_v,x_{\bar v}) \ 
\chi_{\De^j_{\bar v}}(x_{\bar v})  
\ee
and $\chi_\De(x)$ is the characteristic function of $\De$, defined 
by: $\chi_\De(x)=1$ if $x\in \De$ and $\chi_\De(x)=0$ otherwise. 
The interpolated matrix element, for any  $j_c^v=j$ is then 

\be
 M_{v c; {\bar v} {\bar  c}} (h^{j_c^v}_{{\tilde \De},{\tilde \De}'}\  ) 
=\de_{j_{c}^v,j_{\bar c}^{\bar v}}  \
\left [ 
h^{j}_{{\tilde \De}^j_v,{\tilde \De}^j_{\bar v}}\  C^{j}_{\De^j_v,
\De^j_{\bar v}}(x_v,x_{\bar v}) \right]_{j=j_c^v}
\ee
where we defined ${\tilde \De}^j_v$ as the unique generalized 
cubes containing $\De^j_v$, and write for simplicity 
$h^{j}_{{\tilde \De}^j_v,{\tilde \De}^j_{\bar v}}$
instead of  $h^{\fr^h_j}_{{\tilde \De}^j_v,{\tilde \De}^j_{\bar v}}$.

\subsection{Tree and root selection}

\paragraph{Localization of the $gh$-connections}

We now fix, for each
field $h$ or antifield ${\bar h}$
hooked to a vertex $v$, whether it belongs or not to a 
propagator derived by the horizontal expansions (since this costs only
a factor 2 per field or antifield, hence a factor 16 per vertex). 
As we know the position of $\De_v$ for
any $v$, we know exactly for each ${\tilde\De}$ in $y_k^j$ 
the set of $h$, ${\bar h}$ that form at scale
$j$ (as $j^b_h=j$) the propagators of the tree
$T_{jk}$. We denote this set by $b({\tilde\De})$.

The first, rather trivial step, consists in replacing each 
$gh$-connection between generalized cubes by an ordinary $h$-link
between ordinary cubes. This means, in the propagator 
$\chi_{\tilde  \De}C^{j}\chi_{\tilde \De '}$ corresponding
to the  $gh$-connection, that we expand the characteristic functions
as $\chi_{\tilde \De}=\sum_{\De \in {\cal D}_{j}, \De \subset \tilde \De}
\chi_{\De}$, 
and $\chi_{\tilde \De'}=\sum_{\De' \in {\cal D}_{j}, \De' \subset \tilde \De'}
\chi_{\De'}$. Accordingly the $gh$-connection is localized into
an ordinary connection, or $h$-link between $\De$ and $\De'$
\footnote{The corresponding sums are bounded below in two steps:
in the first step, at the beginning of section III.4, the set
$b$ of the fields for the $h$-links is chosen (and paid in section IV.7.3), 
and in section IV.6 the contraction between these fields is performed 
(construction of $T_{jk}$). 
Since in section III.4 the position of all the fields is known,
together these two steps pay for the localization of 
$gh$-connections into ordinary connections.}.

\paragraph{Choice of the roots}

Remember that at each scale $j$ each connected subpolymer $y^j_k$
is actually made of a set of disjoint generalized cubes 
$\tilde \De$.We want now to choose one generalized cube ${\tilde\De}_{root}$
in each $y^j_k$, called the root of the subpolymer, and one particular
cube $\De_{root}$ in each generalized cube $\tilde \De$  
called the root of the generalized cube.

The root cube in ${\tilde\De}_{root}$
is special: it will correspond to the root cube of the whole subpolymer, 
therefore we will denote it by $\De^0_{root}$.

Finally, in each $y^j_k$, for each  
${\tilde\De}\neq {\tilde\De}_{root}$,
we want to choose one field or antifield in $b({\tilde\De})$ as the 
one contracting towards the
root in $T_{jk}$ and we call it $h_{root}$
(the vertex to which it is hooked being called $v_{root}$).
We call then $R_{root}$ the set of all $h_{root}$ for
all generalized cubes at all the different scales.

Remark that the choice of the set 
$R_{root}$ can be performed only after the choice of ${\tilde \De}_{root}$.
The set of  remaining fields in  $b({\tilde\De})$ is denoted by 
 $l_b({\tilde\De})$ (and called the leaves for ${\tilde\De}$).
Remark that for  ${\tilde\De}_{root}$ all fields are leaves:
 $b({\tilde\De})=l_b({\tilde\De})$. 

The roots are chosen inductively scale by scale, from bottom up, 
starting by the biggest index scale $M_Y$ of the polymer
and going up until the smallest index $m_Y$, 
To break translation invariance, we need to assume from now on
that the polymer $Y$ contains a particular point, namely the origin $x=0$. 

At the biggest scale we have only one connected component, 
that must contain the origin $x=0$. Therefore we choose 
${\tilde\De}_{root}$ as the unique 
${\tilde\De}$ containing $x=0$, and $\De_{root}=\De^0_{root}$   
as the unique cube  $\De\in{\tilde\De}_{root}$ containing $x=0$.
Now for each ${\tilde\De}\neq {\tilde\De}_{root}$ 
we define  $\De_{root}$ as the (necessarily unique) cube $\De\in {\tilde\De}$
containing a field $h_{root} \in R_{root}$ of that scale.

With these definitions we can introduce the general 
inductive rule. 
We assume that all  $\De_{root}$ and ${\tilde\De}_{root}$
have been defined until the scale $j$. We now want
to define the roots at scale $j-1$.

Remark that each connected component $y_k^{j-1}$ actually
corresponds to some generalized cube ${\tilde\De}_0$ 
at scale $j$. We denote by $\De_0$ its
root cube. Now we distinguish two cases:
\begin{itemize}

\item{} there exists a cube $\De_1\in y_k^{j-1}$ with $\De_1\subseteq \De_0$
which contains either 0 or one $h_{root}$ at some scale $j'\geq j$.
Remark that this $\De_1$ must be unique.
Then we define as  ${\tilde\De}_{root}$ for  $y_k^{j-1}$
the unique  ${\tilde\De}$ with  $\De_1\subseteq {\tilde\De}$.
Now for all  ${\tilde\De}\neq {\tilde\De}_{root}$
we introduce $h_{root}$ and $\De_{root}$ exactly 
as in the case of the lowest band $M_Y$.
Finally for ${\tilde\De}_{root}$ we choose 
$\De_1$ as root cube: $\De_1=\De^0_{root}$.

\item{} there is no cube  $\De_1\in y_k^{j-1}$ with  
$\De_1\subseteq \De_0$ with $0\in \De_1$ or
 $\De_{v_{root}}\subseteq \De_1$  for some $h_{root}$ at
a lower scale. Therefore 
we choose as root one of the  ${\tilde\De}\in y^{j-1}_k$ satisfying 
${\tilde\De}\cap \De_0\neq \emptyset$ (remark that there must be at
least one of such ${\tilde\De}$ by construction).
For all  ${\tilde\De}\neq {\tilde\De}_{root}$
we introduce $h_{root}$ and $\De_{root}$ exactly 
as in the case of the lowest band $M_Y$.
Finally for ${\tilde\De}_{root}$ we choose as 
$\De^0_{root}$  one of the cubes satisfying
$\De\subseteq \De_0$ (there must be at least one by construction).

\end{itemize}

\begin{figure}
\centerline{\psfig{figure=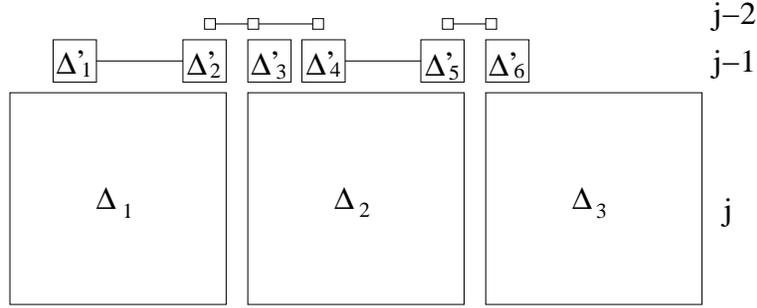,width=10cm}}
\caption{Construction of roots}
\label{had2}
\end{figure}

For an example see Fig.\ref{had2}, where cubes of
three scales are shown. The lines connecting two
cubes are are $h$-links. The union $\De_1\cup\De_2\cup\De_3$
is a generalized cube at scale $j$ (corresponding to  ${\tilde\De}_0$
above). From the figure one can see that
there are three generalized cubes at scale $j-1$:
\bqa
{\tilde\De}_1 &=& \De'_1\no\\
{\tilde\De}_2 &=& \De'_2\cup \De'_3\cup  \De'_4 \no\\
{\tilde\De}_3 &=& \De'_5\cup \De'_6
\eqa
Now, let us say that  
${\tilde\De}_0$ is a root at scale $j$, $\De_2$ 
is the corresponding root cube and $0\not\in \De_2$ and 
no $h_{root}$ has vertex in $\De_2$. Then we have two
equivalent choices for ${\tilde\De}'_{root}$ as 
${\tilde\De}'_2\cap \De_2\neq \emptyset$  and 
${\tilde\De}'_3\cap \De_2\neq \emptyset$. Let
us take  ${\tilde\De}'_{root}={\tilde\De}'_2$. Now inside 
${\tilde\De}'_2$ we have again  two equivalent
choices for $\De'_{root}$ as $\De'_3$ and $\De'_4\subset \De_2$. 

\paragraph{Choice of the $v$-links and $f$-links}

Remember that in order
to avoid loops, each time several cubes in  $y^j_k$ 
have the same ancestor we must choose only one of them in the block to 
bear a link (either of $v$ or $f$ type). The choice of this cube is 
completely arbitrary (for instance choose the first ones in some
lexicographic ordering of the cubes), except for one constraint. 
Actually, for each connected
subpolymer $y$ the root cube  $\De_{root}^0$
acts as root for $y$, therefore we decide to always choose
as vertical link $(\De_{root}^0,{\cal A}(\De_{root}^0))$. 
All other choices are arbitrary.
This constraint is useful because in the following all the 
vertical power counting for $y^j_k$ will be concentrated on 
this special vertical link  $(\De,{\cal A}(\De))$
($\De=\De_{root}^0$).

At the end of this selection process we have therefore an ordinary tree
of either $v$, $f$ or $h$ links connecting together all cubes
of $Y$. 

\subsection{Result of the expansion}

As a  result of this inductive process 
we obtain the following expression

\bqa
&&\hspace{-0.6cm} Z_\La^{u} = \sum_{n=0}^\infty \frac{\la^n}{n!} 
\sum_{\De_V}
  \sum_{\fr} \sum_{V_d, \al_{V_d}} \sum_{a,b,R}
\sum_{l_{V_d}} \sum_{\{J_h^a\},\{J_{\bar h}^a\}}
\sum_{\{j_h^b\},\{j_{\bar h}^b\}} \sum_{C_b}  \vep_\fr 
\left [\prod_{v} \int_{\De_v} d^4x_v \right] \ \no\\ 
&& \hspace{-0.4cm} 
 \left [\prod_{j=0}^{j_M+1}
\lp \prod_{l\in hL_j}\int_0^1 dw_{l}
 \rp\right] \  
\left [\prod_{j=1}^{j_M+1}
\lp \prod_{l\in vL_j}\int_0^1 dw'_{l}
 \rp\right] \
\left [\prod_{j=1}^{j_M+1}
\lp \prod_{l\in fL^6_j}\int_0^1 dw''_{l}
 \rp\right] \no\\
&&\hspace{-0.4cm}  \left [\prod_{j=0}^{j_M+1}\lp
\prod_{l\in hL_j} 
C^{j}_{\De_{l}{\bar \De}_{l}}\lp x_l,{\bar x}_l\rp\rp 
\right ]\;
\left[
\prod_{v\in V_d} \lp  \sum_{n_{v}\si_{v}\rho_v}
\sum_{i_{v}\in I_v} \sum_{ J^{'}_{v}}\rp\right] \ 
\left[
\prod_{v\in {\bar V}_d} \sum_{ J_{v}}\right]  \no\\
&& \hspace{-0.4cm} \left[\prod_{v\in V_d}\lp
\up^{> j_m(v)}(j_{\al_v}^v) \ 
\up^{l_v}(j_{\al_v}^v)\  
\prod_{j=0}^{l_v-1}
\Up_j(v,\al_v) \rp\right] 
\no\\&& \hspace{-0.4cm} 
\left[\prod_{v\in V_d}
 \prod_{c\neq \al_v} \ \lp 
\up^{> j_m(v)}(j_{c}^v) \ 
\prod_{j=0}^{l_v} \ 
\Up_j(v,c)\rp\right] \ 
\no\\&& \hspace{-0.4cm}
\left[ \prod_{v\in {\bar V}_d}  
  \prod_{c=1}^4 \lp
\up^{> j_m(v)}(j_{c}^v) \ 
\prod_{j=0}^{j_M} 
\Up_j(v,c)\rp \right]
\no\\&& \hspace{-0.4cm} 
\left [\prod_{v\in V_d}  \prod_{c\not\in \si_v} \lp
\prod_{j=0}^{j_c^v-1} s_j(v,c)\rp\right] \ \ 
\det M'\lp \{w_l\}\rp
\label{biggdef}
\eqa
where 
\begin{itemize}

\item{} $V_d = \ \{ v\in V\  |\  \exists$   one $v$-link  
associated to $v\  \}$ and ${\bar V}_d= V \backslash V_d$;

\item{} $a = \ \{ h_c^v \ | \ v\in V_d $ and $h_c^v$ is associated to
some $f$-links at one or several scales\};  

\item{} $b = \ \{ h_c^v \ | \ h_c^v $  is associated to one  h-link $\}$;  

\item{} $R\ =\ R_{root} = \ \{ h_c^v \ | \ h_c^v $  
is a root field or antifield $\}$;

\item{} $l_{V_d} = \{l_v\ | \ v\in V_d\}$ where
$l_v+1$ is the scale of the $v$-links associated to  $v$ (they are all
at the same scale);

\item{} $J^a_h$ is the set of scales $j$ where the field $h$ is associated
to a $f$-link: for each $j\in J_h^a$ $h_c^v$ is external field for $y_v^j$.
The same definition holds for ${\bar h}$;

\item{} $j^b_h$  is the scale of the $h$-link  associated to $h$.
 The same definition holds for ${\bar h}$;

\item{} $C_b$ fixes the pairs $h-{\bar h}$ that form the $h$-links;

\item{} $\vep_\fr$ is a sign coming from the horizontal forest
formulas;

\item{} $hL_j$ is the set of $h$-links of scale $j$ in $\fr_j$. For 
each $h$-link $l$ we denote the corresponding field, antifield
by $h_l$, ${\bar h}_l$. The  vertices are denoted by 
$v(l)$ and ${\bar v}(l)$,
their  positions by $x_l$ (${\bar x}_l$) and 
the cubes of the link containing 
them by $\De_{l}$ and ${\bar \De}_{l}$. 

\item{} $vL_j$ is the set of vertical links  of scale $j$
associated to a vertex. We recall that each such 
vertex corresponds to a set of
$v$-links in $\fr_j$ connecting some subset $y$ at scale $j-1$
(which is already connected by $\fr_{j-1}$) to its ancestor;

\item{} $fL_j^p$ is the set of vertical links  of scale $j$
associated to $p$ external fields. We recall that each such 
set of external fields corresponds to a set of
$f$-links  of scale $j$ and order $p$ ($p=2,4,6$)
in $\fr_j$ connecting some subset $y$ at scale $j-1$
(which is already connected by $\fr_{j-1}$) to its ancestor;

\item{} $w'_l= w'_{y_k^j}$ where $l$ is the $v$-links connecting $y_k^j$
to its ancestor. The same definition holds for $w''_l$;

\item{} 
Defining 
\be
\hspace{-0.3cm}
\left\{
\begin{array}{ll}
j_m(v)=\max \{j\ | \ y_v^j {\rm \ connected\ 
to}\ {\cal A}(y_v^j) {\rm \   by\ a\ } f{\rm -link} \}
& \mbox{if $v \in \bar V_d$,} \\
j_m(v)=\max \{j<l_v\ |\  y_v^j {\rm \  connected\ 
to}\ {\cal A}(y_v^j) {\rm \   by\ a\ } f{\rm -link} \} & 
\mbox{if $v \in V_d$,} \\
\end{array} \right.
\label{defjmv}
\ee
we must have, for all $h^v_c$, $j_c^v > j_m(v)$.
This bound can be understood as follows: 
a vertex $v$ cannot have $i_v\leq j_m(v)$.
Indeed otherwise it 
would be internal for $y_v^{j_m(v)}$, and would have been chosen at
that scale instead of the $f$-link connecting $y_v^{j_m(v)}$ 
to its ancestor. We remark that for $v\in V_d$ this argument
only applies for scales $j< l_v$, since after 
$l_v$  the vertex can no longer be selected as a vertical connection.
This explains the definition (\ref{defjmv}).
All these constraints are expressed in formula (\ref{biggdef})
by the function $\up^{> j_m(v)}(j_{c}^v)$.

Moreover, for each $v\in V_d$ we have inserted an additional sum
\be
\sum_{\rho_v} = \prod_{\{h^v_c\ | \ c \neq c_v\}} \sum_{\rho_h}
\ee
where we recall that $c_v = \min \{c\in \si_v\}$ (\ref{iv}), and
we define $\rho_h=1$ if $i_v\leq j_h\leq l_v$ and 
$\rho_h=2$ if $l_v< j_h$. Remark that for $c\in \si_v$ and 
$c\neq c_v$, or for $c=\al_v$, we must have $\rho_{h_c^v}=1$
by construction (Recall that $\al_v$ is defined in (\ref{alphav})).  
On the other hand, if $h\in a$ we must have 
$\rho_{h}=2$ by construction.
\item{}
the values of $s_j$ depend on the $f$-links:  
\begin{itemize}
\item[-] $s_j(v,c) = 1$ if $y_v^j$ is connected to its ancestor
by a $v$-link or if $j\in J^a_{h_c^v}$ (which means $h_c^v$ is associated
to a $f$-link connecting  $y_v^j$  to its ancestor);
\item[-] $s_j(v,c) = w''_{y_v^j}$ if  $y_v^j$ is connected to its ancestor
by a $f$-link of order 6 and $j\not\in J^a_{h_c^v}$;
\item[-] $s_j(v,c) = 0$ if $y_v^j$ is connected to its ancestor
by a $f$-link of order 2 or 4 and  $j\not\in J^a_{h_c^v}$. 
\end{itemize}

\item{} finally  $\det'$ is the determinant remaining after the 
propagators corresponding to $h$-links have been extracted. The matrix
element is 
\be
M'_{vc;{\bar v}{\bar c}}\lp \{w_l\}\rp =
  \de_{j_{c}^v,j_{\bar c}^{\bar v}}\   
\left [ h_{\De^j_v,\De^j_{\bar v}}^{\fr^h_{j}}(w)  
 C^{j}_{\De^j_v,
\De^j_{\bar v}}(x_v,x_{\bar v}) \right]_{j=j_c^v}
\ee
where the weakening factor
$h_{\De^j_v,\De^j_{\bar v}}^{\fr^h_{j}}(w)$ is defined in 
(\ref{w-factor}), substituting in the formulae 
the general forest $\fr$ with 
the horizontal forest $\fr^h_j$.

\end{itemize}

\paragraph{Constrained attributions}
The non zero contributions are given by the
following attributions:

\noindent {\bf -} for $v\in V_d$ and $c\in\si_v$ we must have 
\be
i_v\in I_v^c = [1+j_m(v)\ , \ l_v ]
\ee

\noindent {\bf -} for $v\in V_d$ and $c\not\in\si_v$ we must have
\bqa
j_c^v \in J^{'c}_v &=&  [i_v \ , \ l_v ] \quad {\rm for } 
\quad c=\al_v \no\\
j_{c}^v \in J^{'c}_v &=&  [j_m(v,c) \ , \ j_M(v,c) ] \quad c\neq \al_v 
\eqa
where
\bqa
j_m(v,c) &=& 1+ i_v \quad {\rm if} \quad h_c^v\not\in a\; {\rm and}\;
\rho_{h_c^v}=1\no\\
j_m(v,c) &=& 1+ l_v \quad {\rm if} \quad h_c^v\not\in a\; {\rm and}\;
\rho_{h_c^v}=2\no\\
j_m(v,c) &=& 1+ 
\max \{j\in J^a_{h_c^v} \} 
 \quad {\rm if} \quad h_c^v\in a
\eqa
and  $j_M(v,c)= l_v$ if $h_c^v\not\in a$ and $\rho_{h_c^v}=1$, otherwise
$j_M(v,c) = \min \{ j>i_v\ | \ y_v^j$ is connected to its 
ancestor by a $f$-link of order $p=2,4 \ \}$, with the convention
that $\min \emptyset = j_M+1$. Remark that $\al_v$ satisfies  a
special constraint because this is the field derived in order 
to extract a $v$-link at scale $l_v+1$, therefore it must satisfy
$j_{\al_v}^v\leq l_v$;

\noindent {\bf -} finally, for $v\in {\bar V}_d$ we must have
\be
j_c^v \in J^{c}_v =  [j_m(v)+1 \ , \ j_M+1 ] .
\ee

\paragraph{Reinserting attribution sums inside the determinant}
This is a key step for later bounds.
We observe that for all $v\in {\bar V}_d$ the constraints $\up^j$
and $\up^{>j}$ on the
attributions for each field hooked to $v$ are independent.
Therefore we can reinsert all the sums inside the determinant
(bringing with them the corresponding vertical weakening factors
$w'$ and $w''$).

On the other hand, for  $v\in V_d$, the sum over attributions
for $h_c^v$ with $c\not\in\si_v$ are independent from each other
but are all dependent from $i_v$. Therefore we can reinsert in
the determinant the sums for  $c\not\in\si_v$ (with their vertical
weakening factors), but we must keep the sum over $i_v$ outside
the determinant. The weakening factors for all $c\neq c_v$ are inserted
in the determinant. On the other hand for the particular
field $h_{c_v}^v$ we keep outside the determinant the weakening 
factors $w'$, as they will be used to perform certain sums, and
reinsert the others in the determinant.

Therefore we can write the partition function as
\bqa
&&\hspace{-0.5cm} Z_\La^{u} = \sum_{n=0}^\infty \frac{\la^n}{n!} 
\sum_{\De_V}
 \sum_{\fr} \sum_{V_d, \al_{V_d}} \sum_{a,b,R} 
\sum_{l_{V_d}} \sum_{\{J_h^a\},\{J_{\bar h}^a\}}
\sum_{\{j_h^b\},\{j_{\bar h}^b\}} \sum_{C_b}  \vep_\fr 
\left [\prod_{v} \int_{\De_v} d^4x_v \right] \ \no\\ 
&& \hspace{-0.4cm} 
 \left [\prod_{j=0}^{j_M+1}
\lp \prod_{l\in hL_j}\int_0^1 dw_{l}
 \rp\right] \  
\left [\prod_{j=1}^{j_M+1}
\lp \prod_{l\in vL_j}\int_0^1 dw'_{l}
 \rp\right] \
\left [\prod_{j=1}^{j_M+1}
\lp \prod_{l\in fL^6_j}\int_0^1 dw''_{l}
 \rp\right] \no\\
&&\hspace{-0.4cm} 
\left[
\prod_{v\in V_d} \lp  \sum_{n_{v}\si_{v}\rho_v}
\sum_{i_{v}\in I^c_v} \rp\right]  
 \left [\prod_{v\in V_d\atop c_v\neq \al_v} 
 \prod_{j= i_v}^{l_v} 
w'_{y_v^j}\right]   
\left [\prod_{v\in V_d\atop c_v= \al_v} 
 \prod_{j= i_v}^{l_v-1} 
w'_{y_v^j}\right]  
\no\\
&&\hspace{-0.4cm}
\left [\prod_{j=0}^{j_M+1}\lp
\prod_{l\in hL_j} 
C^{j}_{\De_{l}{\bar \De}_{l}}
\lp x_l,{\bar x}_l,\{w'_{l'}\},\{w''_{l'}\}\rp\rp 
\right ]
\det M''\lp \{w_l\}, \{w'_l\},\{w''_l\}\rp\no\\
&& 
\eqa
where 
the matrix element is
\bqa
\lefteqn{M''_{vc;{\bar v}{\bar c}}\lp \{w_l\},\{w'_l\},\{w''_l\}  \rp =
  \left [ \sum_{j_c^v\in {\cal I}_c^v}  W_{vc}(j_c^v) \right] }\\   
&&\de_{j_{c}^v,j_{\bar c}^{\bar v}}\ 
\left [
 h_{\De^j_v,\De^j_{\bar v}}^{\fr^h_{j}}(w)  \ 
C^{j}_{\De^j_v,
\De^j_{\bar v}}(x_v,x_{\bar v}) \right]_{j=j_c^v}
\left [ \sum_{j_c^v\in {\cal I}_{\bar c}^{\bar v}}   
W_{{\bar v}{\bar c}}(j_{\bar c}^{\bar v})\right] \no
\eqa
and the horizontal propagator is
\be
C^{j}_{\De_{l}{\bar \De}_{l}}\lp x_l,{\bar x}_l,\{w'_{l'}\},\{w''_{l'}\}\rp =
W_{v_lc_l}(j) \  C^{j}_{\De_{l}{\bar \De}_{l}}\lp x_l,{\bar x}_l\rp \
W_{{\bar v}_l{\bar c}_l}(j)
\ee
and $v_l,c_l$ and ${\bar v}_l,{\bar c}_l$ identify respectively
the field and the antifield of the link.
We defined
\bqa
{\cal I}_c^v &=& \{i_v\} \qquad  v\in V_d, c\in \si_v\no\\
{\cal I}_c^v &=&  J^{'c}_{v}  \qquad  v\in V_d, c\not\in \si_v\no\\
{\cal I}_c^v &=&  J^c_{v} \qquad  v\in {\bar V}_d \label{Idef}
\eqa
and $I^c_v$, $J^{'c}_v$ and $J^c_v$ is the set of band attributions
with the constraints due to the forest structure that we introduced above.
Finally the definitions for the factors $W_{vc}$ are given below.

\paragraph{Vertical weakening factors}
The expression for $W_{vc}(j_c^v)$
is given by the $\Up_j(v,c)$ and $s_j$ functions.
Remark that 
\bqa
\Up_j(v,c) &=& 1 \qquad \quad  {\rm if}\ j< j_c^v\no\\
\Up_j(v,c) &=& w'_{y_v^j} \qquad {\rm if} \ j\geq j_c^v
\eqa
Actually we have to distinguish different cases.

If $v\in V_d$, $c=\al_v$ and $c\neq c_v$  
\be
W_{v\al_v}(j_{\al_v}^v) = \left [ \prod_{j=j_{\al_v}^v}^{l_v-1}
w'_{y_v^j}\right] \ \left [ \prod_{j=i_v}^{j_{\al_v}^v-1} s_j(v,\al_v)\right] 
\ .
\label{W1}\ee 

If $v\in V_d$, $c\neq\al_v$ and $c\neq c_v$   
\be
W_{v\al_v}(j_{c}^v) = \left [ \prod_{j= j_{c}^v}^{l_v} 
w'_{y_v^j}\right] \ \left [ \prod_{j=i_v}^{j_c^v-1} s_j(v,c)\right]\ . 
\ee 

If $v\in V_d$ and $c = c_v$  
\be
W_{vc_v}(j_{c_v}^v) =  
\left [ \prod_{j=i_v}^{j_{c_v}^v-1} s_j(v,c_v)\right] \ .
\ee 

Finally if $v\in {\bar V}_d$ 
\be
W_{v\al_v}(j_{c}^v) = \left [ \prod_{j= j_{c}^v}^{j_M} 
w'_{y_v^j}\right] \ \left [ \prod_{j=i_v}^{j_c^v-1} s_j(v,c)\right] 
\label{W2}\ee
where we take the convention that a void product is 1.
Therefore for $v\in V_d$ and $\rho_h=1$ 
the product over $s_j$ is reduced to 1 and for $v\in V_d$ and $\rho_h=2$
the product over $w'$ is reduced to 1.

\subsection{Connected components}

Now, at each order $n$ we can factorize the connected components,
namely the polymers. The forest $\fr$ is connected if
at the highest slice index (hence the lowest energy scale)
there is only one connected component. Remark that $\fr$
could have no link for any $j>j_\fr$. In this case the forest
is connected if $\fr_{j_\fr}$ has only one connected component.

The partition function is written as
\be
Z_\La^{u} =\sum_{k_Y=0}^\infty \frac{1}{k_Y!}\sum_{Y_1,...,Y_{k_Y}
\atop \cup_q Y_q = {\cal D},\; Y_q\cap Y_{q'}=\emptyset}
\prod_q A(Y_q)
\label{zetaf}\ee
where $k_Y$ is the number of different connected polymers
$Y_q$ and the amplitude for a polymer $Y$ is defined as
\bqa
&&\hspace{-0.5cm}A(Y)=   \sum_{n=0}^\infty \frac{\la^n}{n!} 
\sum_{\De_V}  
 \sum_{\fr^c_{M_Y}} \sum_{V_d, \al_{V_d}} \sum_{a,b,R} 
\sum_{l_{V_d}} \sum_{\{J_h^a\},\{J_{\bar h}^a\}}
\sum_{\{j_h^b\},\{j_{\bar h}^b\}} \sum_{C_b}  \vep_\fr 
\left [\prod_{v} \int_{\De_v} d^4x_v \right] \  \ \no\\ 
&& \hspace{-0.7cm} 
\left [\prod_{j=m_Y}^{M_Y}
\lp \prod_{l\in hL_j}\int_0^1 dw_{l}
 \rp\right] \hspace{-0.15cm} 
\left [\prod_{j=m_Y+1}^{M_Y}
\lp \prod_{l\in vL_j}\int_0^1 dw'_{l}
 \rp\right]  \hspace{-0.15cm} 
\left [\prod_{j=m_Y+1}^{M_Y}
\lp \prod_{l\in fL^6_j}\int_0^1 dw''_{l}
 \rp\right]
\no\\
&&\hspace{-0.4cm} 
 \left[
\prod_{v\in V_d} \lp  \sum_{n_{v}\si_{v}\rho_v}
\sum_{i_{v}\in I^c_v} \rp\right]\
 \left [\prod_{v\in V_d\atop c_v\neq \al_v} 
 \prod_{j= i_v}^{l_v} 
w'_{y_v^j}\right]   
\left [\prod_{v\in V_d\atop c_v= \al_v} 
 \prod_{j= i_v}^{l_v-1} 
w'_{y_v^j}\right]   \no\\
&&\hspace{-0.4cm}  
\left [\prod_{j=m_Y}^{M_Y}\lp
\prod_{l\in hL_j} 
C^{j}_{\De_{l}{\bar \De}_{l}}
\lp x_l,{\bar x}_l,\{w'_{l'}\},\{w''_{l'}\}\rp\rp 
\right ]
%\no\\
%&&\hspace{-0.4cm}
\det M''\lp \{w_l\}, \{w'_l\},\{w''_l\}\rp\no\\
&& \label{Y}
\eqa
where $\fr^c_{M_Y}$ is any connected forest over $Y$\footnote{
The constraint that $Y$ must be connected 
implies that the term at order $n$ is zero unless
$n$ is big enough (in order to be able to connect $Y$).}.
The spatial integral for each $v$ is still written in terms of
cubes in ${\cal D}_0$, but 
all sums are restricted to the polymer.
This means that $\De\in {\cal D}_j$ becomes 
 $\De\in {\cal D}_j\cap Y$ and  so on.
Remark that $\fr^c_{M_Y}$ has no link at scale $j<  m_Y$.

\subsection{Main result}

Now we have nearly succeeded in computing
the logarithm of ${\cal Z}$. Actually  (\ref{zetaf}) 
would be  the exponential of $A(Y)$, if there was no  constraint
$Y_q\cap Y_{q'}= \emptyset$, $\cup_q Y_q={\cal D}$. 
Taking out these conditions and computing the logarithm is
the purpose of the so called Mayer expansion [R]. 

By translation invariance, 
a Mayer expansion converges  essentially if the following condition holds:
\be
\sum_{Y\atop 0\in Y} |A(Y)| e^{|Y|} \leq 1
\label{summayer}
\ee
(where $|Y|$ is the cardinal of $Y$, hence the total
number of cubes of all scales forming $Y$).
If we perform power counting, we find that all sub-polymers of $Y$,
$Y_k^j$, with $|E(Y_k^j)|=2,4$ need renormalization. 
This is postponed to a future publication\footnote{In this future publication,
we plan in fact to renormalize only the 2-point function, and to bound
the logarithmic divergence of the 4-point functions by the condition
$\la |\log T| \le K$, like in [DR2].}.
To start with a simpler situation, in this paper 
we restrict ourselves
to the case  $|E(Y_k^j)|>4$ for all  $j<j_M+1$. We call 
this subset the convergent attributions for $Y$
and we denote the corresponding amplitudes by
$A_c(Y)$. Remark that $A_c(Y)$ contains only $f$-links of order 6.
We therefore prove the following theorem, which is a 3-d analog of 
[FMRT] and [DR1].
\paragraph{\Large Theorem}
{\em For any $L>0$, there exists $K>0$, such that if
\be
|\la \ln T| \leq K
\ee
we have
\be
\sum_{Y\atop 0\in Y} |A_c(Y)| L^{|Y|} \leq 1
\label{Y1}\ee 
The sum is performed over all polymers that contain the position
$x=0$,
and $A_c(Y)$ is the amplitude of $Y$ restricted to the convergent
attributions.}
\medskip

The rest of the paper is devoted to the proof of this theorem,
and from now on we further assume $K\le 1$.

\section{Proof}

The general idea is to bound the determinant by a Hadamard inequality,
and to sum over the horizontal structures using the horizontal
propagators decay. The Hadamard inequality generally 
costs a   factor 
\be
n^n\ |\ln T|^{|{\bar V}_d\backslash V_b|+(1-\vep)|V_d\cup V_b|} 
\ee
where $0<\vep<1$ and $V_b$ is the set of vertices hooked to some
horizontal link:
\be
V_b = \{v\in V \ | \ h_c^v\in b\  {\rm for}\  {\rm some}\ c\}.
\ee 
The factor $n^n$ is bounded by the global $1/n!$ symmetry factor
of the vertices, up to a factor $e^n$ by Stirling
formula, which is absorbed in the constant $K'$ (see however
the remark in the Introduction).  
The logarithm is bounded by a fraction of the small coupling constant
$\la^n$. 
A delicate point is to prove that the factor $\vep$ is 
strictly positive $\vep>0$, since we need to spare a fraction 
of $\la$ at each derived vertex $v\in V_d\cup V_b$ 
in order to  extract a small factor
per cube. This factor is necessary to bound the last sum over the 
polymer size and shape.

In the following we will denote fields only by $h$ (not $h_c^v$)
and antifields by ${\bar h}$. The corresponding vertex is $v_h$,
$v_{\bar h}$, the field index is $c_h\in C$, the antifield 
index $c_{\bar h}\in {\bar C}$
($C$ and ${\bar C}$ are introduced in section II.5),
 their slice indices are $j_h$, $j_{\bar h}$ 
 and their vertex 
position is $x_h$, $x_{\bar h}$. 

In order to bound the amplitude of a polymer $A(Y)$ we must
introduce the auxiliary slice decoupling of section II.4.
For each propagator extracted from the determinant we write
\bqa
 C^{j}_{\De_{l}{\bar \De}_{l}}\lp x_{l},{\bar x}_{l},\{w'_{l'}\},
\{w''_{l'}\}
\rp = && \sum_{k=0}^{k_M(j)} 
C^{jk}_{\De_{l}{\bar \De}_{l}}\lp x_{l},{\bar x}_{l},\{w'_{l'}\},
\{w''_{l'}\}\rp \\
 = && \sum_{k_{h_l} k_{{\bar h}_l}}\ 
 \de_{k_{h_l}, k_{{\bar h}_l}} 
 C^{jk_{h_l}}_{\De_{l}{\bar \De}_{l}}\lp x_{l},{\bar x}_{l},\{w'_{l'}\},
\{w''_{l'}\}\rp \no
\eqa
where 
\be
C^{jk}_{\De_{l}{\bar \De}_{l}}\lp x_{l},{\bar x}_{l},\{w'_{l'}\},
\{w''_{l'}\}\rp = 
W_{v_lc_l}(j) \  C^{jk}_{\De_{l}{\bar \De}_{l}}\lp x_l,{\bar x}_l\rp \
W_{{\bar v}_l{\bar c}_l}(j)
\ee
$C^{jk}$ is defined in (\ref{ascale1}) and  $W_h(j)$ corresponds to the 
function $W_{vc}(j)$ defined in (\ref{W1}-\ref{W2}). 
The matrix element is written as
\bqa 
&& \hspace{-0.7cm} M''_{h;{\bar h}}
\lp \{w_l\},\{w'_l\},\{w''_l\}  \rp 
=  \  \sum_{k_h k_{\bar h}} \ \ 
\sum_{j\in {\cal I}_h\cap {\cal I}_{\bar h}\cap J(k_h)} \    
\de_{k_{h},k_{\bar h}} 
\\
&& 
\left [  W_{h}(j) \right ] 
\
\left [ h_{\De^j_{h},\De^j_{{\bar h}}}^{\fr^h_{j}}(w)  \right ] 
 C^{j k_h}_{\De^j_{h},
\De^j_{{\bar h}}} (x_h,x_{\bar h}) \ 
 \left [W_{\bar h}(j) \ \right ] 
 \no
\eqa
where we have exchanged the sums over $j_h$ and $k_h$, 
$J(k)$ is defined in (\ref{ascale3})
and the interval 
 ${\cal I}_h$ corresponds to
the interval ${\cal I}_c^v$ defined in (\ref{Idef}).
Finally  we denote by $\De^j_{h}$ the cube $\De^j_{v_h}$. The
same definitions hold for ${\bar h}$. The sums over $k_h$ and
$k_{\bar h}$ are extracted from the determinant by 
multilinearity. We need now to reorganize the sum over $Y$ according
to a tree structure analogous to the ``Gallavotti-Nicol{\'o} tree'' [GN]. 
that is called here $S$.

\subsection{The $S$ structure}

Let  $M_Y$ be the lowest scale of the polymer. $S$ 
is a rooted tree that pictures the inclusion relations 
for the connected components of $Y$ at each scale and the type of
vertical connection (vertex or field).
In this rooted tree the extremal 
leaves are pictured as dots and the other vertices as circles.
A circle at layer $l$ represents
a connected subpolymer at scale $j=M_Y-l$. A leaf at layer $l$
by convention represents 
an {\it extremal summit cube}, that is a cube such that
$Ex(\De)=\De$ (no cube above), whose scale is $M_Y-l+1$.
The highest layer fixes the scale $m_Y$: $l_{max}= M_Y-m_Y+1$
(as at scale $M_Y-l_{max}$ there are only leaves, hence no cubes)
and satisfies $l_{max}-1\leq j_M$. 

There are two types of links in $S$: the {\it leaf-links} which join
a leaf to a circle, and the {\it circle-links} which join
two circles. To each circle-link corresponds a vertical block in the
multiscale expansion, and we can  
associate to it a label $f$ or $v$ depending if
this block is associated to a vertex or to external fields\footnote{ 
We remark that the circles at level $l$ connected only to leaves at 
level $l+1$ must be connected to the previous circle
at level $l-1$ by a $v$-circle-link. Indeed each of the extremal summit cubes
forming that circle must contain at least one vertex.}.

\begin{figure}
\centerline{\psfig{figure=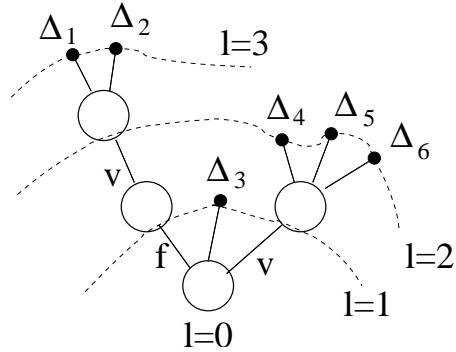,width=6cm}}
\caption{Example of $S$}
\label{had11}
\end{figure}

\begin{figure}
\centerline{\psfig{figure=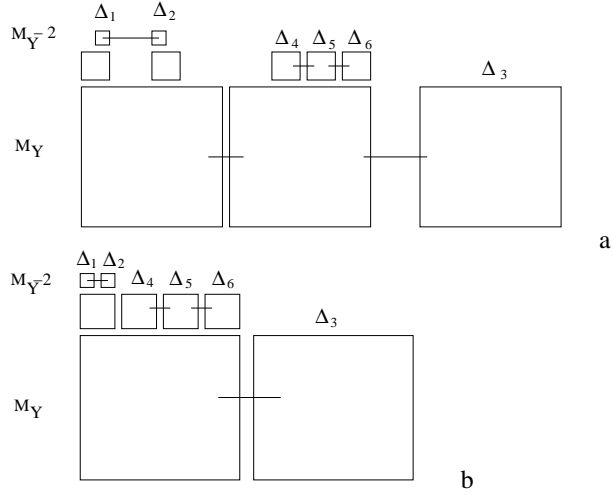,width=8cm}}
\caption{Two possible polymers corresponding to $S$}
\label{had14}
\end{figure}

An example of $S$ structure is given in  Fig.\ref{had11} and  
two possible polymers corresponding to this structure are given in 
Fig.\ref{had14} {\bf a} and {\bf b}. 
We remark that $S$ fixes in a unique way the
number and scales of the extremal summit cubes, but 
that several polymers, with different total number
of cubes, may correspond to the same structure $S$.

%\begin{figure}
%\centerline{\psfig{figure=had13.eps,width=8cm}}
%\caption{Example of $S$}
%\label{had13}
%\end{figure}

\begin{figure}
\centerline{\psfig{figure=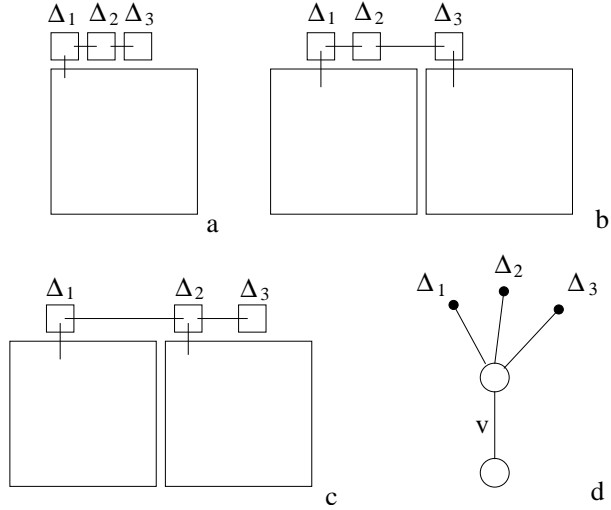,width=8cm}}
\caption{a,b,c: three  polymers corresponding to the same $S$
shown in d: $VL$ can distinguish a from b, but not b from c}
\label{had15}
\end{figure}

In order to fix this total number of cubes, we introduce
for each circle-link of $S$ a further number which fixes  
the number of vertical links (wich are $v$-links or $f$-links depending
of the type of the circle-link) selected in the block in section III.3.
Since there is one vertical link per ancestor cube,
this number is the number of ancestor cubes of the connected
component $y$ corresponding to the circle at the top
of the circle-link. We call this collection of indices $VL$.  
$S$ and $VL$ together fix the number $|Y|$  of cubes in $Y$.
For instance the 
situations in Fig\ref{had15}{\bf a} and {\bf b}.
correspond to the same $S$,  shown in Fig.\ref{had15} {\bf d}.
But the case a) corresponds to an index $VL=\{1\}$ for the 
unique circle-link and to
4 cubes in $Y$, whether the case b) corresponds to an index $VL=\{2\}$
and to 5 cubes in $Y$, 

Finally when $S$ and $VL$ are given,
we can label all the cubes of $Y$,
and we fix the subset $B_S$ of those cubes of $Y$ which are summit cubes.
They are those with non-zero exposed volume: $|Ex(\De)|>0$\footnote{
Actually $B_S$ only really fixes the non-extremal summit
cubes $\De$ (with $0< |Ex(\De)| < |\De|$)
since the extremal summit cubes with $Ex(\De)=\De$ were already known from the data in $S$.}.
Nevertheless we remark that there is still some ambiguity,
as even $VL$ and $B_S$ cannot distinguish between  
Fig.\ref{had15}{\bf b} and {\bf c}, and the position of the cubes
of $Y$ is not yet fixed.

\subsection{The reorganized sum}

The sum (\ref{Y1}) is then reorganized in terms of the structure $S$ as
\bqa
&&\hspace{-0.7cm}
\sum_{Y\atop 0\in Y} |A_c(Y)| L^{|Y|}\  
\leq \ 
\sum_{M_Y}
 \sum_{S}\sum_{VL} L^{|Y|} \sum_{B_{S}} \ 
\sum_{\{x_\De\}^c}\ 
 \left |\ 
\sum_{n=0}^\infty \frac{\la^n}{n!} 
\sum_{V_d, \al_{V_d}} \sum_{a,b,R} 
\sum_{\{v_l\}_{l\in vL}} \ 
 \right .
 \no\\
&& \hspace{-0.5cm}
\sum_{n_{V_d}\si_{V_d}\rho_{V_d}}\ 
\sum_{\{n_\De\}_{\De\in B_S}}\  \sum_{\De^c_{{\bar V}_d}} 
\left [\prod_{v\in V_d} 
\sum_{i_{v}\in I^c_v} \sum_{\De_v\in {\cal D}_{i_v}\cap Y}  \right ]\ 
\sum_{\{J_h^a\}, \{ J_{\bar h}^a\}} \  
\sum_{\{j_h^b\},\{j_{\bar h}^b\}} \     \sum_{\{k_h\}, \{k_{\bar h}\}}
\label{Ac} \no \\ 
&& \hspace{-0.5cm}
\left [\prod_{j=m_Y}^{M_Y}  \prod_{k=1}^{c_j} \sum_{T_{jk}} \right ]
\left [\prod_{v\in {\bar V}_d} \int_{Ex(\De_v)} dx_v \right ] \
 \left [\prod_{v\in V_d} \int_{\De_v} dx_v \right ] \
\vep_\fr
\\
&&\hspace{-0.7cm} 
 \left [\prod_{j=m_Y}^{M_Y}
\lp \prod_{l\in hL_j}\int_0^1 dw_{l}
 \rp\right]  \hspace{-0.15cm}  
\left [\prod_{j=m_Y+1}^{M_Y}
\lp \prod_{l\in vL_j}\int_0^1 dw'_{l}
 \rp\right] \hspace{-0.15cm} 
\left [\prod_{j=m_Y+1}^{M_Y}
\lp \prod_{l\in fL^6_j}\int_0^1 dw''_{l}
 \rp\right] \no\\               
&&\hspace{-0.5cm}   \left [ \prod_{j=m_Y}^{M_Y} 
\prod_{k=1}^{c_j} \left [
\prod_{l\in T_{jk}} 
C^{j k_{h_l}}_{\De_{l}{\bar \De}_{l}}\lp 
x_l,{\bar x}_l,\{w'_{l'}\},\{w''_{l'}\}
\rp \right ] \de_{k_{h_l} k_{{\bar h}_l}} \right ]
\no\\
&&\hspace{-0.5cm} \left . 
\left [
\det M''\lp \{w_l\}, \{w'_l\},\{w''_l\},
\{k_{h,{\bar h}}\}\rp \right ]\ \ 
\left [\prod_{v\in V_d\atop c_v\neq \al_v} 
 \prod_{j= i_v}^{l_v} 
w'_{y_v^j}\right]   
\left [\prod_{v\in V_d\atop c_v= \al_v} 
 \prod_{j= i_v}^{l_v-1} 
w'_{y_v^j}\right]  \right |
\no
\eqa
where 
\begin{itemize}

\item{} $\{x_\De\}^c$ chooses the position of 
each cube in the polymer, constrained by $S$, $VL$ and
$B_S$, with the additional constraint that at the lowest level
$M_Y$ there is one cube containing the origin $x=0$.

\item{} $v_l$ is the vertex $v\in V_d$ associated to the vertical 
link $l\in vL$ where
$vL=\cup_j vL_j$. Remark that once we know $v_l$ for each $l\in vL$,
we automatically know $l_v$ for all $v\in V_d$. The vertices of
$V_d$ are from now on said to be localized in the cube $\De_{i_v} \in 
{\cal D}_{i_v}$ to which they belong. 

\item{} $n_{B_S}=\{n_\De\}_{\De\in B_S}$ 
gives the number of vertices in ${\bar V}_d$ localized
in each summit cube (recall (\ref{locsumcub}): 
$n_{B_S} = \{n_\De | \De\in B_S\}$ 
with the constraint 
$\sum_{\De\in B_S} n_\De = |{\bar V}_d |= n - |V_d|$.

\item{} $n_{V_d},\si_{V_d},\rho_{V_d}$ are the assignments
$n_{v},\si_{v},\rho_v$ $\forall v\in V_d$.

\item{} $\De^c_{{\bar V}_d}$ chooses which vertices $v\in {\bar V}_d$ 
are localized in each summit cube: 
$\De^c_{{\bar V}_d}= \{\De_v \}_{v\in {\bar V}_d}$ with the constraint
$\#\{v\ |\ v\in {\bar V}_d,\ \De_v= \De\}= n_\De$, $\forall \De\in B_S$.
The spatial integral for each $v\in {\bar V}_d$ is then performed over 
the exposed volume of the corresponding cube $Ex(\De_v)$ (see 
(\ref{locsumcub})). 

\item{} $k_h$ fixes the value of an auxiliary scale 
(defined in section II.4) that will
be used in the propagator analysis;
$k_{\bar h}$ is the same thing for antifields.

\item{} $T_{jk}$ chooses the  tree connecting 
the generalized cubes   ${\tilde \De}\in y_j^k$ by 
$h$-links of scale $j$. To fix  $T_{jk}$ one has to
choose the $h$-links and the corresponding fields. As the
fields (antifields) that must contract at scale $j$ in order to
create $T_{jk}$ are already fixed by $b$, $j_h^b$ and $j_{\bar h}^b$,  
we only have to fix  the field-antifield 
pairing $C_b$ restricted
to $y_j^k$. 
 
\end{itemize}

\subsection{Bounding the determinant}

In order to bound the main determinant we apply the following 
 
\medskip
\noindent{\bf Hadamard inequalities}\ {\em
If $M$ is a $n\times n$ matrix with elements $M_{ij}$, its determinant
satisfies the following bounds
\bqa
{\bf Hr:}\qquad |\det M| & \leq &
\prod_{i=1}^n \ \left [ \sum_{j=1}^n\  |M_{ij}|^2 \ 
\right ]^{\frac{1}{2}} \label{hadr} \\
{\bf Hc:}\qquad |\det M| & \leq &
\prod_{j=1}^n \ \left [\sum_{i=1}^n \ |M_{ij}|^2 \ 
\right ]^{\frac{1}{2}} \label{hadc}
\eqa
where  Hr is  obtained by considering each row as a n-component vector, 
and Hc by considering each column as a n-component vector. }

We remark that these two inequalities are both true, but not identical.
In our case it is crucial to optimize as much as possible 
our bounds, and to use either the row or the column inequality depending
of the kind of fields involved and of various scaling and occupation
factors.

Before expanding the determinant in (\ref{Ac})
we distinguish therefore five different types of fields (antifields) 
denoted by an index $\al_h$, $\al_{\bar h}$:
\bqa
\al_h =1 && {\rm if} \qquad  
v_h\not\in V_d  \no\\
\al_h =2 && {\rm if} \qquad v_h\in V_d,\  c_h\neq c_v   
 \ {\rm and}\  \rho_h=1 \\
\al_h =3 && {\rm if} \qquad v_h\in V_d,\  c_h\neq c_v   ,\ h\not\in a   
 \ {\rm and}\  \rho_h=2 \no \\
\al_h =4 && {\rm if} \qquad  v_h\in V_d,\ h\in a   \no \\
\al_h =5 && {\rm if} \qquad v_h\in V_d \      
 {\rm and}\  c_h= c_v  \no
\eqa
The same definitions hold for antifields ${\bar h}$.
The case $\al_h =1$ is the most general one.
This is a partition, since neither the fields with $\rho_h=1$ and $c_h\neq c_v$ 
nor the special fields $h$ with $v_h\in V_d$ and $c_h= c_v $ can belong to $a$.

We now define for each field $h$
a weight $I_h$ which depends of the type of the field as follows: 
\bqa
 \al_h=1: \quad  I_h &=&  n_{\De_{h}}   M^{-4i_{\De_{h}}} f_{\De_{h}}^{-1} 
\no\\
 \al_h=2: \quad  I_h  &=& \quad \  M^{-4i_{v_h} } \no\\
 \al_h=3: \quad  I_h  &=& \quad \  M^{-4l_{v_h} } \no\\
 \al_h=4: \quad  I_h  &=& \quad \  M^{-4i_h}  \no\\
 \al_h=5: \quad  I_h  &=& \quad \  M^{-4i_{v_h} } 
 \label{Ifef1}
\eqa
where $\De_{h}$ is the cube where the vertex $v_h$ is localized.
For $\De\in B_S$ we defined 
$f_\De$ as the exposed fraction of the volume $|\De|=M^{4i_\De}$,
and $n_{\De}$ as the number of vertices in ${\bar V}_d$ localized
in the summit cube $\De$. Finally, for each $h\in a$  the scale $i_h$ 
is defined as 
\be
i_h = \max J_h^a\ .
\ee
We remark that actually $h\in a$ can only have attributions
$j\geq 1+i_h$. The same definitions hold for ${\bar h}$.

The Hadamard inequality will be either of the row or of the column type
depending on whether the ratio of weights of the fields involved
is larger or smaller than 1. In fact we need to discretize
these ratios in order to transfer some factors from fields
to antifields and conversely and to obtain a correct bound. 
To implement this program
we introduce an auxiliary expansion called the weight expansion.

\subsubsection{The weight expansion}

We expand 
\be
h = \sum_{\bt_h=1}^5 h^{\bt_h}
\ee
where $h^{\bt_h}$ means that $h$ can contract only with ${\bar h}$
such that $\al_{\bar h}=\bt_h$. The same holds for the antifields.

Finally, we expand each $h^{\bt_h}$ (${\bar h}^{\bt_{\bar h}}$) as
\be
h^{\bt_h} = \sum_{r\in \ZZ} h^{\bt_h}(r)  
\ee
where $h^{\bt_h}(r)$ 
means that $h$ can contract only with ${\bar h}$ such that
\be
\frac{I_h}{I_{\bar h}}\in I_r \ ;\ \ I_0 = [1], 
\ I_r= ] 2^{r-1},2^{r}]\ {\rm if}\ r>0, \ \
I_r=[2^{r},2^{r+1}[\ \ {\rm if}\ r<0
\label{defir}
\ee
We remark that the intervals $I_r$ are disjoint with $\cup_{r\in \ZZ} I_{r} = ]0,+\infty[$ 
and that with this definition $h(r)$ can contract only with 
antifields ${\bar h}(r')$ with $r'=-r$.
The same holds for the antifields.

The special fields or antifields of type 5 require an additional
expansion. We define for each such field $h$ an occupation number
$n(h)$ which is the number of derived vertices localized in the same
cube than $h$
\be
n(h) \ = \ n_d(\De_{i_{v_h}}) \  =  \ |\{
{\rm \ vertices \  in \ } V_d {\rm \ localized \  in \  the  \ cube \ } 
\De_{i_{v_h}} \}| \label{defnh}
\ee
We remark that $ n_d(\De_{i_{v_h}})$ has nothing to do with 
$n_{\De_{v_h}}$ 
in general, since these numbers concern respectively $V_d$ and $\bar V_d$.
We recall that the vertices  $v \in V_d$ are localized in the cube
of ${\cal D}_{i_v}$ to which they belong, whether the vertices of $\bar V_d$
are localized in the summit cube to which they belong.

By convention, for any field not of type 5 we put
\be
n(h)=1
\label{defnhg}
\ee
The same definitions hold for the antifields.
Now we expand each field as
\be
h^\bt(r) = \sum_{s\in \ZZ} h^\bt(r,s)
\ee
where $h^{\bt_h}(r,s)$ 
means that $h$ can contract only with ${\bar h}$ such that
\be
\frac{n(h)}{n({\bar h})} \in I_s \label{additexp}
\ee
where $I_s$ is defined like $I_r$ in (\ref{defir}).
We remark that this additional 
$s$ expansion is trivial (reduced to the term $s=0$)
unless $\al$ or $\bt$ equals 5, and that for $\al \ne 5$ $\bt =5$,
$s$ is negative: $s\le 0$. 
Symmetrically for  $\al = 5$ $\bt \ne 5$, $s$ is positive:
$s\ge 0$. 

Summarizing all constraints, 
the field  $h^{\bt_h}(r,s)$ contracts only with antifields 
${\bar h}^{\bt_{\bar h}}(r',s')$ such that  
$\bt_h=\al_{\bar h}$, $\bt_{\bar h}=\al_h$, $k_h=k_{\bar h}$,
$r'=-r$ and $s'=-s$. Therefore we have   
\be
\big| \{h^\bt(r,s)\ | \ \al_h=\al,\  k_{h}=k\} \big| = 
\big| \{{\bar h}^\al(-r,-s)\ | \ \al_{\bar h}=\bt, \
k_{\bar h}=k \}\big|\ .
\ee

The determinant in (\ref{Ac}) is now written as 
\bqa
\hspace{-0.8cm}\det M'' &=&  \sum_{\{\bt_h\}\{\bt_{\bar h}\}} 
\sum_{\{r_h\},\{r_{\bar h}\} }    
\sum_{\{s_h\},\{s_{\bar h}\} } 
%\no\\&&
 \left[ \prod_{r,s \in \ZZ}\  
\det  {\cal M}_{r,s}\lp \{\bt_h\}\{\bt_{\bar h}\}\rp\right]
\eqa
where the sums over $r_h$, $r_{\bar h}$,  $s_h$, $s_{\bar h}$,
$\bt_h$ and  $\bt_{\bar h}$
are extracted from the determinant by multilinearity, 
and ${\cal M}_{r,s}$ is the matrix containing only fields 
with $r_h=r$ (therefore only antifields with $r_{\bar h}=-r$)
and $s_h=s$ (therefore   only antifields with
 $s_{\bar h}= -s $) 
We take the convention that ${\cal M}_{r,s}=1$ if there is no field 
with $r_h=r$ and $s_h =s$.  We recall that the sums over $s_h $ and 
$s_{\bar h}$ are restricted by some constraints: 
$s=0$ unless $\bt_h$ or $\bt_{\bar h}$ equals 5, $s\le 0$ for
$\bt_h = 5$, $\bt_{\bar h} \ne 5$, and $s\ge 0$  for
$\bt_h \ne 5$, $\bt_{\bar h} = 5$.

Now we can insert absolute values inside the sums and
(\ref{Ac}) can be bounded by

\bqa
&&\hspace{-0.7cm}
\sum_{Y\atop 0\in Y} |A_c(Y)| L^{|Y|}\  
\leq \ 
\sum_{M_Y}
 \sum_{S}\sum_{VL} L^{|Y|} \sum_{B_{S}} \ 
\sum_{\{x_\De\}^c}\  
\sum_{n=0}^\infty \frac{|\la|^n}{n!} 
\sum_{V_d, \al_{V_d}} \sum_{a,b,R} 
\sum_{\{v_l\}_{l\in vL}} \ 
 \no\\
&& \hspace{-0.5cm}
\sum_{n_{V_d}\si_{V_d}\rho_{V_d}}
\sum_{\{n_\De\}_{\De\in B_S}}\  \sum_{\De^c_{{\bar V}_d}} 
\left [\prod_{v\in V_d} 
\sum_{i_{v}\in I^c_v} \sum_{\De_v\in {\cal D}_{i_v}\cap Y}  \right ]\ 
\sum_{\{J_h^a\}, \{ J_{\bar h}^a\}} \  
\sum_{\{j_h^b\},\{j_{\bar h}^b\}} \     \sum_{\{k_h\}, \{k_{\bar h}\}}
\label{Ac1} \no \\ 
&& \hspace{-0.5cm}
\left [\prod_{j=m_Y}^{M_Y}  \prod_{k=1}^{c_j} \sum_{T_{jk}} \right ]
\left [\prod_{v\in {\bar V}_d} \int_{Ex(\De_v)} dx_v \right ] \
 \left [\prod_{v\in V_d} \int_{\De_v} dx_v \right ] \
\\
&&\hspace{-0.7cm} 
 \left [\prod_{j=m_Y}^{M_Y}
\lp \prod_{l\in hL_j}\int_0^1 dw_{l}
 \rp\right]  \hspace{-0.15cm} 
\left [\prod_{j=m_Y+1}^{M_Y}
\lp \prod_{l\in vL_j}\int_0^1 dw'_{l}
 \rp\right] \hspace{-0.15cm} 
\left [\prod_{j=m_Y+1}^{M_Y}
\lp \prod_{l\in fL^6_j}\int_0^1 dw''_{l}
 \rp\right] \no\\               
&&\hspace{-0.5cm}   \left [ \prod_{j=m_Y}^{M_Y} 
\prod_{k=1}^{c_j} \left [
\prod_{l\in T_{jk}} \left |
C^{j k_l}_{\De_{l}{\bar \De}_{l}}\lp x_l,{\bar x}_l,\{w'_{l'}\},\{w''_{l'}\}
\rp\right | \right ] \de_{k_{h_l} k_{{\bar h}_l}} \right ]
\sum_{\{\bt_h\}\{\bt_{\bar h}\}} \ \sum_{\{r_h\},\{r_{\bar h}\} } \no\\
&&\hspace{-0.5cm}     
\sum_{\{s_h\},\{s_{\bar h}\} }\left[ \prod_{r,s \in {\ZZ}} 
\ \left |\det  
{\cal M}_{r,s}\lp \{\bt_h\}\{\bt_{\bar h}\}
\rp\right |\right] 
%\no\\&&\hspace{-0.5cm}  
\left [\prod_{v\in V_d\atop c_v\neq \al_v} 
 \prod_{j= i_v}^{l_v} 
w'_{y_v^j}\right]   
\left [\prod_{v\in V_d\atop c_v= \al_v} 
 \prod_{j= i_v}^{l_v-1} 
w'_{y_v^j}\right] \no
\eqa

%\subsubsection{Bound on $\det{\cal M}_{r,s,{\bar s}}$}

Now, for each $r$, $s$ we distinguish between three cases.
\begin{itemize}
\item{} If $r> 0$
(which means $r_h=r> 0$ and $r_{\bar h}=-r <0$), 
then $I_h> I_{\bar h}$ for any $h$, ${\bar h}$ in 
${\cal M}_r$. In this case we apply 
the row inequality (\ref{hadr}).

\item{} If $r< 0$
(which means $r_h=r< 0$ and $r_{\bar h}=-r > 0$),
then $I_h< I_{\bar h}$ for any $h$, ${\bar h}$ in 
${\cal M}_r$. This case is similar to the first case,
exchanging the role of fields and antifields, so we apply 
the column inequality (\ref{hadc}).

\item{} If $r=0$
(which means $r_h=r=0$ and $r_{\bar h}=-r=0$), 
then $I_h = I_{\bar h}$ for any $h$, ${\bar h}$ in 
${\cal M}_r$. In this case we must analyze in more detail
the subdeterminants as will be explained later.

\end{itemize}

%\noindent 
With these conventions the fixed index (field or antifield) in 
the sum $\sum_{j=1}^n\  |M_{ij}|^2$ for Hr or
$\sum_{i=1}^n\  |M_{ij}|^2$ for Hc  
is always the one with the  highest weight $I$. 
This is essential in the following bounds.

\subsubsection{Case {\bf $r> 0$} (and  {\bf $r< 0$})}

As remarked above we treat only the case $r>0$, the other case being
similar, exchanging fields and antifields, hence rows and columns.
In that case we apply the  row inequality  (\ref{hadr}):
\be
\left |\det  
{\cal M}_{r,s}\lp \{\bt_h\}\{\bt_{\bar h}\}\rp\right |
  \leq 
\prod_{\left \{{h\not \in b, r_h=r
\atop s_h=s }\right \}} \lp 
\sum_{\left \{{{\bar h} \not\in b | \bt_{h}=\al_{\bar h},  
\al_{h}=\bt_{\bar h}, \atop 
  k_{\bar h}=k_h, r_{\bar h}=-r,\ s_{\bar h}=-s 
}\right \} } |M_{h,{\bar h}}|^2 
\rp^{\frac{1}{2}}
\label{haddetr}\ee
where $h\not \in b$ is the set of fields that are not extracted from
the determinant to give some $h$-link. Now 
\bqa
&& \hskip-0.5cm |M_{h,{\bar h}}|^2 = 
\left | \sum_{j\in {\cal I}_h\cap {\cal I}_{\bar h}\cap J(k_h)} \    
\de_{k_{h},k_{\bar h}} 
\left [  W_{h}(j) \right ] 
\left [ h_{\De^j_{h},\De^j_{{\bar h}}}^{\fr^h_{j}}(w)  \right ] 
 C^{j k_h}_{\De^j_{h},
\De^j_{{\bar h}}} (x_h,x_{\bar h}) 
 \left [W_{\bar h}(j) \right ] \right |^2 \no\\
&&\hskip-0.5cm
\leq 
 \de_{k_{h},k_{\bar h}}  \sum_{j\in {\cal I}_{h}} \left |  C^{j k_h}_{\De^j_h,
\De^j_{\bar h}}(x_h,x_{\bar h}) 
\right |^2
\eqa
where the weakening factors $W_h(j)$,  $W_{\bar h}(j)$ and 
 $h_{\De^j_{h},\De^j_{{\bar h}}}^{\fr^h_j}(w)$ are  bounded
by one,  the sum over $j$ is performed
over the larger set  
${\cal I}_h\cap {\cal I}_{\bar h}\cap J(k_h)\subset {\cal I}_h$,
which is an upper bound, and we applied
the identity
\be
C^{j k_h}_{\De^j_h,
\De^j_{\bar h}}(x_h,x_{\bar h}) \ 
C^{j' k_h}_{\De^{j'}_h,
\De^{j'}_{\bar h}}(x_h,x_{\bar h}) = 0 \qquad {\rm if} \ j\neq j'
\ee
which is true by construction.
For any  ${\bar h}$ in the sum, its weight
satisfies
\be
I_h \ 2^{-r}\  \leq \ I_{\bar h}\  < \ I_h \ 2^{-r+1}.
\label{I1}\ee
Before going on we prove the following lemma
\paragraph{Lemma.}
{\em If $r>0$, the only non zero contributions are
for $\al_h<5$.}
\paragraph{Proof} 
Actually if there exists $\al_h=5$ we must have
\be
M^{-4i_{v_h}} > 2^{r-1} \  I_{\bar h}\  \geq I_{\bar h}\ .
\ee
But this is impossible.
Indeed let us consider for instance the case $\al_{\bar h}=1$. Then
\be
M^{-4i_{v_h}} > \  \ n_{\De_{\bar h}}   
M^{-4i_{\De_{\bar h}}} f_{\De_{\bar h}}^{-1} 
\geq   M^{-4i_{\De_{\bar h}}}
\ee
which implies $i_{v_h}< i_{\De_{\bar h}}$. But to contract 
$h$ with ${\bar h}$ we must also have  $i_{v_h}\geq i_{\De_{\bar h}}$,
which is a contradiction. The other cases are verified in the same way.

\qed

\medskip
Now the first step is to estimate the sum over ${\bar h}$ 
\be
\Si_{\bar h}=: 
\sum_{\left \{{{\bar h} \not\in b |  
\bt_{h}=\al_{\bar h},  \al_{h}=\bt_{\bar h},  
\atop k_{\bar h}=k_h,
 r_{\bar h}=-r, s_{\bar h}=-s}   \right \} } |M_{h,{\bar h}}|^2 
\ee
For this purpose we  distinguish five cases.

\paragraph{1.)} 

$\bt_{h}=1$ which means that $h$ can contract only with
${\bar h}$ of type 1 ($\al_{\bar h}=1$). Therefore for any ${\bar h}$
the weight $I$  is 
\be
I_{\bar h} = \  n_{\De_{\bar h}} \   M^{-4i_{\De_{\bar h}} }
\ f_{\De_{\bar h}}^{-1}
\label{I2}\ee
Therefore the sum $\Si_{\bar h}$ is bounded by
\be
\Si_{\bar h} \leq 
 \sum_{j\in {\cal I}_{h}} \sum_{\De\in {\cal D}_{j}} 
\left |  C^{j k_h}(x_{\De_h^j},x_{\De}) \right |^2
\sum_{\De'\in B_S, \De'\subset\De } 2n_{\De'}
\label{sum2}\ee
where $ 2n_{\De'}$ is the maximal number of antifields (two
for each vertex) localized in $\De'$. 
We remark that the vertex position in the propagator is 
substituted by the cube center $x_\De$.
By (\ref{I1}) and (\ref{I2}) we see that
\be
n_{\De'} <  I_h \ 2^{-r+1}\   M^{4i_{\De'}} f_{\De'}
\ee
therefore  (\ref{sum2}) is bounded by
\be 
 \Si_{\bar h} \  \leq \ 2\  I_h \ 2^{-r+1} \ 
\sum_{j\in {\cal I}_{h}} \sum_{\De\in {\cal D}_{j}} 
\left |  C^{j k_h}(x_{\De_h^j},x_{\De}) \right |^2
\sum_{\De'\in B_S , \De'\subset\De }  M^{4i_{\De'}} f_{\De'}
\ee
Now we observe that $M^{4i_{\De'}} f_{\De'}$  is the exposed volume
of $\De'$ and that, for any cube $\De$, $\cup_{\De'\subset \De} Ex(\De')$ 
is a partition of $\De$. Therefore we have
\be
\sum_{\De'\in B_S, \De'\subset\De }  M^{4i_{\De'}} f_{\De'} = M^{4j}
\ee
hence  $\Si_{\bar h}$  is bounded by
\be
\Si_{\bar h} \leq  \  I_h \ 2^{-r+2}\   
\sum_{j\in {\cal I}_{h}}  M^{4j}  \sum_{\De\in {\cal D}_{j}} 
\left |  C^{j k_h}(x_{\De_h^j},x_{\De}) \right |^2  
\ee
Finally the sum over $\De$ is bounded by
\be
\sum_{\De\in {\cal D}_{j}} 
\left |  C^{j k_h}(x_{\De_h^j},x_{\De}) \right |^2 
\leq  C \  M^{\frac{16}{3}}\  M^{-4j} \  M^{-\frac{4}{3}k_h} \ 
\sum_{\De\in {\cal D}_{j}}  \chi_{j,k}(x_{\De_h^j},x_{\De})
\label{hadb}\ee
where from now on we use $C$ as generic name for a constant independent
of $M$ which can be tracked but whose numerical precise value is inessential.
We applied the scaled decay (\ref{sdec1})-(\ref{sdec3}), and 
the function $\chi_{j,k}$ is different from zero only for
$|\vec{x}_{\De_h^j}-\vec{x}_{\De}|\leq M^j$ and
$|t_{\De_h^j}-t_{\De}|\leq M^{j+k}$
(actually for $k>0$ we have
$|\vec{x}_{\De_h^j}-\vec{x}_{\De}|
\simeq M^{j-\frac{k}{3}+\frac{1}{3}}\leq M^j$). Now, for 
$x_{\De_h^j}$ fixed, the number of cubes such that their center
$x_{\De}$ satisfies these bounds is at most $26 (2M^{k_h})$ where 
$26$ is the number of nearest neighbors of $\De_h^j$ in the position
space, and $2M^{k_h}$ is the number of choices in the time direction.
Therefore
 \be
\sum_{\De\in {\cal D}_{j}} 
\left |  C^{j k_h}(x_{\De_h^j},x_{\De}) \right |^2 
\leq C M^{\frac{16}{3}}\ M^{-4j} \  M^{-\frac{k_h}{3}}
\ee

Remark that the case $j=j_M+1$ needs a different treatment.
Actually in this case we have
\bqa
&&\hspace{-2.1cm}\sum_{\De\in {\cal D}_{j}} 
\left |  C^{j_M+1,  0}(x_{\De},x_{\De'}) \right |^2 
\leq C_{p}\    M^{-4j_M} \   
\sum_{\De\in {\cal D}_{j}}  \frac{
\chi(|t_{\De}-t_{\De'}| \leq M^{j_M})}{  
(1+M^{-j_M}|\vec{x}_\De-\vec{x}_{\De'}|)^{2p}}\no\\
&&\hspace{-2.1cm}
\leq   M^{-4j_M} \hspace{-0.5cm}  \sum_{n_1,n_2,n_3\in \ZZ}  
\frac{C_{p}}{(1+M^{-j_M} M^{j_M}(|n_1|+|n_2|+|n_3|))^{2p}} 
\leq  C \    M^{-4j_M}.
\eqa
The sum $\Si_{\bar h}$ is finally bounded by
\bqa
\Si_{\bar h} & \leq &  \  C\  M^{\frac{16}{3}} \ 
 I_h \ 2^{-r}\  \sum_{j\in {\cal I}_{h}} M^{4j}\  
M^{-4j} \  M^{-\frac{k_h}{3}} \leq  C \  M^{\frac{16}{3}}   
I_h \ 2^{-r} 
\ M^{-\frac{k_h}{3}} \  |{\cal I}_{h}| \no\\
&\leq & C \  M^{\frac{16}{3}}   
I_h \ 2^{-r} \ M^{-\frac{k_h}{3}} \  j_M 
\eqa
where $|{\cal I}_{h}|$ is the number of elements in
the interval ${\cal I}_{h}$, the numerical constants have been absorbed in 
$C$ and we bounded $ |{\cal I}_{h}|$ by $ j_M$.

\paragraph{2.)} 
$\bt_{h}=2$ which means that $h$ can contract only with
${\bar h}$ of type 2 ($\al_{\bar h}=2$). Therefore all ${\bar h}$
must be hooked to some vertex in $V_d$ and must have
scale attribution $i_{v_{\bar h}}\leq j_{\bar h}\leq l_{v_{\bar h}}$. 
The weight $I_{\bar h}$ is 
\be
I_{\bar h} \ =\    M^{-4i_{v_{\bar h}} }\  =  
\ M^{-4i_r(h)}
\ee
where $i_r(h)$ is the unique scale for which (\ref{I1})
is satisfied. Now 
\be
\Si_{\bar h}\leq 
 \sum_{j\in {\cal I}_{h}\atop j\geq i_r(h) }\  \sum_{\De\in {\cal D}_{j}} \ 
\left |  C^{j k_h}(x_{\De_h^j},x_{\De}) \right |^2 2(j_M+2-j)
\ee
where $j_M+2-j\leq 2j_M$ is the maximal number of cubes at scale
$j_M+1\geq j'\geq j$ containing $\De$. As 
$j=j_{\bar h}\leq l_{v_{\bar h}}$, only vertices localized in
these cubes can contribute.  The factor 2 appears because
there is only one vertex localized in each cube and at most
2 antifields hooked to that vertex. The sum over $\De$ is performed 
as in the case {\bf 1.)}. Therefore
\be
\Si_{\bar h}\leq  \ C \ j_M\    
M^{\frac{16}{3}} \  M^{-\frac{k_h}{3}}
\sum_{j\in {\cal I}_{h}}\  M^{-4j}
\ee
Now 
\be
 M^{-4j}=  M^{-4(j-i_r(h))} \  M^{-4i_r(h)}
\leq  M^{-4(j- i_r(h))} \ 2^{-r+1}\ I_h
\ee
The sum over $j$ is performed with the decay $ M^{-4(j-i_r(h))}$
\be
 \sum_{j\in {\cal I}_{h}\atop j\geq i_r(h) }\ M^{-4(j- i_r(h))}
\leq C
\ee
Finally 
\be
\Si_{\bar h}\  \leq  \ C \  M^{\frac{16}{3}} \  
I_h \ 2^{-r}\  M^{-\frac{k_h}{3}}   j_M\
\ee
where all constants have been inserted in $C$. 

\paragraph{3.)} 
$\bt_{h}=3$ which means that $h$ can contract only with
${\bar h}$ of type 3 ($\al_{\bar h}=3$). Therefore all ${\bar h}$
in the sum  are hooked to some $v\in V_d$ and have
$j_{\bar h}> l_v$.
The weight is 
\be
I_{\bar h} = \ M^{-4l_{v_{\bar h}}} =  \ M^{-4i_r(h)}
\label{I3}\ee
where $i_r(h)$ is the unique scale for which (\ref{I1})
is satisfied. Then 
\be
\Si_{\bar h}\leq 
 \sum_{j\in {\cal I}_{h}}\  \sum_{\De\in {\cal D}_{j}} \ 
\left |  C^{j k_h}(x_{\De_h^j},x_{\De}) \right |^2
\sum_{\De'\in {\cal D}_{i_r(h)}, \De' \subset\De } 2
\ee
where 2 is the maximal number of antifields with $\De_v\subseteq \De'$
that are hooked to the vertex $v_l$ of the  vertical link
$l\in vL_{i_{\De'}}$  connecting 
the connected component $y_k^{j}$ ($j=i_{\De'}$)  containing $\De'$
to its ancestor.
Now we observe that
\be
\sum_{\De'\in {\cal D}_{i_r(h)}, \De' \subset\De } 
2 \ \leq \ 2 \ M^{4j}\  M^{-4i_r(h)}
\ee
where $ M^{4j-4i_r(h)}$ is the number of cubes of scale $i_r(h)$
contained in a cube of scale $j$.  By (\ref{I1})-(\ref{I3})
we see that
\be
 M^{-4i_r(h)} \leq  I_h \ 2^{-r+1} 
\ee
hence $\Si_{\bar h}$ is bounded by
\be
\Si_{\bar h} \ \leq \ 2  I_h \ 2^{-r+1}  
\sum_{j\in {\cal I}_{h}} M^{4j}\  
\sum_{\De\in {\cal D}_{j}} 
\left |  C^{j k_h}(x_{\De_h^j},x_{\De}) \right |^2
\ee
The sum over $\De$ is bounded as in the case {\bf 1.)} above. Therefore
\bqa
\Si_{\bar h} & \leq &  \  C \  M^{\frac{16}{3}} \    
I_h \ 2^{-r}\  
\sum_{j\in {\cal I}_{h}} M^{4j}\  
M^{-4j} \  M^{-\frac{k_h}{3}} \leq C \  
M^{\frac{16}{3}} \  I_h \ 2^{-r} 
\ M^{-\frac{k_h}{3}} |{\cal I}_{h}| \no\\
&\leq & C \  
M^{\frac{16}{3}} \  I_h \ 2^{-r}
\ M^{-\frac{k_h}{3}} j_M
\eqa
where $ |{\cal I}_{h}|\leq j_M+1\leq 2j_M$
and all constant factors are absorbed in $C$.

\paragraph{4.)} 
$\bt_{h}=4$ which means that $h$ can contract only with
${\bar h}$ of type 4 ($\al_{\bar h}=4$). Therefore all ${\bar h}$
in the sum  are associated to some $f$-link of order 6 and its weight is 
\be
I_{\bar h} =\   M^{-4i_{\bar h}} =  \  M^{-4i_r(h)}
\ee
where $i_r(h)$ is the unique scale for which (\ref{I1})
is satisfied. Then 
\be
\Si_{\bar h}\leq 
 \sum_{j\in {\cal I}_{h}}\  \sum_{\De\in {\cal D}_{j}} \ 
\left |  C^{j k_h}(x_{\De_h^j},x_{\De}) \right |^2
\sum_{\De'\in {\cal D}_{i_r(h)}, \De' \subset\De } 6
\ee
where 6 is the maximal number of antifields with $\De_v\subset \De'$
that have been derived by a $f$-link of order 6 at scale $i_{\De'}$ for 
the connected component $y_k^{j}$ ($j=i_{\De'}$)  containing $\De'$.
Now we can apply the same analysis as for the case {\bf 3.)} 
except that instead of a factor 2 we have a factor 6.
Hence we obtain
\be
\Si_{\bar h} \leq  C \  
M^{\frac{16}{3}} \  I_h \ 2^{-r}
\ M^{-\frac{k_h}{3}} j_M
\ee

\paragraph{5.)} 
$\bt_{h}=5$ which means that $h$ can contract only with
${\bar h}$ of type 5 ($\al_{\bar h}=5$). 
Therefore all ${\bar h}$
in the sum  are hooked to some $v\in V_d$ and have
$j_{\bar h}= i_v$. The weight is 
\be
I_{\bar h} = \ M^{-4i_{v_{\bar h}}} =  \ M^{-4i_r(h)} \label{I5}
\ee
where $i_r(h)$ is the unique scale for which (\ref{I1})
is satisfied. There is no sum over $j$ to compute, as
we have only $j=i_r(h)$. 
\be
\Si_{\bar h}\leq 
\sum_{\De\in {\cal D}_{i_r(h)}} \ 
\left |  C^{i_r(h) k_h}(x_{\De_h^{i_r(h)}},x_{\De}) \right |^2
n({\bar h})
\ee
We know that $s$ is negative, and 
by (\ref{additexp}) (and the fact that $n(h)=1$), we obtain 
$n({\bar h}) \le 2^{-s}= 2^{|s|}$. Therefore
\be
\Si_{\bar h}\leq   2^{|s|}
\sum_{\De\in {\cal D}_{i_r(h)}} \ 
\left |  C^{i_r(h) k_h}(x_{\De_h^{i_r(h)}},x_{\De}) \right |^2
\ee
The sum over $\De$ is performed as in the other cases then
\be
\Si_{\bar h}\leq  \   C \  2^{|s|}
M^{\frac{16}{3}} \ M^{-\frac{k_h}{3}}  \ M^{-4i_r(h)}
\ee
Applying (\ref{I5}) we have
\be
\Si_{\bar h}\leq  C \    2^{|s|} \ 2^{-r}\ I_h
M^{\frac{16}{3}} \ M^{-\frac{k_h}{3}}  
\ee
\medskip

Now we can insert all these bounds in (\ref{haddetr}):
\bqa
\left |\det {\cal M}_{r,s}\lp \{\bt_h\}\{\bt_{\bar h}\}\rp\right |
&\leq  & \lp C\ M^{\frac{8}{3}}\rp^{n_{r,s}}
\prod_{\left \{{h\not \in b, r_h=r\atop s_h=s, \bt_h=1,...,4}\right \}}
\left [  \   I_h^{\frac{1}{2}} \ 
2^{-\frac{r}{2}}\  j_M^{\frac{1}{2}}  
\ M^{-\frac{k_h}{6}}  \right ]\no\\
&& \prod_{\left \{{h\not \in b, r_h=r \atop s_h=s, \bt_h=5}\right \}}
\left [  \   I_h^{\frac{1}{2}} \ 
2^{-\frac{r}{2}}\  2^{\frac{|s|}{2}}  
\ M^{-\frac{k_h}{6}}  \right ]\label{dethadr1}
\eqa
where $C$ is a constant and $n_{r,s}$ is the number of fields 
belonging to the matrix ${\cal M}_{r,s}$.
Now we observe that
\bqa
&& \prod_{\left \{{h\not \in b, r_h=r\atop s_h=s,  \bt_h=\bt}\right \}}
\left [ I_h^{\frac{1}{2}} \  2^{-\frac{r}{2}} \right ]
\leq  
\prod_{\left \{{h\not \in b, r_h=r \atop s_h=s,  \bt_h=\bt}\right \}}
\left [ I_h^{\frac{1}{4}} \  2^{\frac{r}{4}}\   2^{-\frac{r}{2}} \right ]
\prod_{\left \{{{\bar h}\not \in b, r_{\bar h}=-r\atop s_{\bar h}=-s,  
\al_{\bar h}=\bt}\right \}}
\left [ I_{\bar h}^{\frac{1}{4}}  \right ]\no\\
&&= \prod_{\left \{{h\not \in b, r_h=r\atop s_h=s,  \bt_h=\bt}\right \}}
\left [ I_h^{\frac{1}{4}} \  2^{-\frac{r}{8}}  \right ]
\prod_{\left \{{{\bar h}\not \in b, r_{\bar h}=-r\atop s_{\bar h}=-s,  
\al_{\bar h}=\bt}\right \}}
\left [ I_{\bar h}^{\frac{1}{4}} \  2^{-\frac{r}{8}}  \right ]
\eqa
where we applied the relation (\ref{I1}) and the fact that 
$| \{h\ | r_h=r, s_h=s, \bt_h=\bt\} | = 
| \{{\bar h}\ | r_{\bar h}=-r, s_{\bar h}=-s, \al_{\bar h}=\bt\} | $ for 
$\bt=1,...,5$. Moreover
\be
\prod_{\left \{{h\not \in b, r_h=r\atop s_h=s}\right \}}
\left [  \ M^{-\frac{k_h}{6}} \right ] = 
 \prod_{\left \{{h\not \in b,r_h=r \atop s_h=s}\right \}}
\left [  \ M^{-\frac{k_h}{12}} \right ] 
\prod_{\left \{{{\bar h}\not \in b, r_{\bar h}=-r\atop s_{\bar h}=-s}\right \}}
\left [  \ M^{-\frac{k_{\bar h}}{12}} \right ] 
\ee
since $|\{h\ | r_h=r,s_h=s, k_h=k\}| = 
|\{{\bar h}\ | r_{\bar h}=-r, s_{\bar h}=-s,k_{\bar h}=k\}|$
for any $k\geq 0$ and
\bqa
&&
\prod_{\left \{{h\not \in b, r_h=r\atop s_h=s, \bt_h=1,...,4}\right \}}
\left [ j_M^{\frac{1}{2}} \right ]  
= \prod_{\left \{{h\not \in b, r_h=r\atop s_h=s, \bt_h=1,...,4}\right \}}
\left [ j_M^{\frac{1}{4}} \right ]  \ 
\prod_{\left \{{{\bar h}\not \in b, r_{\bar h}=-r\atop s_{\bar h}=-s, 
\al_{\bar h}=1,...,4}
\right \}}
\left [ j_M^{\frac{1}{4}} \right ] \no\\
&& 
\leq  \prod_{\left \{{h\not \in b, r_h=r\atop s_h=s, \al_h=1,...,4}\right \}}
\left [ j_M^{\frac{1}{4}} \right ]  \ 
\prod_{\left \{{{\bar h}\not \in b, r_{\bar h}=-r\atop s_{\bar h}=-s, 
\al_{\bar h}=1,...,4}
\right \}}
\left [ j_M^{\frac{1}{4}} \right ]  
\eqa
where we  applied the relation 
$| \{h\ | \   r_h=r, s_h=s, \al_h=1,...,4\} |
= | \{{\bar h}\ | \   r_{\bar h}=-r, s_{\bar h}=-s, \al_{\bar h}=1,...,4\} | +
| \{{\bar h}\ | \   r_{\bar h}=-r, s_{\bar h}=-s, \al_{\bar h}=5\} |$
which is true because $\al_h<5$ $\forall h$.
Now, for any $h$ with $\bt_h=5$ there is no factor $j_M$ therefore
we write $1\leq j_M^{\frac{1}{4}}$.

Finally we observe that (see (\ref{defnh})):
\bqa
&&\prod_{\left \{{h\not \in b, r_h=r\atop s_h=s, \bt_h=5}\right \}}
2^{ |s| /2} = 
\prod_{\left \{{{\bar h}\not \in b, r_{\bar h}=-r\atop s_{\bar h}=-s, 
\al_{\bar h}=5}\right \}}
2^{|s| / 2} \leq 
\prod_{\left \{{{\bar h}\not \in b, r_{\bar h}=-r,\atop s_{\bar h}=-s, 
\al_{\bar h}=5}\right \}}
2^{-|s| / 2 }
\ 2 n_d(\De_{i_{v_{\bar h}}}) 
\no\\
&& 
\le \prod_{\left \{{ h \not \in b, r_{h}= r,\atop s_{h}= s }\right \}}
2^{-|s| / 4 }
\prod_{\left \{{{\bar h}\not \in b, r_{\bar h}=-r,\atop s_{\bar h}=-s, 
\al_{\bar h}=5}\right \}}
2^{-|s| / 4 }
\ 2 n_d(\De_{i_{v_{\bar h}}})
\eqa
where we apply the inequality $2^{|s|}\leq 2  n_d(\De_{i_{v_{\bar h}}}) $.
The determinant $|\det {\cal M}_{r,s}|$ is then bounded by
\bqa
&& \left |\det  
{\cal M}_{r,s}\right |\ 
\leq \  \lp C \ M^{\frac{8}{3}}\rp^{n_{r,s}}\
\prod_{\left \{{h\not \in b, r_h=r\atop s_h=s, \al_h<5}\right \}}
\left [  \  I_h^{\frac{1}{4}} \ 
2^{-\frac{r}{8}}\ 2^{- \frac{|s|}{4}}
M^{-\frac{k_h}{12}} j_M^{\frac{1}{4}}  \right ]
\no\\
&& \prod_{\left \{{{\bar h}\not \in b, r_{\bar h}=-r\atop s_{\bar h}=-s, 
\al_{\bar h}<5}\right \}}
\left [  \   I_{\bar h}^{\frac{1}{4}} \ 
2^{-\frac{r}{8}}\ 2^{- \frac{|s|}{4}} j_M^{\frac{1}{4}}  
\ M^{-\frac{k_{\bar h}}{12}} \right ] 
\\&&
\prod_{\left \{{{\bar h}\not \in b, r_{\bar h}=-r\atop s_{\bar h}=-s, 
\al_{\bar h}=5}\right \}}
\left [  \   I_{\bar h}^{\frac{1}{4}} \ 
2^{-\frac{r}{8}}\   2^{-\frac{ |s|}{4}}
M^{-\frac{k_{\bar h}}{12}}
\ n_d(\De_{i_{v_{\bar h}}})    \right ]\no
\eqa
where $n_{r,s}$ is the number of fields in the determinant.
Inserting the definitions for $I$ we can write the  determinant
as
\bqa
&& \hspace{-0.7cm}\left |\det  
{\cal M}_{r,s}\right |\ 
\leq \    \lp C \ M^{\frac{8}{3}}\rp^{n_{r,s}}\
\prod_{\left \{{h\not \in b, r_h=r\atop s_h=s, \al_h=1}\right \}}
\left [   M^{-i_{\De_{h}}}  
n_{\De_{h}}^{\frac{1}{4}}
\frac{1}{f_{\De_{ h}}^{\frac{1}{4}}}
2^{-\frac{r}{8}} 2^{-\frac{ |s|}{4}} M^{-\frac{k_h}{12}} 
\  j_{M}^{\frac{1}{4}}
\right ]\no\\
&& \hspace{-0.7cm}
\prod_{\left \{{h\not \in b, r_h=r\atop s_h=s, \al_h=2}\right \}}
\left [     M^{-i_{v_{h}}}      
2^{-\frac{r}{8}} 2^{ - \frac{|s|}{4}}
 M^{-\frac{k_h}{12}}\  j_M^{\frac{1}{4}} \right ]
%\no\\&& 
\hspace{-0.1cm}
\prod_{\left \{{h\not \in b, r_h=r\atop s_h=s, \al_h=3}\right \}}
\left [   M^{-l_{v_h}}     
2^{-\frac{r}{8}} 2^{-\frac{|s|}{4}}
M^{-\frac{k_h}{12}}\   j_{M}^{\frac{1}{4}}
\right ]
\no\\
&& \hspace{-0.7cm}
\prod_{\left \{{h\not \in b,r_h=r \atop s_h=s, \al_h=4}\right \}}
\left [   M^{-i_{ h}}   2^{-\frac{r}{8}}  2^{-\frac{ |s|}{4}}
M^{-\frac{k_h}{12}}  \   j_{M}^{\frac{1}{4}}
\right ]
\\
&&  \hspace{-0.7cm}
\prod_{\left \{{{\bar h}\not \in b,r_{\bar h}=-r \atop s_{\bar h}=-s, 
\al_{\bar h}=1}\right \}}
\left [  
 M^{-i_{\De_{\bar h}}}  
n_{\De_{\bar h}}^{\frac{1}{4}}
\frac{1}{f_{\De_{\bar h}}^{\frac{1}{4}}}
2^{-\frac{r}{8}}   2^{- \frac{|s|}{4}}
 M^{-\frac{k_{\bar h}}{12}}  \   j_M^{\frac{1}{4}} \right ]\no\\
&& \hspace{-0.7cm}
\prod_{\left \{{{{\bar h}\not \in b, r_{\bar h}=-r \atop s_{\bar h}=-s,} \atop
\al_{\bar h}=2}\right \}}\hspace{-0.2cm}
\left [    
  M^{-i_{v_{\bar h}}}  
2^{-\frac{r}{8}} 2^{-\frac{ |s|}{4}}
 M^{-\frac{k_{\bar h}}{12}}    j_M^{\frac{1}{4}}   \right ]\hspace{-0.2cm}
\prod_{\left \{{{{\bar h}\not \in b, r_{\bar h}=-r \atop s_{\bar h}=-s,} \atop
\al_{\bar h}=3}\right \}}\hspace{-0.2cm}
\left [  
  M^{-l_{v_{\bar h}}}  
2^{-\frac{r}{8}}\ 2^{-\frac{ |s|}{4}}
 M^{-\frac{k_{\bar h}}{12}}    j_M^{\frac{1}{4}} \right ]
\no\\
&& \hspace{-0.7cm}
\prod_{\left 
\{{{\bar h}\not \in b, r_{\bar h}=-r \atop s_{\bar h}=-s, 
\al_{\bar h}=4}\right \}}
\left [    
 M^{-i_{\bar h}}  
2^{-\frac{r}{8}} 2^{-\frac{ |s|}{4}}
 M^{-\frac{k_{\bar h}}{12}}     j_{M}^{\frac{1}{4}}  
\right ]
\no\\
&& \hspace{-0.7cm}
\prod_{\left \{{{\bar h}\not \in b, r_{\bar h}=-r \atop s_{\bar h}=-s, 
\al_{\bar h}=5}\right \}}
\left [   
  M^{-i_{v_{\bar h}}}  
2^{-\frac{r}{8}}   2^{-\frac{|s|}{4}} 
 M^{-\frac{k_{\bar h}}{12}} 
 n_d(\De_{i_{v_{\bar h}}}) 
\right ]
\no
\eqa
where we  applied the fact that  for $r>0$, $\al_h<5$ $\forall h$.

\subsubsection{Case $r=0$}

The subdeterminant for $r=0$ actually needs a more detailed 
analysis. We can write it as
\bqa
&& \det  
{\cal M}_{0,s}\lp \{\bt_h\}\{\bt_{\bar h}\}\rp 
= \det {\cal M}_{0,s} (\leq 5, <5)\\ 
&&
\det {\cal M}_{0,s}(<5, =5)\det {\cal M}_{0,s}(=5, =5) \no
\eqa
where  the first subdeterminant contains contractions between
fields with $\al_h<5$ (which corresponds to $\bt_{\bar h}<5$)
and any antifield (which corresponds to $\bt_h\leq 5$), the second 
 subdeterminant contains contractions between
fields with $\al_h=5$ (which corresponds to $\bt_{\bar h}=5$)
and antifields  with $\al_{\bar h}<5$ 
(which corresponds to $\bt_h<5$). Finally 
the third subdeterminant contains contractions between
fields with $\al_h=5$ (which corresponds to $\bt_{\bar h}=5$)
and antifields  with $\al_{\bar h}=5$ 
(which corresponds to $\bt_h=5$).
 
In the first case ( $\al_h<5$) we apply exactly the same bound as for
$r>0$. In the second case ( $\al_h=5$, $\bt_h<5$) we apply the column
inequality (\ref{hadr}) and everything goes as in
the case $r>0$ exchanging fields and antifields.  

Finally in the third case we have some field with 
$\al_h=5$ contracting with some antifield with $\al_{\bar h}=5$.
Here again we optimize the Hadamard inequalities depending
on the sign of $s$. If $s\ge 0$ we apply the row inequality,
and symmetrically\footnote{This second optimization is not really necessary,
but nicer.}.

The two main weights are equal
\be
I_{\bar h} =  M^{-4i_{v_{\bar h}}} =   M^{-4i_{v_h}} = I_h
\ee
Remark  that there is no sum over $j$ to compute, as
we have only $j=i_{v_h}$. 
\be
\Si_{\bar h}\leq \sum_{\De\in {\cal D}_{i_{v_h}}} \ 
\left |  C^{i_{v_h} k_h}(x_{\De_h^{i_{v_h}}},x_{\De}) \right |^2
n({\bar h})
\ee
We know that $n({\bar h}) < 2^{-s+1}n(h) $ therefore
\be
\Si_{\bar h}\leq   2^{-|s|+1} n(h)
  \sum_{\De\in {\cal D}_{i_{v_h}}} \ 
\left |  C^{i_r(h) k_h}(x_{\De_h^{i_r(h)}},x_{\De}) \right |^2
\ee
The sum over $\De$ is performed as in the other cases and we get
\be
\Si_{\bar h}\leq   2^{-|s|+1} n(\De_{v_h})
 C \  
M^{\frac{16}{3}} \ M^{-\frac{k_h}{3}}  \ M^{-4i_{v_h}} 
=  C \    2^{-|s|} n(\De_{v_h}) \ I_h
M^{\frac{16}{3}} \ M^{-\frac{k_h}{3}}  
\ee
where the constant $2$ has been inserted into $C$.
As before we will distribute the factor $ 2^{-|s|}$ on both sides
of the determinant which gives again factors   $ 2^{-|s|/4}$
for each field or antifield of this determinant after the
Hadamard inequality. The other factors are unchanged.

\subsubsection{Result of the weight expansion}

The global determinant is bounded by

\bqa
&& \hspace{-0.5cm} \prod_{r,s} \left |\det  
{\cal M}_{r,s}\right |\ 
\leq \  C^{n}\   M^{\frac{16n}{3}}\
\prod_{ h\not \in b }
2^{-\frac{|r_h|}{8}-\frac{|s_h|}{4}}\ M^{-\frac{k_h}{12}} 
\prod_{ {\bar h}\not \in b }
2^{-\frac{|r_{\bar h}|}{8}-\frac{|s_{\bar h}|}{4}}\ M^{-\frac{k_{\bar h}}{12}} 
\no\\
&& \hspace{-0.4cm}
\prod_{\left \{{h\not \in b,\atop  \al_h=1}\right \}}
\left [    M^{-i_{\De_{h}}}  
n_{\De_{h}}^{\frac{1}{4}}
f_{\De_{h}}^{-\frac{1}{4}}
j_{M}^{\frac{1}{4}}  
\right ]
%\no\\&& \hspace{-0.4cm}
\prod_{\left \{{h\not \in b,\atop  \al_h=2,3}\right \}}
\left [  \    M^{-i_{v_{h}}}     \ j_{M}^{\frac{1}{4}}  \right ]
\prod_{\left \{{h\not \in b,\atop \al_h=4}\right \}}
\left [  \   M^{-i_{ h}}     \ j_{M}^{\frac{1}{4}}  
\right ]\no\\
&& \hspace{-0.4cm}
\prod_{\left \{{h\not \in b,\atop  \al_h=5}\right \}}
\left [  \    M^{-i_{v_{h}}}     \ 
 n_d(\De_{i_{v_h}}) \right ]
 \prod_{\left \{{{\bar h}\not \in b,\atop  
\al_{\bar h}=1} \right \}}
\left [  M^{-i_{\De_{\bar h}}}  
n_{\De_{\bar h}}^{\frac{1}{4}} f_{\De_{\bar h}}^{-\frac{1}{4}}
 j_M^{\frac{1}{4}}  \right ]
\prod_{\left \{{{\bar h}\not \in b,\atop 
\al_{\bar h}=2,3}\right \}} \left [  \  M^{-i_{v_{\bar h}}}   \ 
 j_M^{\frac{1}{4}}   \right ]
\no\\&&\hspace{-0.4cm}
\prod_{\left \{{{\bar h}\not \in b,\atop  \al_{\bar h}=4}\right \}}
\left [  \  M^{-i_{\bar h}}   \ j_{M}^{\frac{1}{4}}  \right ]
\prod_{\left \{{{\bar h}\not \in b,\atop 
\al_{\bar h}=5}\right \}}
\left [  \    M^{-i_{v_{\bar h}}}    \  
 n_d(\De_{i_{v_{\bar h}}})  \right ]\label{bigequa}
\eqa
where we have applied   $\sum_{r,s} n_{r,s}\leq 2n$, 
and all numerical factors have been absorbed in the constant $C$.
The factors $M^{-l_{v_h}}$ have been moreover bounded by
$M^{-i_{v_h}}$.
Inserting this result inside (\ref{Ac1}) we have

\bqa
&&\hspace{-0.7cm}
\sum_{Y\atop 0\in Y} |A_c(Y)| L^{|Y|} \leq 
\sum_{M_Y}
 \sum_{S}\sum_{VL} L^{|Y|} \sum_{B_{S}} \ 
\sum_{\{x_\De\}^c}\ 
 \sum_{n=0}^\infty 
\frac{|\la|^n  \lp C M^{\frac{16}{3}}\rp^n}{n!} 
\sum_{V_d, \al_{V_d}} \sum_{a,b,R}
\no\\ 
&& \hspace{-0.7cm} \sum_{\{v_l\}_{l\in vL}} \sum_{n_{V_d}\si_{V_d}\rho_{V_d}}
\sum_{\{n_\De\}_{\De\in B_S}}  \sum_{\De^c_{{\bar V}_d}} 
\left [\prod_{v\in V_d} 
\sum_{i_{v}\in I^c_v} \sum_{\De_v\in {\cal D}_{i_v}\cap Y}  \right ]\ 
\sum_{\{J_h^a\}, \{ J_{\bar h}^a\}}  
\sum_{\{j_h^b\},\{j_{\bar h}^b\}}     \sum_{\{k_h\}, \{k_{\bar h}\}}
\no\\ 
&& \hspace{-0.5cm}
\lp\prod_{j=m_Y}^{M_Y}  \prod_{k=1}^{c_j} \sum_{T_{jk}} \rp
\lp\prod_{v\in V_d} \int_{\De_v} dx_v \rp \
\lp\prod_{v\in V_b\backslash V_d} \int_{Ex(\De_v)} dx_v \rp \
\lp\prod_{v \in {\bar V}_d\backslash V_b} M^{4i_{\De_v}} \rp   
\no\\
&&\hspace{-0.6cm}
\left [ \prod_{j=m_Y}^{M_Y} 
\prod_{k=1}^{c_j} \left [
\prod_{l\in T_{jk}} 
\left | C^{j k_l}_{\De_{l}{\bar \De}_{l}}\lp x_l,{\bar x}_l\rp 
\right|\right ] 
\de_{k_{h_l} k_{{\bar h}_l}} \right ]
 \sum_{\{r_h\},\{r_{\bar h}\} }
\sum_{\{s_h\},\{s_{\bar h}\} } 
\\
&&\hspace{-0.7cm}  
\left [
\prod_{\{h\not \in b \}}   2^{-\frac{|r_h|}{8}-\frac{|s_h|}{4}}
\prod_{\{{\bar h}\not \in b\}}
2^{-\frac{|r_{\bar h}|}{8}-\frac{|s_{\bar h}|}{4}}\right ]  \hspace{-0.1cm} 
\left [\prod_{j=m_Y+1}^{M_Y}
\lp \prod_{l\in vL_j}\int_0^1 dw'_{l} \rp\right] \hspace{-0.1cm} 
\left [\prod_{v\in V_d\atop c_v\neq \al_v} 
\prod_{j= i_v}^{l_v} w'_{y_v^j}\right]   
\no\\
&& \hspace{-0.5cm}  
\left [\prod_{v\in V_d\atop c_v= \al_v} 
 \prod_{j= i_v}^{l_v-1} 
w'_{y_v^j}\right] \left [
\prod_{\{h\not \in b \}} \left [  M^{-\frac{k_h}{12}} \right ] \
\prod_{\{{\bar h}\not \in b \}} \left [  M^{-\frac{k_{\bar h}}{12}}
\right ]\right ] \
\prod_{v\in V_d\cup V_b} 
|\ln T|^{3/4} 
\no\\
&& \hspace{-0.5cm}  
\prod_{v \not \in V_d\cup V_b} |\ln T|  
\sum_{\{\bt_h\}\{\bt_{\bar h}\}} \ \left [  
\prod_{\left \{{ h\not \in b \atop \al_h=1}\right \}}
\left [   M^{-i_{\De_h}}  
n_{\De_h}^{\frac{1}{4}}
f_{\De_h}^{-\frac{1}{4}}\right ] \
\prod_{\left \{{{\bar h}\not \in b,\atop \al_{\bar h}=1}\right \}}
\left [   M^{-i_{\De_{\bar h}}}  
n_{\De_{\bar h}}^{\frac{1}{4}}
f_{\De_{\bar h}}^{-\frac{1}{4}}\right ] \right ]  \no\\
&&\hspace{-0.5cm} \left [ 
\prod_{\left \{{h\not \in b,\atop \al_h=2,3}\right \}}
 M^{-i_{v_h}}  \
\prod_{\left \{{{\bar h}\not \in b,\atop  
\al_{\bar h}=2,3}\right \}} M^{-i_{v_{\bar h}}} \right ] \
\left [ 
\prod_{\left \{{h\not \in b,\atop \al_h=4}\right \}}
 M^{-i_h}  \
\prod_{\left \{{{\bar h}\not \in b,\atop  
\al_{\bar h}=4}\right \}} M^{-i_{\bar h}} \right ] \
\no\\
&& \hspace{-0.5cm} 
\left [ 
\prod_{\left \{{h\not \in b,\atop \al_h=5}\right \}}
 M^{-i_{v_h}} n_d(\De_{i_{v_h}}) \hspace{-0.2cm}  
\prod_{\left \{{{\bar h}\not \in b,\atop  
\al_{\bar h}=5}\right \}} M^{-i_{v_{\bar h}}}  
n_d(\De_{i_{v_{\bar h}}}) \right ] 
\no
\eqa
where 
\be
\left |C^{j k_l}_{\De_{l}{\bar \De}_{l}} \lp x_{l},{\bar x}_{l},\{w'_{l'}\},
\{w''_{l'}\}
\rp\right | \leq \left |C^{j k_l}\lp x_{l},{\bar x}_{l}\rp\right |  \ .
\ee
To get the factor $\prod_{v\in V_d\cup V_b} \left [ 
|\ln T|^{3/4} \right ] 
\prod_{v \not \in V_d\cup V_b} |\ln T|$ in
this bound we collected the factors $j_M^{1/4}$, which are bounded 
by $ |\ln T|^{1/4}$ (\ref{jM}), and we used the fact that a vertex
$v\in V_d \cup V_b$ either is in $V_d$, hence has a field or antifield
of type 5 hooked to it, which has no  $j_M^{1/4}$ factor,
or is in  $V_b - V_d$, hence has at least a
field or antifield in $b$ which does not appear in the products
of (\ref{bigequa}).

The integrals over the weakening factors
$w_l$ and $w''_l$ have been bounded by one, but the ones over
$w'_l$ are kept preciously since they are used below. 
Now we observe that
\be
\sum_{\{r_h\},\{r_{\bar h}\} \atop \{s_h\},\{s_{\bar h}\} } 
\left [
\prod_{\{h\not \in b \}} \left [  2^{-\frac{|r_h|}{8}-\frac{|s_h|}{4}}\ 
\right ]
\prod_{\{{\bar h}\not \in b\}}
\left [ 2^{-\frac{|r_{\bar h}|}{8}-\frac{|s_{\bar h}|}{4}}\right ] \right ] 
\leq C^n 
\ee

The logarithms are bounded using the relation
$|\ln T| |\la|   \leq K $. Hence we can write (since we assumed $K\le 1$)   
\be
\prod_{v\in V_d\cup V_b} \ |\la|\ 
|\ln T|^{3/4}  \prod_{v\not \in V_d\cup V_b} 
\ |\la|\ |\ln T| \ 
 \leq \  K^{|{\bar V_d}\backslash V_b|} \ \prod_{v\in V_d\cup V_b} \ 
 |\la|^{\frac{1}{4}} 
\ee
The $n_\De$ and $n(\De)$ 
factors coming from the Hadamard bound can be estimated
using Stirling's formula
as follows:
\bqa
&& \prod_{\left \{{h\not \in b,\atop \al_h=1}\right \}}
n_{\De_h}^{\frac{1}{4}} 
\prod_{\left \{{{\bar h}\not \in b,\atop \al_{\bar h}=1}\right \}}
n_{\De_{\bar h}}^{\frac{1}{4}} \leq \prod_{\De\in B_S} n_\De^{n_\De}
\leq \prod_{\De\in B_S} n_\De! \ e^{n_\De} = e^{n} \ 
\prod_{\De\in B_S} n_\De! \no\\
&& \prod_{\left \{{h\not \in b,\atop \al_h=5}\right \}}
n_d(\De_{i_{v_h}}) 
\prod_{\left \{{{\bar h}\not \in b,\atop \al_{\bar h}=5}\right \}}
n_d(\De_{i_{v_{\bar h}}}) \leq \prod_{\De\in Y} n_d(\De)^{n_d(\De)}
\leq  e^{n} \ 
\prod_{\De\in Y} n_d(\De)! \no\\
\eqa
Inserting all these results  and absorbing all constants
except $K$ in the global factor $C^n$ we have

\bqa
&&\hspace{-0.7cm}\sum_{Y\atop 0\in Y} |A_c(Y)| L^{|Y|}\  
\leq \ 
\sum_{M_Y}
 \sum_{S}\sum_{VL} L^{|Y|}\sum_{B_{S}} \ 
\sum_{\{x_\De\}^c}\  
\sum_{n=0}^\infty \frac{\lp C M^{\frac{16}{3}} M^4\rp^n}{n!} 
K^{|{\bar V_d}\backslash V_b|} \sum_{V_d, \al_{V_d}}
\no\\ 
&& \hspace{-0.5cm}
 \sum_{a,b,R} \sum_{\{v_l\}_{l\in vL}} \ 
\sum_{n_{V_d}\si_{V_d}\rho_{V_d}}
%\no\\ 
%&& \hspace{-0.5cm}
\sum_{\{n_\De\}_{\De\in B_S}}\  \sum_{\De^c_{{\bar V}_d}} 
\left [\prod_{v\in V_d} 
\sum_{i_{v}\in I^c_v} \sum_{\De_v\in {\cal D}_{i_v}\cap Y}  \right ]\ 
\sum_{\{J_h^a\}, \{ J_{\bar h}^a\}} \  
\sum_{\{j_h^b\},\{j_{\bar h}^b\}} \  
\no\\ 
&& \hspace{-0.5cm}
\sum_{\{k_h\}, \{k_{\bar h}\}} \sum_{\{\bt_h\}\{\bt_{\bar h}\}} \ 
\left [ \prod_{v\in V_d\cup V_b} \ 
|\la|^{\frac{1}{4}}  \right ]
\left [ \prod_{v \not\in V_d\cup V_b} M^{4i_{\De_v}} \right ]  
\left [\prod_{\De\in Y} n_d(\De)!\right ]   \ 
\no\\
&&\hspace{-0.5cm}
\left [\prod_{\De\in B_S} n_\De! \right ] \left [
\prod_{\{h\not \in b \}}  M^{-\frac{k_h}{12}}  \
\prod_{\{{\bar h}\not \in b \}}   M^{-\frac{k_{\bar h}}{12}}
\right ] \
\left [  
\prod_{\left \{{h\not \in b,\atop \al_h=1}\right \}}
M^{-i_{\De_h}} \
\prod_{\left \{{{\bar h}\not \in b,\atop \al_{\bar h}=1}\right \}}
M^{-i_{\De_{\bar h}}} \right ]  
\no\\
&&\hspace{-0.5cm} 
\left [ 
\prod_{\left \{{h\not \in b,\atop \al_h=2,3,5}\right \}}
 M^{-i_{v_h}}  \
\prod_{\left \{{{\bar h}\not \in b,\atop  
\al_{\bar h}=2,3,5}\right \}} M^{-i_{v_{\bar h}}} \right ] 
\left [ 
\prod_{\left \{{h\not \in b,\atop \al_h=4}\right \}}
 M^{-i_h}  \
\prod_{\left \{{{\bar h}\not \in b,\atop  
\al_{\bar h}=4}\right \}} M^{-i_{\bar h}} \right ] 
\no\\
&&\hspace{-0.5cm} 
\lp\prod_{j=m_Y}^{M_Y}  \prod_{k=1}^{c_j} \sum_{T_{jk}} \rp
\lp\prod_{v\in V_d} \int_{\De_v} dx_v \rp \
\lp\prod_{v\in V_b\backslash V_d} \int_{Ex(\De_v)} dx_v \rp \
\no\\
&&\hspace{-0.5cm} 
\left [ \prod_{j=m_Y}^{M_Y} 
\prod_{k=1}^{c_j} \left [
\prod_{l\in T_{jk}} 
\left | C^{j k_l}_{\De_{l}{\bar \De}_{l}}\lp x_l,{\bar x}_l\rp 
\right|\right ] 
\de_{k_{h_l} k_{{\bar h}_l}} \right ]  \no\\
&&\hspace{-0.5cm} \left [\prod_{j=m_Y+1}^{M_Y}
\lp \prod_{l\in vL_j}\int_0^1 dw'_{l}
 \rp\right] \
\left [\prod_{v\in V_d\atop c_v\neq \al_v} 
 \prod_{j= i_v}^{l_v} 
w'_{y_v^j}\right]   
\left [\prod_{v\in V_d\atop c_v= \al_v} 
 \prod_{j= i_v}^{l_v-1} 
w'_{y_v^j}\right] 
\eqa
where we applied $f_\De\geq M^{-4}$ to bound
every factor
$f_{\De_h}^{-\frac{1}{4}}$ by  $M $.
As we have at most four fields of type 1 per vertex
$v\in {\bar V}_d$ we obtain at most the factor $M^{4n}$.

\subsection{Extracting power counting}

In order to extract the power counting for $h$-links we define

\be
 C^{j k_l}_{\De_{l}{\bar \De}_{l}}\lp x_l,{\bar x}_l\rp =
M^{-j^b_h} \ M^{-j^b_{\bar h}} 
M^{-\vep k_h}\ M^{-\vep k_{\bar h}}\  D^{j k_l}_{\De_{l}{\bar \De}_{l}}
\lp x_l,{\bar x}_l\rp 
\label{defpropd}\ee
where $h$ and ${\bar h}$ are the field, antifield contracted
to form the propagator and $\vep$ is some small constant
 $0<\vep<1$ that will be determined later.
Remark that  $j^b_h=j^b_{\bar h}=j$,  
$k_h=k_{\bar h}=k_l$ by construction. The factor $M^{-\vep k_h}$
is necessary to sum over $k_h$ and extract a small
factor per cube. The factor
$M^{-j^b_h}$ corresponds to a kind of power counting for the field.

Now we can write
\bqa
&&\hspace{-0.7cm}\sum_{Y\atop 0\in Y} |A_c(Y)| L^{|Y|}\  
\leq \ 
\sum_{M_Y}
 \sum_{S}\sum_{VL} L^{|Y|}\sum_{B_{S}} \ 
\sum_{\{x_\De\}^c}\  
\sum_{n=0}^\infty \frac{\lp C M^{\frac{28}{3}} \rp^n}{n!} 
K^{|{\bar V_d}\backslash V_b|} \sum_{V_d, \al_{V_d}}
\no\\ 
&& \hspace{-0.5cm}
 \sum_{a,b,R} \sum_{\{v_l\}_{l\in vL}} \ 
\sum_{n_{V_d}\si_{V_d}\rho_{V_d}}
%\no\\ 
%&& \hspace{-0.5cm}
\sum_{\{n_\De\}_{\De\in B_S}}\  \sum_{\De^c_{{\bar V}_d}} 
\left [\prod_{v\in V_d} 
\sum_{i_{v}\in I^c_v} \sum_{\De_v\in {\cal D}_{i_v}\cap Y}  \right ]\ 
\sum_{\{J_h^a\}, \{ J_{\bar h}^a\}} \  
\sum_{\{j_h^b\},\{j_{\bar h}^b\}} \ 
\no\\ 
&& \hspace{-0.5cm}
\sum_{\{k_h\}, \{k_{\bar h}\}} \sum_{\{\bt_h\}\{\bt_{\bar h}\}} \ 
\left [ \prod_{v\in V_d\cup V_b} \ 
|\la|^{\frac{1}{4}}   
\right ]
\left [ \prod_{v \not\in V_d\cup V_b} M^{4i_{\De_v}} \right ]  
\left [\prod_{\De\in Y} n_d(\De)!\right ]   \ 
\no\\
&&\hspace{-0.5cm}
\left [\prod_{\De\in B_S} n_\De! \right ] \left [
\prod_{\{h\not \in b \}}  M^{-\frac{k_h}{12}}  \
\prod_{\{{\bar h}\not \in b \}}   M^{-\frac{k_{\bar h}}{12}}
\right ] \
 \left [
\prod_{\{h\in b \}}  M^{-\vep k_h}  \
\prod_{\{{\bar h} \in b \}}   M^{-\vep k_{\bar h}}
\right ] \
\no\\
&&\hspace{-0.5cm} 
\left [  
\prod_{\left \{{h\not \in b,\atop \al_h=1,2,3,5}\right \}}
M^{-i_{\De_{h}}} \
\prod_{\left \{{{\bar h}\not \in b,\atop \al_{\bar h}=1,2,3,5}\right \}}
M^{-i_{\De_{{\bar h}}}} \right ]  
\left [ 
\prod_{\left \{{h\not \in b\ |\  \al_h=4, 
\atop {\rm or \ } h\in b }\right \}}
 M^{-i_h}  \hspace{-0.5cm}
\prod_{\left \{{{\bar h}\not \in b\ | \   
\al_{\bar h}=4, \atop {\rm or \ } 
{\bar h}\in b}\right \}} M^{-i_{\bar h}} \right ] \ 
\no\\
&&\hspace{-0.5cm} 
\lp\prod_{j=m_Y}^{M_Y}  \prod_{k=1}^{c_j} \sum_{T_{jk}} \rp
\lp\prod_{v\in V_d\cup V_b} \int_{\Om_v} dx_v \rp \
\left [ \prod_{j=m_Y}^{M_Y} 
\prod_{k=1}^{c_j} \left [
\prod_{l\in T_{jk}} 
\left | D^{j k_l}_{\De_{l}{\bar \De}_{l}}\lp x_l,{\bar x}_l\rp 
\right|\right ] 
\de_{k_{h_l} k_{{\bar h}_l}} \right ]
  \no\\
&&\hspace{-0.5cm} \left [\prod_{j=m_Y+1}^{M_Y}
\lp \prod_{l\in vL_j}\int_0^1 dw'_{l}
 \rp\right] \
\left [\prod_{v\in V_d\atop c_v\neq \al_v} 
 \prod_{j= i_v}^{l_v} 
w'_{y_v^j}\right]   
\left [\prod_{v\in V_d\atop c_v= \al_v} 
 \prod_{j= i_v}^{l_v-1} 
w'_{y_v^j}\right] 
\eqa
where we defined $i_h=j^b_h$ if $h\in b$, 
and we defined $\Om_v=\De_{i_v}$ if $v\in V_d$ and 
 $\Om_v=Ex(\De_{v})$ if $v\in V_b\cap {\bar V_d}$.
We also defined $\De_h=\De_{v_h}\in B_S$ if $\al_h=1$
and   $\De_h=\De_{i_{v_h}}$  if $\al_h=2,3,5$.
Now for each $h$ with $i_h>i_{\De_h}$ we can write
\be
M^{-i_h} = M^{-i_{\De_h}} M^{-(i_h-i_{\De_h})}.
\ee
The same formulas hold for ${\bar h}$.
The factor $M^{-i_{\De_h}}$ will be used to compensate the 
integration over $x_v\in \De_{v_h}$. 

To extract power counting for a $v$-link associated to the vertex 
$v$ we extract a fraction $|\la|^{\frac{1}{8}}$ for
each vertex in $V_d$:
\be
 \prod_{v\in V_d} \ 
|\la|^{\frac{1}{4}}  \ =
 \prod_{v\in V_d} \ 
|\la|^{\frac{1}{8}} |\la|^{\frac{1}{8}}  \ .
\ee
Now, for each $y^j_k$ connected to its ancestor
by a $f$-link, there are 6 external fields. One of these may
be the field $h_{root}$. For this field we keep the
vertical decay $M^{-(j^b_h-i_{\De_h})}$  untouched,
in order to perform later the sum over the tree structure.

The vertical decay for the remaining five external
fields, together with the factors
$\la^{\frac{1}{8}}$  are  necessary for several purposes:
\begin{itemize}
\item{} to ensure a factor $M^{-4}$ to sum the root cube
for any $y^j_k$ inside a cube at scale $j+1$;
\item{} to sum over  $J^a_h$ and $j^b_h$;
\item{} to extract one small factor per cube;
\item{} to sum over the tree structure.
\end{itemize}
Therefore we write the vertical decay for each of
the five fields  as follows:
\be
 M^{-(i_h-i_{\De_h})} =  M^{-\frac{\vep'}{2}(i_h-i_{\De_h})}\  
 M^{-\frac{\vep'}{2}(i_h-i_{\De_h})}  \ M^{-(1-\vep')(i_h-i_{\De_h})} 
\ee
where $0<\vep'<1$ is some small constant that will be
chosen later.
One of the two fractions $\vep'/2$ is necessary to sum
over  $J^a_h$ and $j^b_h$, and   
to extract one small factor per cube. The other
fraction will be used to reconstruct some vertical decay
in order to sum over the tree.
Now we call $GF$ the set of subpolymers $y^j_k$ 
connected to their ancestor by a $f$-link, and
$GV$ the set of subpolymers $y^j_k$ 
connected to their ancestor by a $v$-link. 
Therefore we can write

\bqa
&&\hspace{1cm}
\prod_{\left \{{h\not \in b\ |\   \al_h=4,  {\ \rm or \ }
\atop h\in b \ {\rm and\ } h\not\in R_{root}}\right \}}
%\hspace{-0.5cm}
  M^{-(1-\vep')(i_h-i_{\De_h})}  M^{-\frac{\vep'}{2}(i_h-i_{\De_h})}
%\hspace{-0.5cm}
\\
&&\hspace{-1.5cm}\prod_{\left \{{{\bar h}\not \in b\ | \   
\al_{\bar h}=4, \  {\rm or \ } \atop h\in b\ {\rm and\ }
 {\bar h}\not\in R_{root} }\right \}} 
\hspace{-0.5cm}
M^{-(1-\vep')(i_{\bar h}-i_{\De_{\bar h}})} 
M^{-\frac{\vep'}{2}(i_{\bar h}-i_{\De_{\bar h}})}   \ 
\leq  \ \prod_{g_k^j\in GF}  M^{-5(1-\vep')} \ 
M^{-5\frac{\vep'}{2}}\no
\eqa
where we applied the equation
\be
M^{-(1-\vep')(i_h-i_{\De_h})}  = \prod_{j=i_h}^{i_{\De_h}-1}
 M^{-(1-\vep')[j-(j-1)]}
\ee

On the other hand for subpolymers with $v$-links we write
\be
\prod_{v\in V_d} |\la|^{\frac{1}{8}} \ = \ \prod_{y_k^j\in GV}
 |\la|^{\frac{1}{8}} \leq  
\left  [ 
\prod_{y_k^j\in GV} M^{-5(1-\vep')}
\right  ] \left [   
\prod_{v\in V_d} |\la|^{\frac{\vep'}{8}}
\right ]
\ee 
where from now on we assume
\be
 |\la|^{\frac{1}{8}} \leq  M^{-5}.
\ee

Finally we observe that for each $h\in R_{root}$ 
we can reconstruct a fraction of 
the vertical decay $j_h^b-i_{\De_h}$. 
This is possible because any cube $\De$ in the set 
\be
A_h =: \{\De\ |\ j_h^b> i_\De \geq i_{\De_{h}}\
{\rm and}\  \De_{h}\subseteq \De\}
\ee
 must be $\De^0_{root}$ for some connected 
component at scale $i_\De$, with $\De_h\subseteq \De$,
 and we can extract a fraction 
 $ |\la|^{\frac{\vep'}{16}}$ or  $M^{-5\frac{\vep'}{2}}$
of its vertical decay.
Remark that no field $h'\neq h \in R_{root}$ can hook
to any cube in $A_h$, because they are all cubes of type
 $\De^0_{root}$, therefore  $A_{h}\cap A_{h'}= \emptyset$ 
for any  $h'\neq h \in R_{root}$. This means 
that the same  $\De$ is never used for more than one
$h\in R_{root}$.
Therefore we can write
\be
 \left [ \prod_{v\in V_d} |\la|^{\frac{\vep'}{8}}\right ]\
 \left [ \prod_{y_k^j\in GF}  M^{-5\frac{\vep'}{2}}\right ]\ \leq 
 \left [  \prod_{h\in R_{root}}
M^{-5\frac{\vep'}{2}(i_h-i_{\De_h})}\right ]\  \left [
\prod_{v\in V_d}  |\la|^{\frac{\vep'}{16}} \right ]\
\ee
One of this fractions can be used to sum over $j^b_h$, the others
will be used to sum over the tree.
Inserting all these bounds  we have

\bqa
&&\hspace{-0.7cm}\sum_{Y\atop 0\in Y} |A_c(Y)| L^{|Y|}\  
\leq \ 
\sum_{M_Y}
 \sum_{S}\sum_{VL} L^{|Y|} \sum_{B_{S}} \ 
\sum_{\{x_\De\}^c}\  
\sum_{n=0}^\infty \frac{\lp C M^{\frac{28}{3}} \rp^n}{n!} 
K^{|{\bar V_d}\backslash V_b|} 
\no\\ 
&& \hspace{-0.5cm}
\sum_{V_d, \al_{V_d}} \sum_{a,b,R} \sum_{\{v_l\}_{l\in vL}} \ 
\sum_{n_{V_d}\si_{V_d}\rho_{V_d}}
\sum_{\{n_\De\}_{\De\in B_S}}\  \sum_{\De^c_{{\bar V}_d}} 
\left [\prod_{v\in V_d} 
\sum_{i_{v}\in I^c_v} \sum_{\De_v\in {\cal D}_{i_v}\cap Y}  \right ]\ 
\sum_{\{J_h^a\}, \{ J_{\bar h}^a\}} \  
\no\\ 
&& \hspace{-0.5cm}
\sum_{\{j_h^b\},\{j_{\bar h}^b\}} \  
\sum_{\{k_h\}, \{k_{\bar h}\}} \sum_{\{\bt_h\}\{\bt_{\bar h}\}} \ 
\left [ \prod_{v\in V_d\cup V_b} \ 
|\la|^{\frac{1}{16}}   
\right ] \left [ \prod_{v\in V_d\cup V_b} \ 
|\la|^{\frac{1}{16}}   
\right ] 
\no\\
&& \hspace{-0.5cm} 
\left [ \prod_{v \in V_d\cup V_b} M^{-4i_{\De_v}} \right ] 
\left [\prod_{\De\in Y} n_d(\De)!\right ] 
\left [\prod_{\De\in B_S} n_\De! \right ] \left [
\prod_{\{h\not \in b \}}  M^{-\frac{k_h}{12}}  \
\prod_{\{{\bar h}\not \in b \}}   M^{-\frac{k_{\bar h}}{12}}
\right ] \
\no\\
&&\hspace{-0.5cm} 
\no\\ 
&& \hspace{-0.5cm}
 \left [
\prod_{\{h\in b \}}  M^{-\vep k_h}  \
\prod_{\{{\bar h} \in b \}}   M^{-\vep k_{\bar h}}
\right ] \
\left [ 
\prod_{\{h\in R_{root} \}}
 M^{-(i_h-i_{\De_h})}  \
\prod_{\{ {\bar h}\in R_{root} \}} 
 M^{-(i_{\bar h}-i_{\De_{\bar h}})} \right ] \ 
\no\\
&&\hspace{-0.5cm} 
\left [ \prod_{v\in V_d} \ 
|\la|^{\frac{\vep'}{16}}\right ]
\left [ 
\prod_{\left \{{h\not \in b\ |\  \al_h=4
\atop {\rm or \ } h\in b}\right \}}
 M^{-\frac{\vep'}{2}(i_h-i_{\De_h})}  \
\prod_{\left \{{{\bar h}\not \in b\ | \   
\al_{\bar h}=4, \atop {\rm or \ } {\bar h}\in b}\right \}} 
 M^{-\frac{\vep'}{2}(i_{\bar h}-i_{\De_{\bar h}})}   \right ] \ 
\no\\
&&\hspace{-0.5cm} 
\left [ 
\prod_{\{h\in R_{root} \}}
 M^{-4\frac{\vep'}{2}(i_h-i_{\De_h})}  \
\prod_{\{ {\bar h}\in R_{root} \}} 
 M^{-4\frac{\vep'}{2}(i_{\bar h}-i_{\De_{\bar h}})} \right ] \ 
\left [ \prod_{j=m_Y}^{M_Y-1}  \prod_{k=1}^{c_j} M^{-5(1-\vep')} 
\right ]
\no\\
&& \hspace{-0.5cm} 
\lp\prod_{j=m_Y}^{M_Y}  \prod_{k=1}^{c_j} \sum_{T_{jk}} \rp
\lp\prod_{v\in V_d\cup V_b} \int_{\Om_v} dx_v \rp \
\left [ \prod_{j=m_Y}^{M_Y} 
\prod_{k=1}^{c_j} \left [
\prod_{l\in T_{jk}} 
\left | D^{j k_l}_{\De_{l}{\bar \De}_{l}}\lp x_l,{\bar x}_l\rp 
\right|\right ] 
\de_{k_{h_l} k_{{\bar h}_l}} \right ]
  \no\\
&&\hspace{-0.5cm} \left [\prod_{j=m_Y+1}^{M_Y}
\lp \prod_{l\in vL_j}\int_0^1 dw'_{l}
 \rp\right] \
\left [\prod_{v\in V_d\atop c_v\neq \al_v} 
 \prod_{j= i_v}^{l_v} 
w'_{y_v^j}\right]   
\left [\prod_{v\in V_d\atop c_v= \al_v} 
 \prod_{j= i_v}^{l_v-1} 
w'_{y_v^j}\right] \label{bigbig}
\eqa
where we applied
\be
\prod_{v\not\in V_d\cup V_b} M^{4i_{\De_v}} 
\prod_{v\not\in V_d\cup V_b} M^{-4i_{\De_v}} = 1. 
\ee
and
\be
\prod_{y^j_k\in GV} M^{-5(1-\vep')} 
\prod_{y^j_k\in GF} M^{-5(1-\vep')}  = 
\left [ \prod_{j=m_Y}^{M_Y-1}  \prod_{k=1}^{c_j} M^{-5(1-\vep')} 
\right ].
\ee
where we have extracted from $|\la|^{\frac{1}{8}}$ a fraction
 $|\la|^{\frac{1}{16}}$ that will be used to extract a small factor
per cube. The factors $M^{-5(1-\vep')}$  
will be used to sum over the cube positions and to perform the last sum 
over $S$. Remember that $\De_v$ is a cube in $B_S$ if 
$v\in V_b\cap {\bar V_d}$, but if $v\in V_d$, the localization cube
$\De_{i_v}$  of $v$ may not be a summit cube.

\subsection{Extracting a small factor per cube}

Now, before bounding the polymer structure, we must
extract a small factor $g$ for each cube, in order to
obtain a factor $g^{|Y|}$. 

First we still need to extract some fractions of vertical decay.
Actually, we will need also a fraction of the $k$
decay for tree lines. Therefore we write
\be
 M^{-\vep k_h}  \ = \  M^{-\frac{\vep}{2} k_h} \  
 M^{-\frac{\vep}{2} k_h}
\ee
for each $h,{\bar h}\in b$. One fraction will 
be used to sum over $k_h$, and the remaining fraction is
used to extract a small factor per cube.
Finally we need to extract a fraction $\vep'/4$ of the
vertical decay $M^{-\frac{\vep'}{2}(i_h-i_{\De_h})}$ for each
$h,{\bar h}\not \in b$ with $ \al_h=4$, and for each 
$h,{\bar h}\in b$.
One fraction will be used to sum over $J^a$ and $j^b$.
The remaining fraction is bounded by
\be
\left [ 
\prod_{\left \{{h\not \in b\ |\  \al_h=4
\atop {\rm or \ } h\in b}\right \}} \hspace{-0.5cm}
 M^{-\frac{\vep'}{4}(i_h-i_{\De_h})}  \hspace{-0.5cm}
\prod_{\left \{{{\bar h}\not \in b\ | \   
\al_{\bar h}=4, \atop {\rm or \ } {\bar h}\in b}\right \}} 
 \hspace{-0.5cm} 
M^{-\frac{\vep'}{4}(i_{\bar h}-i_{\De_{\bar h}})}   \right ]  
\hspace{-0.1cm} \left [ \prod_{v\in V_d}  
|\la|^{\frac{\vep'}{16}}\right ]
\leq   
\left [ \prod_{j=m_Y}^{M_Y-1}  \prod_{k=1}^{c_j} M^{-5\frac{\vep'}{4}} 
\right ].
\ee

Now we can prove the following lemma. 
\medskip

\noindent{\bf Lemma.} {\em One can extract from (\ref{bigbig}) 
at least one small
factor $g<1$ for each cube in $Y$, where $g$ is defined by
\be
g \ = \ \max [\  |\la|^{\frac{1}{32}}\ , \ M^{-5\frac{\vep'}{4d}} \
, \  M^{-\frac{\vep}{2d}} \ ]
\ee 
where $d= 3^4= 81$ is the number of nearest neighbors for each cube
(including itself).}
\medskip

\paragraph{Proof.} We will proof the following inequality
\be
\left [ \prod_{v\in V_d\cup V_b} \ 
|\la|^{\frac{1}{16}}\right ] \ 
\left [ \prod_{j=m_Y}^{M_Y-1}  \prod_{k=1}^{c_j} M^{-5\frac{\vep'}{4}} 
\right ] \ 
\left [ \prod_{h\in b} \ 
 M^{-\frac{\vep}{2} k_h} \  \prod_{{\bar h}\in b} \ 
 M^{-\frac{\vep}{2} k_{\bar h}}\right ] \ 
\leq \ g^{|Y|} \label{proof}
\ee
which is enough to prove the lemma. 

First we make some remarks. 
\paragraph{1)} For all extremal summit cube $\De\in Y$ ($Ex(\De)=\De$), 
there must be at least one vertex 
$v  \in V_d\cup V_b$ with $\De_v=\De$, as this cube must be connected to
the polymer by a horizontal or vertical link. 
For this vertex we have a factor $|\la|^{\frac{1}{16}}\leq g^2$.
Therefore we a factor $g^2$ for each  
$\De\in Y$ with $Ex(\De)=\De$. 

\paragraph{2)}  For all  $\De\in Y$ such that
$\De=\De^0_{root}$ for some connected subpolymer $y^j_k$,
there is a vertical link connecting $\De$ to its ancestor
and we have a fraction of the vertical
decay $ M^{-5\frac{\vep'}{4}}\leq g^d$.

\paragraph{3)}  For each tree line $C^{jk}$ connecting 
some $\De, \De'\in {\cal D}_j$, we can write the vertical decay 
$ M^{-\frac{\vep}{2} k}$ for the corresponding 
$h$ and ${\bar h }$ ($k_h=k_{\bar h}=k$) as 
\be
 M^{-\frac{\vep}{2} k} M^{-\frac{\vep}{2} k}
= \prod_{j'=j}^{j+k-1}  M^{-\frac{\vep}{2}} M^{-\frac{\vep}{2}}.
\ee
Therefore for all $\De''\in {\cal D}_{j'}$ with
$j\leq j'\leq j+k-1$ such that $\De\subseteq \De''$ or
 $\De'\subseteq \De''$ we have a factor $M^{-\frac{\vep}{2}}\leq g^d$.
\medskip

With these remarks we can now prove (\ref{proof}) by induction.
Actually we will prove that, if at the scale $j$ we have a factor
$g^2$ for any $ \De\in {\cal D}_j\cap Y$ then we can rewrite 
this factors in such a way to have a factor $g$ 
for any $\De\in {\cal D}_j\cap Y$ and a factor
$g^2$ $\forall\ \De\in {\cal D}_{j+1}\cap Y$.

\paragraph{Inductive hypothesis:} At scale $j$ we have a factor
$g^2$ for any $\De\in {\cal D}_j\cap Y$.

This is certainly true for  
the highest scale $m_Y$, because at this scale all cubes 
are extremal summit cubes therefore by remark {\bf 1)} they have
a factor $g^2$.

\paragraph{Proof of the induction.} Now we must prove that,
given a factor $g^2$ for any $\De\in {\cal D}_j\cap Y$,
we have a factor  $g$ for any $\De\in {\cal D}_j\cap Y$
and  a factor $g^2$ for any $\De\in {\cal D}_{j+1}\cap Y$.

We consider a connected component $y^{j+1}_k$. This is
made from a set of generalized cubes connected by a tree.
Let us consider one particular generalized cube 
${\tilde \De}$ which is made of cubes of scale $j+1$
connected by links of higher scales.
Now we consider each cube in ${\tilde \De}$. 
For each such $\De$ we denote
by $s_\De$ the number of cubes above that is 
$s_\De= \{ \De'\in {\cal D}_{j}\cap Y\ | \ \De'\subset \De \}$
We distinguish three situations.

\paragraph{a)} If $|s_\De|=0$ then we are in the special
case $Ex(\De)=\De$ therefore the extremal summit cube $\De$ has a factor $g^2$.

\paragraph{b)} If $|s_\De|\geq 2$ then we have
\be
g^{2|s_\De|} = g^{|s_\De|}\  g^{|s_\De|}\leq  g^{|s_\De|}\ g^2
\ee 
therefore we can keep a factor $g$ for each $\De'\in s_\De$
and we have a factor $g^2$ for $\De$.

\paragraph{c)} The case $|s_\De|=1$ is the most difficult one.
We call the unique element of $s_\De$ $\De'$.
Again we distinguish three cases:
\begin{itemize}

\item{} there is no tree line of any scale 
connecting $\De'$ to some other $\De''\in {\cal D}_j$
(see Fig.\ref{had9} a). Therefore
$\De={\tilde \De}$ and $\De'$ must be $\De^0_{root}$ for some connected
component at scale $j$, therefore there is
a vertical link connecting $\De'$ to $\De$, and, by 
{\bf 3)} we have a factor $g^d$.
Hence we can write
\be
g^2\ g^d = g\ g^{d+1}\leq g\ g^2
\ee    
and we can keep a factor $g$ for $\De'$ and assign a factor
$g^2$ to $\De$.

\item{} there is at least one tree line
$C^{j'k}$ at some scale $j'\leq j$
connecting $\De'$ to some $\De''\subset \De_1$ ($\De_1\in {\tilde \De}$)
and $\De'$ is not nearest neighbor of $\De''$
(see Fig.\ref{had9} b).

\begin{figure}
\centerline{\psfig{figure=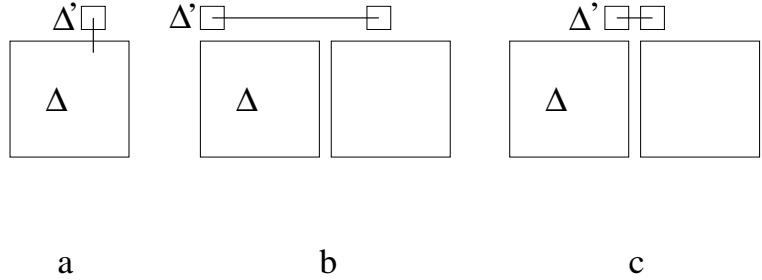,width=10cm}}
\caption{Three possible cases for $|s_\De|=1$.}
\label{had9}
\end{figure}

Then, $\left |t_{\De'}-t_{\De''}\right |\geq M^{j+1}$
(in the space directions they must always be nearest neighbors)
and the propagator must have $j'+k\geq j+1$. Therefore as
$j'\leq j$ $k$ cannot be zero and by remark {\bf 3)} we can associate
to $\De'$ a factor  $g^d$ in addition to $g^2$. Hence we can write
\be
g^2\ g^d = g\ g^{d+1}\leq g\ g^2
\ee    
and we can keep a factor $g$ for $\De'$ and assign a factor
$g^2$ to $\De$.

\item{} there is at least one tree line
$C^{j'k}$ at some scale $j'\leq j$
connecting $\De'$ to some $\De''\subset \De_1$ ($\De_1\in {\tilde \De}$)
and $\De'$ is nearest neighbor of $\De''$
(see Fig.\ref{had9} c). 

%\begin{figure}
%\centerline{\psfig{figure=had5.eps,width=4cm}}
%\caption{}
%\label{had5}
%\end{figure}

In this case $j'+k\geq j$ and no factor can be extracted from the
$k$ decay. Remark that, if $\De'=\De^0_{root}$ for
some connected component at scale $j$, then there is a vertical
link and everything works as in the case of Fig 10 a). 
On the other hand, if  $\De'\neq\De^0_{root}$,  there
is still a vertical link connecting $\De'$ to $\De$ but
it does not have any vertical decay associated. 
In this case we have to distinguish three possible 
situations:
\begin{itemize}
\item[{\bf a'}]\hskip-.1cm) there is no other tree line connecting 
$\De'$ or $\De''$ to some other cube in ${\cal D}_j$. Therefore
 $\De''$ must be $\De^0_{root}$ for some connected component 
at scale $j$ and the corresponding vertical link has
a vertical decay associated. Hence we have a factor
$g$ in addition to $g^2$ for each cube nearest neighbor 
(nn) of
$\De''$, hence for each of them we can keep one factor
$g$ and give the remaining $g^2$ to its ancestor.

\item[{\bf b'}]\hskip-.1cm) there is a tree line 
connecting $\De''$ to some cube
which is not nn of  $\De''$. Then we have some $k$ vertical
decay from the tree propagator, and we can 
assign a factor $g$ in addition to $g^2$ for each 
cube nn of $\De''$.  
Therefore, as in {\bf a'}, we have a factor
$g$ in addition to $g^2$ for each cube nn of 
$\De''$, hence for each of them we can keep one factor
$g$ and give the remaining $g^2$ to its ancestor.

\item[{\bf c'}]\hskip-.1cm) there is a tree line connecting $\De$ or 
$\De''$ to some cube nn. Then we test case {\bf a'} 
and {\bf b'} again, and we go on  until  {\bf a'} 
 or
 {\bf b'} (see Fig\ref{had10} a,b) is satisfied, or until the chain of 
nn cubes at scale $j$ arrives to a cube at scale $j+1$ that is
not nn of $\De$. In this last case (see Fig\ref{had10} c)
we must have at least $M$ of such cubes, therefore we can write
\be
(g^2)^{M}  \leq g^{M} (g^2)^d
\ee
which means that we keep one factor $g$ for each 
cube at scale $j$ and we give a factor $g^2$ to each nn of
$\De$ at scale $j+1$. This is true if $M$ satisfies:
\be
M \geq 2d .
\ee
\end{itemize}
\end{itemize} 
\qed
\medskip

\begin{figure}
\centerline{\psfig{figure=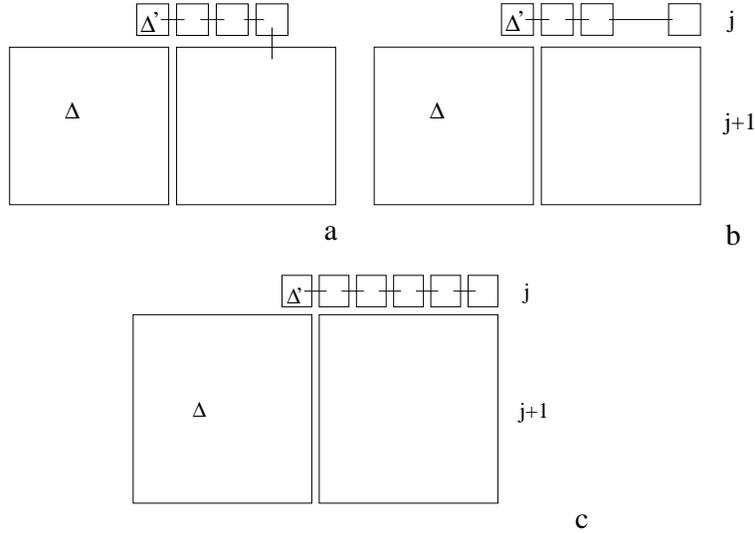,width=10cm}}
\caption{Three possible situations when extracting a small factor $g$}
\label{had10}
\end{figure}

%\begin{figure}
%\centerline{\psfig{figure=had6.eps,width=5cm}}
%\caption{}
%\label{had6}
%\end{figure}

%\begin{figure}
%\centerline{\psfig{figure=had7.eps,width=5cm}}
%\caption{}
%\label{had7}
%\end{figure}

%\begin{figure}
%\centerline{\psfig{figure=had8.eps,width=5cm}}
%\caption{}
%\label{had8}
%\end{figure}

\subsection{Bounding the tree choice}

\paragraph{Construction of $T_{jk}$}
Before summing over the trees we must see how the
tree is built. In the connected 
component $y_{jk}$, for each ${\tilde\De}\neq
{\tilde\De}_{root}$ we have one $h\in R_{root}$ 
and $d_{{\tilde\De}}$ fields in $l_b({\tilde\De})$
(defined in sec. III.3). For  ${\tilde\De}_{root}$
we have no $h\in R_{root}$ but we still have 
 $d_{{\tilde\De}}$ fields in $l_b({\tilde\De})$.
 Each $h\in l_b({\tilde\De})$ can contract only with  
a $h\in R_{root}$ in some ${\tilde\De}\neq{\tilde\De}'$. As there is
only one field $h\in b({\tilde\De}')$  
we only have to choose ${\tilde\De}'$. This last sum is performed 
using the decay of the tree line as we will prove below.
 
Therefore for each $h\in l_b({\tilde\De})$ we have to perform
the following sum
\be
\sum_{{\tilde\De}'\in y^j_k \atop
{\tilde\De}'\neq {\tilde \De}} \int_{\Om_{h'}} dx_{h'} 
\left | C^{jk_{h'}}(x_h,x_{h'}) \right | =
\sum_{\De'_{root}\neq \De_{root}(h),  \De^0_{root}} \int_{\Om_{h'}} dx_{h'} 
\left | C^{jk_{h'}}(x_h,x_{h'}) \right |
\ee
where $h'$ is the unique field in $R_{root}$ hooked to ${\tilde\De}'$,
$\Om_{h'}$ is the localization volume $\Om_{v_{h'}}$ of the vertex
to which $h'$ is hooked, 
$\De'_{root}$ is the corresponding cube in ${\tilde\De}'$
and $\De_{h'}\subseteq \De'_{root}$ is the localization cube 
for $h'$. Finally we denoted by $\De_{root}(h)$ the cube 
$\De_{root}$ for the generalized cube  ${\tilde\De}$
where $h$ is hooked (this contraction is not possible as
it would generate a loop). Remark that the condition
$\De'_{root}\neq \De^0_{root}$ holds because
this last cube does not contain any $h\in R_{root}$.
The sum over the tree $T_{jk}$ is then bounded
by
\be
\prod_{h\in b\backslash R_{root} \atop
j^b_h=j, \ {\rm and}\ \De_h\subseteq y^j_k } 
\sum_{\De'_{root}\neq\De_{root}(h), \De^0_{root}}
  \int_{\Om_{h'}} dx_{h'} 
\left | C^{jk_{h'}}(x_h,x_{h'}) \right |\ .
\ee

\paragraph{Sum over the cube positions and $T_{jk}$}

Now, for fixed $T_{jk}$, we have a multiscale tree structure.
We want to sum over the cube positions following this tree
from the leaves towards
the root (which is the cube $\De^0_{root}$ at scale $M_Y$,
which contain $x=0$). For this purpose we give a direction 
(represented by an arrow) to all links (vertical and
horizontal). 

\begin{figure}
\centerline{\psfig{figure=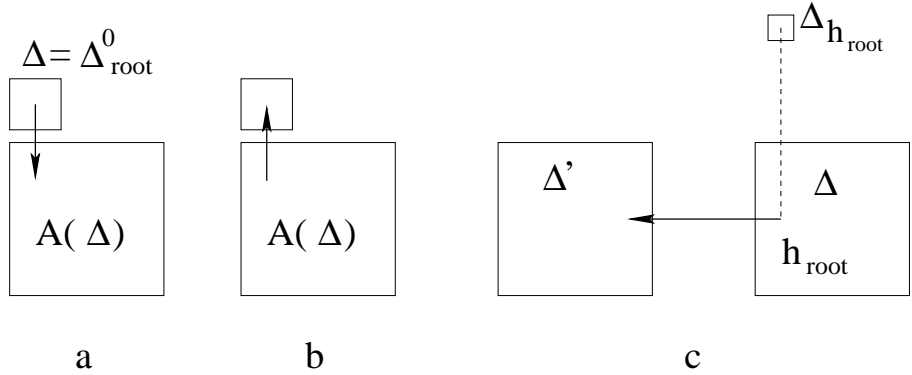,width=12cm}}
\caption{Three types of oriented links}
\label{had17}
\end{figure}

\begin{itemize}
\item{}  For any vertical link connecting some 
 $\De^0_{root}$ to its ancestor we draw an arrow going
from  $\De^0_{root}$ down to its ancestor and we call it a 
$down-link$  (see Fig.\ref{had17}a)

\item{}  For all other  vertical links 
 connecting some $\De$ to its ancestor
 we draw an arrow going
from  its ancestor up to  $\De$ or and we call it a 
$up-link$  (see Fig.\ref{had17}b).

\item{}  For each horizontal link, that is made by the 
contraction of a  field (antifield) in $R_{root}$
with an  antifield ( field) in $b\backslash R_{root}$
we draw  an arrow going from the  
field (antifield) in $R_{root}$ towards the  
antifield ( field) in $b\backslash R_{root}$  
(see Fig.\ref{had17}c).
\end{itemize}

Now we can perform the sums following the tree. We have three
situations.

\begin{itemize}

\item{} If we have a down-link we have to sum over the 
choices for $\De$, for $\De'={\cal A}(\De)$ fixed. Remark that
for each down-link we have the vertical decay $M^{-5(1-\vep')}$.
From this we first extract a fraction  $M^{-5\vep'}$ that will
be used for the last sum. With the remaining  $M^{-5(1-2\vep')}$
assuming $ \vep'\leq 1/10 $ we can write
\be
\sum_{\De\in {\cal D}_j \atop \De\subset \De', i_{\De'}=j+1} M^{-5(1-2\vep')}
= \frac{|\De'|}{|\De|} M^{-5(1-2\vep')} = M^4 M^{-5(1-2\vep')}\leq 1\ .
\ee

\item{} If we have an up-link we have  to sum over the 
choices for $\De'={\cal A}(\De)$  for $\De$ fixed. As there is
only one $\De'$ such that  $\De'={\cal A}(\De)$ there is no sum 
at all.

\item{} If we have an horizontal link the argument is more subtle
and we explain it below.

\end{itemize}

\paragraph{Sum over horizontal links}

For some $h\in R_{root}$ we want to prove that 
\be
 M^{-(i_h-i_{\De_h})}  \
 M^{-4\frac{\vep'}{2}(i_h-i_{\De_h})}  \ 
\sum_{x_\De} \int_{\Om_h} dx_h    \
\left | D^{j k_h}
\lp x_h,x_{h'}\rp \right| \leq C \  M^{11/3} \ 
M^{4i_{\De_h}}
\label{treebound}\ee
where $\De$ is the unique cube at scale $i_h=j^b_h=j$ with
$\De_h\subseteq \De$ (see Fig.\ref{had17}c). From now on we write
$j$ instead of $i_h$. We recall that we defined for  $k>0$ 
(see (\ref{defpropd}))
\bqa
&& \hspace{-0.7cm}\left | D^{j, k_h}
\lp x_h,x_{h'}\rp \right| =  \left | C^{j, k_h}
\lp x_h,x_{h'}\rp \right| M^{2j} M^{2\vep k_h}\\
&& \hspace{-0.5cm}\leq  
C \ M^{8/3}   M^{2\vep k_h} M^{-2k_h/3} 
\chi\lp |\vec{x}_h-\vec{x}_{h'}|\leq M^{j-k_h/3
+1/3}, |t_h-t_{h'}|\leq M^{j+k} \rp\no
\eqa
and for $k=0$
\bqa
\left | D^{j, 0}
\lp x_h,x_{h'}\rp \right| &  = &  \left | C^{j, 0}
  \lp   x_h,x_{h'}\rp \right| M^{2j} M^{2\vep k_h}\\
&\leq & C \ M^{8/3}   \ 
\chi\lp |\vec{x}_h-\vec{x}_{h'}|\leq M^{j}, 
|t_h-t_{h'}|\leq M^{j} \rp\no
\eqa
 
The case $k=0$ is simple as
\bqa
 &  & \hspace{-0.7cm}  
\int_{\Om_h} dx_h  \chi\lp |\vec{x}_h-\vec{x}_{h'}|\leq M^{j}, 
|t_h-t_{h'}|\leq M^{j} \rp\no\\
 &  &  \leq M^{4i_{\De_h}}\ 
 \chi\lp |\vec{x}_\De-\vec{x}_{\De'}|\leq M^{j}, 
|t_\De-t_{\De'}|\leq M^{j} \rp
\eqa
and 
\be
\sum_{x_\De} \chi\lp |\vec{x}_\De-\vec{x}_{\De'}|\leq M^{j}, 
|t_\De-t_{\De'}|\leq M^{j} \rp \leq d
\ee
where $d$ is the number of nearest neighbors. Therefore
\be
 M^{-(j-i_{\De_h})}  \
 M^{-2\vep'(j-i_{\De_h})}
\sum_{x_\De} \int_{\Om_h} dx_h    \
\left | D^{j k_h}
\lp x_h,x_{h'}\rp \right| \leq C \  M^{8/3}\  M^{4i_{\De_h}}\ 
\ee
where the decay $ M^{-(1+2\vep') (j-i_{\De_h})}$
is just bounded by one.

The case  $k>0$ is more difficult.
Now the integral is given by
\bqa
& &\hspace{-0.5cm}  \int_{\Om_h} dx_h \ 
\chi\lp |\vec{x}_h-\vec{x}_{h'}|\leq M^{j-k_h/3 + 1/3}, 
|t_h-t_{h'}|\leq M^{j+k} \rp\\&  &\hspace{-0.5cm}  \leq M^{i_{\De_h}}\lp  
\min\{ M^{i_{\De_h}} , M^{j-k_h/3+ 1/3} \}\rp^3
\chi\lp |\vec{x}_\De-\vec{x}_{\De'}|\leq M^{j}, 
|t_\De-t_{\De'}|\leq M^{j+k} \rp\no
\eqa
and the sum over $x_\De$ gives
\be
\sum_{x_\De} \chi\lp |\vec{x}_\De-\vec{x}_{\De'}|\leq M^{j}, 
|t_\De-t_{\De'}|\leq M^{j+k} \rp \leq d \ 2M^k
\ee
Now we have to distinguish two cases.

\paragraph{1.} If we have 
\be
i_{\De_h} < j-\frac{k_h}{3}+ \frac{1}{3}
\label{case1}\ee
 (\ref{treebound}) is bounded by
\be
C \  M^{8/3}\  M^{4i_{\De_h}}\  M^{k_h}\ 
M^{-\lp 1+2\vep'\rp (j-i_{\De_h})}
 M^{-k_h\lp 2/3 -2\vep\rp} 
\ee
By (\ref{case1}) we have
\be
M^{-\lp 1+2\vep'\rp (j-i_{\De_h})}
\leq M^{-\lp 1+2\vep'\rp k_h /3}
 M^{\lp 1+2\vep'\rp /3}.
\ee
Inserting this bound in the equation above we obtain
\be
\lp C  \ M^{8/3} M^{\lp 1+2\vep'\rp /3}\rp
\  M^{4i_{\De_h}}\  M^{k_h\lp 2\vep- 2\vep'\rp} \leq \
C\ \ M^{11/3} \   M^{4i_{\De_h}}
\ee
for $\vep<\vep'$.

\paragraph{2.} On the other hand, if we have  
\be
i_{\De_h} \geq  j-k_h\frac{1}{3}+\frac{1}{3}
\ \Rightarrow \ 
k_h \geq 3 \lp j -i_{\De_h}  \rp +1
\label{case2}\ee
  (\ref{treebound}) is bounded by
\be
C  \  M^{8/3}\  M^{i_{\De_h}}\ 
M^{3 \lp j-k_h/3+1/3\rp }\ 
 M^{k_h}\ 
M^{-\lp 1+2\vep'\rp (j-i_{\De_h})}
 M^{-k_h\lp 2/3-2\vep\rp} 
\ee
Now we can write
\be
 M^{i_{\De_h}}\ M^{3 \lp j-k_h/3+1/3\rp } M^{k_h}\ 
=M\  M^{-k_h}\ M^{k_h}\ M^{4i_{\De_h}}\ 
 M^{3 \lp j-i_{\De_h}\rp} 
\ee
and  (\ref{treebound}) is bounded by
\be
\lp C \  M^{1+8/3}\rp\  M^{4i_{\De_h}}\ 
M^{\lp 2-2\vep'\rp (j-i_{\De_h})}
 M^{-k_h\lp 2/3-2\vep\rp} \leq C\ M^{11/3}    
M^{4i_{\De_h}}
\ee
if we can prove that
\be
k_h \geq \frac{\lp 1 -\vep'\rp }{\lp 1/3-\vep\rp} (j-i_{\De_h})
\ee
This is true by (\ref{case2}) if
\be
 \frac{\lp 1-\vep'\rp }{\lp 1/3-\vep\rp}\leq 3
\ee
hence for $\vep< \frac{\vep'}{3}$ which is consistent with the
condition we find in the case {\bf 1.} 
\medskip

With all these results we can now write

\bqa 
&& \hspace{-0.5cm} \sum_{\{x_\De\}}
\left [ 
\prod_{\{h\in R_{root} \}}
 M^{-(i_h-i_{\De_h})}  \
\prod_{\{ {\bar h}\in R_{root} \}} 
 M^{-(i_{\bar h}-i_{\De_{\bar h}})} \right ]   
\left [\prod_{v\in V_d\cup V_b} M^{-4i_{\De_v}} \right ]  
\\
&& \hspace{-0.5cm}  
\left [ 
\prod_{\{h\in R_{root} \}}
 M^{-4\frac{\vep'}{2}(i_h-i_{\De_h})}  \
\prod_{\{ {\bar h}\in R_{root} \}} 
 M^{-4\frac{\vep'}{2}(i_{\bar h}-i_{\De_{\bar h}})} \right ] \ 
\left [ \prod_{j=m_Y}^{M_Y-1}  \prod_{k=1}^{c_j} M^{-5(1-2\vep')} 
\right ]
\no\\
&& \hspace{-0.5cm} 
\left [ \prod_{j=m_Y}^{M_Y}  \prod_{k=1}^{c_j} \sum_{T_{jk}} \right ]  
\left [\prod_{v\in V_d\cup V_b} \int_{\Om_v} dx_v  \right ]  \
\left [ \prod_{j=m_Y}^{M_Y} 
\prod_{k=1}^{c_j} \left [
\prod_{l\in T_{jk}} 
\left | D^{j k_l}_{\De_{l}{\bar \De}_{l}}
\lp x_l,{\bar x}_l\rp \right|\right ] 
\de_{k_{h_l} k_{{\bar h}_l}} \right ]  \no\\
&& \hspace{-0.7cm} \leq \  C^{|Y|} 
M^{11n/3}
\prod_{v\in V_d\cup V_b} M^{4i_{\De_v}} M^{-4i_{\De_v}}
\leq   C^{|Y|} 
 M^{11n/3} \no
\eqa
where the constant $ C^{|Y|}$  comes from (\ref{treebound}), 
and we applied
\be
\prod_{h\in R_{root}} M^{4i_{\De_h}} 
\prod_{{\bar h}\in R_{root} }  M^{4i_{\De_{\bar h}}} 
\leq \prod_{v\in V_d\cup V_b} M^{4i_{\De_v}}
\ee
This is true because two $h_{root}$ cannot be hooked to the same 
vertex by construction.

\subsection{Final bound}

Now we can perform all the remaining bounds, namely

\bqa
&&\hspace{-0.7cm}\sum_{Y\atop 0\in Y} |A_c(Y)| L^{|Y|}\  
\leq \ 
\sum_{M_Y}
 \sum_{S}\sum_{VL} L^{|Y|}\sum_{B_{S}} \  g^{|Y|}\ C^{|Y|} 
\sum_{n=0}^\infty C^n M^{13n}
\frac{1}{n!}
 \no\\ 
&& \hspace{-0.5cm}
|\la|^{\frac{|V_d\cup V_b|}{16}} K^{|{\bar V_d}\backslash V_b|}
\sum_{V_d, \al_{V_d}} \sum_{a,b,R} \sum_{\{v_l\}_{l\in vL}} \ 
\sum_{n_{V_d}\si_{V_d}\rho_{V_d}}
\sum_{\{n_\De\}_{\De\in B_S}}\  \sum_{\De^c_{{\bar V}_d}} 
\label{finalbound}\\ 
&& \hspace{-0.5cm}
\left [\prod_{v\in V_d} 
\sum_{i_{v}\in I^c_v} \sum_{\De_v\in {\cal D}_{i_v}\cap Y}  \right ]\ 
\sum_{\{J_h^a\}, \{ J_{\bar h}^a\}} \  
\sum_{\{j_h^b\},\{j_{\bar h}^b\}}  \ 
\sum_{\{k_h\}, \{k_{\bar h}\}} \sum_{\{\bt_h\}\{\bt_{\bar h}\}} \ 
\no\\ 
&& \hspace{-0.5cm}
\left [\prod_{\De\in Y} n_d(\De)!\right ] 
\left [\prod_{\De\in B_S} n_\De! \right ]
 \left [
\prod_{\{h\not \in b \}}  M^{-\frac{k_h}{12}}  \
\prod_{\{{\bar h}\not \in b \}}   M^{-\frac{k_{\bar h}}{12}}
\right ] \
 \left [
\prod_{\{h\in b \}}  M^{-\frac{\vep}{2} k_h}  \right ]
\no\\
&&\hspace{-0.5cm} 
 \left [\prod_{\{{\bar h} \in b \}}   M^{-\frac{\vep}{2} k_{\bar h}}
\right ] \ \left [ 
\prod_{\left \{{h\not \in b\ |\  \al_h=4
\atop {\rm or \ } h\in b}\right \}}
 M^{-\frac{\vep'}{4}(i_h-i_{\De_h})}  \
\prod_{\left \{{{\bar h}\not \in b\ | \   
\al_{\bar h}=4, \atop {\rm or \ } {\bar h}\in b}\right \}} 
 M^{-\frac{\vep'}{4}(i_{\bar h}-i_{\De_{\bar h}})}   \right ]  
 \no\\
&&\hspace{-0.7cm} \left [ \prod_{j=m_Y}^{M_Y-1}  \prod_{k=1}^{c_j} M^{-5\vep'} 
\right ] \left [\prod_{j=m_Y+1}^{M_Y}
\lp \prod_{l\in vL_j}\int_0^1 dw'_{l}
 \rp\right] 
\left [\prod_{v\in V_d\atop c_v\neq \al_v} 
 \prod_{j= i_v}^{l_v} 
w'_{y_v^j}\right]   
\left [\prod_{v\in V_d\atop c_v= \al_v} 
 \prod_{j= i_v}^{l_v-1} 
w'_{y_v^j}\right] \no
\eqa
where $ M^{-5\vep'} $ is the factor we extracted form 
 $ M^{-5(1-\vep')} $ before performing the sum over the
tree choice.
Now we can immediately bound the following sums.
 
\begin{itemize}
\item{} the sum over $\bt_h$ costs only a factor 5 per field, hence
\be
\sum_{\{\bt_h\}\{\bt_{\bar h}\}} 1 \leq 5^{4n}
\ee
\item{} the sum over $k_h$ is performed using the vertical decay
\be
\sum_{\{k_h\}, \{k_{\bar h}\}} 
\prod_{\{h\not \in b \}}  M^{-\frac{k_h}{12}}  \
\prod_{\{{\bar h}\not \in b \}}   M^{-\frac{k_{\bar h}}{12}}
\prod_{\{h\in b \}}  M^{-\frac{\vep}{2} k_h}  \
\prod_{\{{\bar h} \in b \}}   M^{-\frac{\vep}{2} k_{\bar h}} \leq 
C^{n} 
\ee
\item{} the sums over $J_h^a$ and $j_h^b$  are done
using the vertical decay $ M^{-\frac{\vep'}{4}(i_h-i_{\De_h})}$
For  $j_h^b$ we write
\be
\sum_{\{j_h^b\},\{j_{\bar h}^b\}}
\prod_{ \{h\in b \}}
 M^{-\frac{\vep'}{4}(j^b_h-i_{\De_h})}  \
\prod_{ \{{\bar h} \in b \} } 
 M^{-\frac{\vep'}{4}(j^b_{\bar h}-i_{\De_{\bar h}})}
\leq C^{|V_b|}
\ee

For  $J_h^a$ we define $V'_d$ as the set of vertices $v\in V_d$
that have some field (or antifield) $h\in a$. Then we can write
\be
\prod_{v\in V'_d}
\prod_{\{ h_v^c\in a\}}
 \sum_{J_h^a} \ 
  M^{-\frac{\vep'}{4}(i_h-i_{\De_h})}  
\leq C^{n}
\ee
where we applied $i_{\De_h}=i_v$ (as $v\in V_d$) and  
\bqa
 &&\sum_{J_h^a}  M^{-\frac{\vep'}{4}(i_h-i_{v})} 
=  \sum_{i_h>i_v} \sum_{p=0}^{i_h-i_v-1} 
\sum_{i_v<j_1<j_2...<j_p<i_h}  M^{-\frac{\vep'}{4}(i_h-i_v)}\no\\
&&\leq     \sum_{i_h}  M^{-\frac{\vep'}{4}(i_h-i_v)} \sum_{p=0}^{i_h-i_v-1} 
 \  2^{(i_h-i_v)}  \leq    C   \no
\eqa
and we used 
$\sum_{p=0}^m\sum_{0< j_1<j_2...<j_p<m} 1 \leq 2^{m}$ 
\end{itemize}

\subsubsection{Choice of $i_v$ and $\De_{i_v}$}

For each vertex $v_l\in V_d$ associated to the vertical 
link $l\in vL$ we can sum over the choices for
$i_v$ and $\De_{i_v}$ using the weakening factors $w'$.
Actually these factors not only allow to choose 
$i_v$ and $\De_{i_v}$, but they also give a factor $1/n_d(\De)!$ for each 
$\De$, where we recall that $n_d(\De)$ is the number of
vertices $v\in V_d$ localized in $\De$ (\ref{defnh}). 
This is proved in the following lemma.
\paragraph{Lemma IV.7.1a}
{\em The integrals over the weakening factors  $w'$ 
 allow to choose 
$i_v$ and $\De_{i_v}$, and give a factor $1/n_d(\De)!$ for each 
$\De$, namely
\bqa
&&\left [\prod_{v\in V_d} 
\sum_{i_{v}\in I^c_v} \sum_{\De_{i_v}\in {\cal D}_{i_v}\cap Y}  \right ]\ 
 \left [\prod_{j=m_Y+1}^{M_Y}
\lp \prod_{l\in vL_j}\int_0^1 dw'_{l}
 \rp\right] \
 \no\\
&& \left [\prod_{v\in V_d\atop c_v\neq \al_v} 
 \prod_{j= i_v}^{l_v} 
w'_{y_v^j}\right]   
\left [\prod_{v\in V_d\atop c_v= \al_v} 
 \prod_{j= i_v}^{l_v-1} 
w'_{y_v^j}\right] \ \ \leq \ C^{|Y|}\ 
\prod_{\De\in Y} \frac{1}{n_d(\De)!} 
\label{planes}\eqa
}
\paragraph{Proof.}
We perform the sum following the structure of the rooted
tree $S$. We call $T$ the tree obtained by $S$ taking away the leaves 
(dots).  We work with $T$ and not with $S$ because the leaves of
$S$ do not correspond to a connected component but to a void subset.
We denote each vertex of $T$
as $v_T$. Remark that the vertex $v_T$ at the layer $l$ corresponds
to a set of connected cubes in ${\cal D}_j$,  with $j=M_Y-l$. For $VL$
fixed, we know the number of cubes at this scale belonging to $v_T$,
hence also the number of vertices $v\in V_d$ localized in some
cube of  $v_T$. We denote this number by $n(v_T)$. This 
satisfies
$n(v_T)=\sum_{\De\in v_T} n_d(\De)$. 

\begin{figure}
\centerline{\psfig{figure=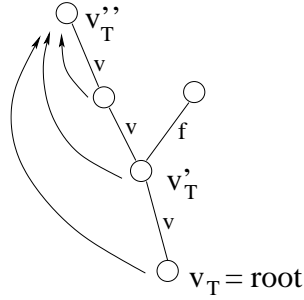,width=4cm}}
\caption{Example of arrow system}
\label{had18}
\end{figure}

Now we visualize the sums using a set of arrows on the rooted tree
$T$. For each link  of type $v$ (which corresponds to a vertical 
link $l\in vL$)  between a vertex $v_T'$ and its ancestor in the tree $v_T$,
we draw an arrow starting at  $v_T$ and going up, and stop
the arrow at the vertex $v''_T$ 
corresponding to the connected subpolymer 
at scale $i_{v_l}$ containing the cube $\Delta_{v_l}$, 
where $v_l$ is the vertex associated to the link (see Fig.\ref{had18}).

Therefore $n(v_T)$ actually corresponds
to the number of arrows which end at vertex $v_T$
in the tree $S$. Let $d(v_T)$ be the number of arrows departing
from $v_T$. For any line $l$ of $T$ 
let us call $t(l)$ the traffic
over $l$, namely the number of arrows flying above line $l$. 
We have obviously at any vertex $v_T$ of the tree a conservation law.
If $l_0(v_T)$ is the trunk arriving at node $v_T$ from below in the tree,
and  $l_1(v_T)$, .... $l_p(v_T)$ the branches going up from 
$v_T$, 
we have $t(l_0) + d(v_T) = n(v_T) + \sum_{i=1}^{p} 
t(l_i(v_T)) $.

Now the integration of
the $w'$ factors gives exactly 
$\prod_{l\in L(T)}  1/t(l)$, where
$L(T)$ is the set of lines in $T$ of type $v$. This
can be seen as each such line corresponds to a vertical
link $l\in vL$ that is to the introduction
of a specific $w'_l$ parameter (it is not every line of $T$,
because there can be $f$ links too, see Fig.\ref{had18}). 
Indeed the power of that $w'$ factor
to integrate is then exactly $t(l)-1$ for that line $l$.
The -1 is there because the derivation with respect to
$w'$ erased the factor $w'$ for the $v$ link created, but it
did not erase all other $w'$ factors for the other $v$ links going
up through that line. 

We can now decide to fix the numbers $n(v_T)$ and $d(v_T)$ of arrows 
arriving and departing at $v_T$.
(\ref{planes}) is then  written as

\be
\left [\prod_{v_T} \sum_{n(v_T)}\sum_{d(v_T)}\right ]\
\sum_{syst}  \prod_{l \in L(T)}  1/t(l)\
\left [\prod_{v_T} \sum_{\{n_d(\De)\}} 
\lp\prod_{v\in v_T}\sum_{\De_v\in v_T}\rp
\right ] \ 1
\ee
$\sum_{syst}$ is the sum over all  systems of
arrows compatible with $n(v_T)$ and $d(v_T)$, $\{n_d(\De)\}$ chooses
the number of vertices localized in each $\De\in v_T$ with the condition
$\sum_{\De\in v_T}n_d(\De)=n(v_T)$ and 
$\prod_{v\in v_T}\sum_{\De_v\in v_T}$ chooses for each vertex localized in 
$v_T$ by the arrow system, the localization cube $\De_v$. 
The sums over  $n(v_T)$ and $d(v_T)$ will be performed later and
will  cost at most $C^{|Y|}$.

Let us perform first the sum over  $\{n_d(\De)\}$  and $\De_v$.
\be
\prod_{v\in v_T} \sum_{\De_v\in v_T}\ 1
 \leq 
   \frac{n(v_T)!}{\prod_{\De\in v_T} n_d(\De) !} \quad
{\rm and}\;
\sum_{\{n_d(\De)\}_{\De\in v_T}}\   \leq   \ 2^{|v_T|+n(v_T)} \  
\label{somme}\ee
where $|v_T|$ is the number of cubes in $v_T$ and  we applied
$\sum_{\{n_d(\De)\}_{\De\in v_T}}\ 1 \leq 2^{|v_T|+n(v_Y)}$
(by a well known combinatoric
trick, $\sum_{i_1,i_2,...i_p | \sum i_j =m} 1 \leq 2^{m+p-1}\leq 2^{2m-1}$).
Therefore
\be
\left [\prod_{v_T}\sum_{\{n_d(\De)\}_{\De\in v_T}}\
\lp\prod_{v\in v_T}\sum_{\De_v\in v_T}\rp
\right ] \ 1 \leq \  C^{|Y|} \    
\frac{\prod_{v_T} n(v_T)!}{\prod_{\De } n_d(\De) !}
\label{deltabound}\ee
Indeed the number of vertical links of type $v$ is at most
$|V_T|-1$ 
where  $ |V_T| $ is the number of vertices in $T$.
Therefore we have
$\sum_{v_T} n(v_T)= m \leq |V_T|-1\leq |Y|$.

Now we perform  the sum over the arrow system. 
Remark that once the numbers 
$n(v_T)$ and $d(v_T)$ are fixed, the traffic
numbers $t(l)$ are also known, since for any line the
traffic $t(l)$ is equal to the sum of all arrows arriving in the subtree
for which $l$ is the trunk, minus the number of arrows 
departing in that subtree (because arrows always go upwards in the tree,
so the ones departing in the subtree have to end there too).

Now, it is easy to check that the complete choice over the system of
arrows consists, for each node $v_T$ of the tree, in choosing
by multinomial coefficients the $n(v_T)$ ones from the arriving traffic
$t(l_0)$ which stop at $v_T$, and then which of the remaining ones
go into which subbranches. This costs exactly a factor
\be
A(v_T)= \frac{t(l_0(v_T))!}{ n(v_T) !
\prod_i t'(l_i(v_T)) !} 
\ee
where $t'(l_i)=t(l_i)-1$ if $l_i \in L(T)$, that is if there
is one departing arrow from
node $v_T$ flying over line $l_i$ corresponding to a vertical
link of type  $v$ 
attaching the vertex $v'_T$ at the upper end of line $l_i$ to its ancestor
$v_T$; and $t'(l_i) = t(l_i)$ otherwise (actually in that last case,
$t'(l_i) = t(l_i)=0$ because 
that link must be of type $f$ therefore 
no vertex from a higher scale can be associated to a
vertical link at a lower scale).
Therefore we have to bound
\be
\left [\prod_{v_T}
 \frac{t(l_0(v_T))!}{ n(v_T) !
\prod_i t'(l_i(v_T)) !} \right ]\left [
 \prod_{l \in L(T)}  1/t(l) \right ]
= \left [\prod_{v_T} \frac{1}{ n(v_T) !}  \right ]
 \left [  A.B \right ]
\ee
  where $A$ is $\prod_{v_T}A(v_T)=
\prod_{v_T} {t(l_0(v_T))! \over \prod_i t'(l_i(v_T)) !}$
and where $B$ is our good factor coming from the $w'$ integrals,
namely  $\prod_{l \in L(T)}  1/t(l)$. 

\paragraph{Lemma IV.7.1b}
{\em For any tree and any
choice of the numbers $n(v_T)$ and  $d(v_T)$ (which determine
the traffic numbers $t(l)$, as said above), we
have $A.B=1$ ({\it exactly!})}
\medskip

\noindent{\bf Proof}
By induction, starting from the leaves of the tree towards
the trunk, we see that this is true.

For instance from a leaf $v_T$ of a tree, 
we have an apparently bad factor  $t(l_0(v_T))!$ in $A$, where 
because we are at a leaf, $t(l_0(v_T))= n(v_T)$ (all arrows must end
at $v_T$, because there is nothing beyond if $v_T$ is a leaf).
But then if at the node $v_T'$ below that line 
$l_0(v_T)$ there is a departing arrow flying over $l_0(v_T)$,
we have a factor $1/t(l_0(v_T))$ from $B$, and $l_0(v_T)=l_i(v_T')$ 
for some $i$. Combining the factor $1/t(l_0(v_T))$ from $B$ and
the factorial $ t'(l_i(v_T')) ! =  [t(l_0(v_T))-1]! $
in $A$ at the next node, we can reconstruct a denominator
$1/t(l_0(v_T))!$, which exactly cancels our bad factor  
$t(l_0(v_T))!$. Doing that for all leaves above $v'_T$, we erase all bad
factors and remain with exactly the numerator of the $A$ factor at node
$v_T'$, namely $[t(l_0(v_T'))]!$. Continuing
this way towards the bottom of the tree, we are finally left with
a single factorial of the traffic, namely   $[t(l_0(v_{T0}))]!$ which
is the last traffic at the trunk. But this traffic is 1!
Therefore $A.B= 1$. This ends the proof of Lemma IV.7.1b\footnote{This 
lemma is a particular
variation on well known combinatoric identities [BF][DR2, Appendix B1].}.
\qed
\medskip

Now, the factor $\prod_{v_T}
 \frac{1}{ n(v_T) !}$ cancels the corresponding factor
on the numerator in (\ref{deltabound}), while
the $\prod_{\De} \frac{1}{ n_d(\De) !}$ is kept outside. 
Finally we  check that 
\be
\left [\prod_{v_T} \sum_{n(v_T)}\sum_{d(v_T)}\right ]\ 1 \ 
\leq \ C^{|Y|} \ .
\ee
As for (\ref{somme}-\ref{deltabound})  we have
$\sum_{v_T} n(v_T)=\sum_{v_T} d(v_T)= m \leq |V_T|-1\leq |Y|$
and we apply
$\sum_{i_1,i_2,...i_p | \sum i_j =m} 1 \leq 2^{m+p-1}\leq 2^{2m-1}$.
This ends the proof of Lemma IV.7.1a.
\qed

\subsubsection{Extracting a global factor $|\la|$}

The last sum over $M_Y$ will cost an extra logarithm.
Therefore, in order to prove $\sum_{Y\atop 0\in Y} |A_c(Y)| L^{|Y|}\leq 1$
we must ensure that we can extract at least one factor
$|\la|$ from the sums\footnote{In fact to perform a Mayer expansion,
we need only to control $\sum_{Y\atop 0\in Y}$ with  $M_Y$ fixed 
in our main result (\ref{summayer}). However we prove the slightly stronger
result (\ref{Y1}) for simplicity, since it is also true.}. 
This is not trivial because we have only
a fraction $|\la|^{\frac{1}{16}}$ per vertex $v\in V_d$.
If $|V_d|\geq 17$ we can extract the factor $|\la|$ to sum over $M_Y$
and keep a remaining small
factor  $|\la|^{|V_d|/(16\times 17)}= |\la|^{|V_d|/272}$
per vertex.
The case  $|V_d|\le 16$ is more delicate. 
Remark that, when  $|V_d|\le 16$, the Hadamard bound is simpler 
in the sense that we do not need to pay any logarithm
(see case 2 in IV.3.2) or any factor $n_\De, n(\De)$
(see case 1 and 5 in IV.3.2) to choose the contractions
as the number of choices to
contract a field with an antifield are bounded by $2\cdot 16$.
The only logarithms appearing are then the ones given by the
sums over possible attributions ( the $\sum_j$ in the Hadamard bound). 

We distinguish two situations:
\begin{itemize}
\item{}  $|V_d|\le 16$ and $|Y|=|V_d|$. In this case we have at most 17
energy scales, therefore any sum over scale attributions  costs
just a factor 17, hence the Hadamard bound does not 
produce any  logarithm. This means that the three fields
(antifields)
of type $\al_h\neq 5$ hooked to the vertex $v\in V_d$ 
still have their factor $|\la|^{\frac{1}{4}}$, therefore we have
a factor  
$|\la|^{\frac{3|V_d|}{4}}|\la|^{\frac{|V_d|}{16}}=
|\la|^{\frac{13|V_d|}{16}} $. Now, for $|V_d|>1$ we can extract a factor
$|\la|$. Otherwise, if  $|V_d|=1$, we have a polymer reduced to one or
two cubes, therefore there is no
logarithms. We can extract the complete coupling constant for the unique
vertex. Remark that in this case we have not extracted a small factor
$g$ for the cube, but only a factor $K$. 
Nevertheless this is only one term of the sum
(only the polymers with $|Y|=1$).

\item{}  $|V_d|\le 16$ and $|Y|>|V_d|$.  In this case we must have 
at least $|Y|-|V_d|$ vertical links of type $f$, therefore
there must be at least 2 vertices with some derived fields hooked:
$|V'_d|\geq 2$. Let us say that the lowest $f$-link is at
scale $j$. At lower scale there can be only $v$-links, therefore
there are at most 16 scales. As $M_Y-j\leq 16$ the 
set of attributions for  six fields
derived to give the $f$-link has at most size  $M_Y-j\leq 16$,
therefore these links do no give any logarithm, and we have a factor
$|\la|^{6/4}<|\la|$. 

\end{itemize}

\subsubsection{Remaining sums}

Now the remaining sum is
\bqa
&&\hspace{-0.5cm}\sum_{Y\atop 0\in Y} |A_c(Y)| L^{|Y|}
\leq |\la|
\sum_{M_Y}
 \sum_{S}\sum_{VL} \sum_{B_{S}} \ ( gLC)^{|Y|}  
\no\\
&&\hspace{-0.5cm}
\sum_{n\geq 1} \lp C M^{13}\rp^n
\frac{1}{n!}
|\la|^{\frac{|V_d\cup V_b|}{272}} K^{|{\bar V_d}\backslash V_b|}
\sum_{V_d, \al_{V_d}} \sum_{a,b,R} \sum_{\{v_l\}_{l\in vL}} \  \no\\ 
&& \hspace{-0.5cm}
\sum_{n_{V_d}\si_{V_d}\rho_{V_d}}
\sum_{\{n_\De\}_{\De\in B_S}}\  \sum_{\De^c_{{\bar V}_d}} 
\left [\prod_{\De\in B_S} n_\De! \right ]
\ \left [ \prod_{j=m_Y}^{M_Y-1}  \prod_{k=1}^{c_j} M^{-5\vep'} 
\right ]
\eqa
where all constants have been inserted in $C$ and the 
factor $\left [\prod_{\De\in Y} n_d(\De)!\right ] $ coming from
(\ref{finalbound}) is compensated by 
$\left [\prod_{\De\in Y} \frac{1}{n_d(\De)!} \right ]$ coming from 
Lemma IV.7.1a.

\paragraph{Sum over  $\{n_\De\}$ and $\De^c_V$.} These sums are bounded 
as follows.
\be
\sum_{\{n_\De\}_{\De\in B_S}}\  \sum_{\De^c_{{\bar V}_d}} 
\prod_{\De\in B_S}\left [   n_\De! \right ] \leq 
\sum_{\{n_\De\}_{\De\in B_S}}\  
 \frac{|{\bar V}_d|!}{\prod_{\De\in B_S} n_\De !}
\prod_{\De\in B_S} n_\De! \leq  |{\bar V}_d|! \ 2^{|Y|+n}
\label{facto1}\ee
where we applied
\be
\sum_{\{n_\De\}_{\De\in B_S}} 1 \leq 2^{|Y|+n}
\ee
as $\sum_{\De\in B_S} n_\De = |{\bar V}_d|\leq n$.

\paragraph{Sum over  $\{v_l\}_{l\in vL}$  }
This sum actually consumes a fraction of the global factorial,
namely

\be
\frac{1}{n!} \sum_{\{v_l\}_{l\in vL}} 1 
\leq \frac{1}{n!}  [n\ (n-1)\ (n-2)\ ...\ (n-|V_d|+1)]
= \frac{1}{|{\bar V}_d|!}
\label{facto2}\ee
where we applied $n-|V_d|=|{\bar V}_d|$.
 
\paragraph{Sum over  $\si_{v_d}, n_{V_d}, a, b, R, V_d, \rho_{V_d}$ 
and  $ \al_{V_d}$.}
The sum over $\si_{v_d}$ costs at most
a factor
$4!$ per vertex, the sum over $n_{V_d}$ at most 
a factor 4 per vertex,   the sums over $a$, $b$ and $R$ a factor
2 per field,  the sum over $V_d$ a factor 2 per vertex,
the sum over  $\rho_{V_d}$ a factor 2 per field and finally 
the sum over $ \al_{V_d}$  a factor 4 per vertex. Therefore

\be
\sum_{V_d, \al_{V_d}} \sum_{a, b,R}  
\sum_{n_{V_d}\si_{V_d}}\ \leq C^n
\ee

The  remaining bound is now
\bqa
&&\hspace{-0.7cm}\sum_{Y\atop 0\in Y} |A_c(Y)| L^{|Y|}\  
\leq \ |\la|\ 
\sum_{M_Y}
 \sum_{S}\sum_{VL} \sum_{B_{S}} \ ( gLC)^{|Y|}   \\
&&\hspace{-0.5cm}
\sum_{n\geq 1} \lp C M^{13}\rp^n
|\la|^{\frac{|V_d\cup V_b|}{272}} K^{|{\bar V_d}\backslash V_b|} 
\left [ \prod_{j=m_Y}^{M_Y-1}  \prod_{k=1}^{c_j} M^{-5(1-2\vep')} 
\right ]\no \eqa
where all constants have been inserted in $C$ and
the factorial  $|{\bar V}_d|!$ in (\ref{facto1}) has been canceled
by  the factor $\frac{1}{|{\bar V}_d|!}$ in (\ref{facto2}).
Now
\bqa
&&\hspace{-1cm}\sum_{n\geq 1} \lp C M^{{13}}\rp^n
|\la|^{\frac{|V_d\cup V_b|}{272}} K^{|{\bar V_d}\backslash V_b|} 
=\\ && \sum_{|V_d\cup V_b|\geq 1}  \lp C M^{{13}}
|\la|^{1/272}\rp^{|V_d\cup V_b|} 
\sum_{|{\bar V}_d\backslash V_b |\geq 0}
 \lp C M^{{13}} K \rp^{|{\bar V_d}\backslash V_b|}  
\leq C \no
\eqa
for $\la$ and $K$ small enough, depending on $M$. 
The choice of $B_{S}$ costs a factor 2 per cube
so finally we have to bound
\be
|\la| \ 
\sum_{M_Y}
 \sum_{S}\sum_{VL}  \  (gLC)^{|Y|}\  
\left [ \prod_{j=m_Y}^{M_Y-1}  \prod_{k=1}^{c_j} M^{-5\vep'} 
\right ]
\ee

\paragraph{Sum over  $S$ and $VL$} These  
sums are performed together. For this purpose 
%Actually the general rooted tree  
%$S$  does not depend on $M_Y$ but only only on the difference
%$l_S=M_Y-m_Y$. Remark that we can reconstruct a fraction of the global
%decay $M^{-\vep'(M_Y-m_Y)}$ extracting a fraction $\vep'$
%from the $ M^{-5\vep'} $ for each connected component. Now 
we  reorganize the sum   as follows:

\bqa
&&
 \sum_{S}\sum_{VL}  \  (gLC)^{|Y|}\ 
\left [ \prod_{j=m_Y}^{M_Y-1}  \prod_{k=1}^{c_j} M^{-5\vep'} 
\right ]
 \leq 
\sum_{p\geq 1} \lp 8 g LC  \rp^{p^0} 
\\
&&  \sum_{d^0\geq 0} 
\prod_{i=1}^{d^0}
\left [   \sum_{p^{1}_i\geq 1} \lp  
8 g LC \rp^{p^1_i} M^{-5\vep'}  \sum_{d_i^1\geq 0} 
\prod_{i'=1}^{d_i^1} 
\left [   \sum_{p^{2}_{i'}\geq 1} \lp  
8 g LC \rp^{p^2_{i'}} M^{-5\vep'}  \sum_{d_{i'}^2\geq 0}\  \cdots\ \right ]
\right ] \no
\eqa
where $p^0$ is the number of cubes in the connected subpolymer 
at the layer $l=0$ (corresponding to the scale $M_Y$), 
$d^0$ the number of connected components at the scale
$M_Y-1$  (circles in the rooted tree) connected to the root,
$p^1_i$ the number of cubes for the connected subpolymer $i$ and so
on. 
The factor $8$ include a factor $2$ to decide, for each vertical
link, whether it is a $v$ or $f$ link, a factor 2 to decide
for any cube of the connected subpolymer if it is going to a give
a dot or not in $S$ at the next layer (see Fig. 7),
and finally a factor $2^p$ to decide the remaining positive
numbers $VL$ for the circle links of $S$ 
(since they are strictly positive and their
sum is $p$). 

The products stop at $p^{M_Y}$ as this is the maximal number
of layers. We remark that for the root we do not have any vertical
link, hence no vertical decay $ M^{-5\vep'}$.

We start computing this formula from leaves, which correspond to
$d=0$. Assuming $gLC\le 1/16$ and $M^{-5\vep'/2}\le 1/2$ we have
\be
\sum_{p\geq 1}  \lp  8 g LC \rp^{p} M^{-5\vep'}
\leq 
\frac{1}{2}M^{-5\vep' /2} 
\ee

Now we can perform the sum over $d$ at the previous layer 
\be
\sum_{d\geq 0}  \lp M^{- 5\vep' /2}\rp^d \le 2
\ee
and at each layer 
we compensate the factor 2 by the new factor $M^{-5\vep'/2}\le 1/2$.

Therefore we can sum over all layers until the root, and the
result is bounded by 2 because the last layer has no  $M^{-5\vep'}$
factor.

\paragraph{Sum over $M_Y$} This sum is finally bounded as announced
by our spared factor $\la$ 
\be
\sum_{Y\atop 0\in Y} |A_c(Y)| L^{|Y|}\ \leq
\ |\la|  \sum_{M_Y} 2 \le 2 |\ln T|  |\la| \leq 2K \le 1 .
\ee
for $|\la\ln T| \leq K$.

This ends the proof of the theorem. To summarize our conditions,
for a given $L$ we compute first the constant $C$, we choose $M$
large enough (and $\la $ small enough)
so that $ g L C \le 1/16$ and $M^{-5\vep'/2}\le 1/2$, and we restrict again
$\la$ so that  $CM^{13}\la^{1/272} \le 1/2$. These restrictions
on $\la$ are therefore enforced solely by taking $K$ small enough
depending on $L$, which is our theorem.

\newpage

\setcounter{section}{1}

\renewcommand{\thesection}{\Alph{section}}

\noindent{\Large {\bf Appendix A }}
\medskip

%\noindent{\large {\bf Optimization of $\al$ in  \ref{Oj1}  }}

\resetequ

In section II.4 we have introduced band decoupling on the
position space, and defined, for each band $j$ the 
characteristic function $\Om_j$. Let us introduce the following
generalization of
(\ref{Oj2}):
\be
\ba{rll}
\Om_j  =&  \{ \ (\vec{x},t) \ | \ M^{j-1} \leq 
(1+|\vec{x}|)^{\frac{1}{2}+\al} \; 
 (1+f(t)+|\vec{x}|)^{\frac{1}{2}-\al} < M^{j} & \} 
\; j\leq j_M\\
  =&  \{ \ (\vec{x},t) \ | \ M^{j_M} \leq 
(1+|\vec{x}|)^{\frac{1}{2}+\al} \; 
 (1+f(t)+|\vec{x}|)^{\frac{1}{2}-\al} & \} 
\; j=j_M \\
\ea
\ee

To select the optimal value for $\al$ we must insert auxiliary scales
as in section II.4 and estimate the scaled decay of the propagator
$C^{jk}$, as a function of $\al$.
We insert auxiliary scale decomposition  as in (\ref{tdec}).

\paragraph{Spatial constraints}
The constraints on spatial positions now are:

\begin{itemize}

\item{} if $j< j_M$ and $k>0$  
there is a non zero contribution only for
\be
 M^{j} M^{-k\lp \frac{1-2\al}{1+2\al}\rp}
M^{-\frac{2}{1+2\al}}\  2^{-\frac{1-2\al}{1+2\al}} \  
\leq \  
(1 + |\vec{x}|) \ \leq  \ 
 M^{j} M^{-k\lp \frac{1-2\al}{1+2\al}\rp}
M^{ \frac{1-2\al}{1+2\al}}
\ee

\item{} for $j\leq j_M$ and $k=0$  
there is a non zero contribution only for
\be
 M^{j}\ 
 M^{-\frac{2}{1+2\al}}\  
2^{-\frac{1-2\al}{1+2\al}} \  
\leq \  
(1 + |\vec{x}|) \ \leq  \ 
M^{j} 
\ee

\item{} for $j= j_M+1$
there is a non zero contribution only for
\be
M^{j_M}\  
2^{-\frac{1-2\al}{1+2\al}} \  
\leq \  (1 + |\vec{x}|) 
\ee
\end{itemize}

\paragraph{Scaled decay of the propagator}
Now for each $j$ and $k$ we can estimate the 
scaled decay of the propagator $C^{j, k}$. We distinguish three 
cases:
\begin{itemize}

\item{} 
for $j< j_M$ and $k>0$ we have 
\be
\left |C^{j, k}(\vec{x},t) \right | \ 
\leq \ M^{-2j} \ 
M^{-k\lp \frac{4\al}{1+2\al}\rp}
M^{ \frac{3+2\al}{1+2\al}}\ 
2^{\frac{1-2\al}{1+2\al}} \  
\chi_{j, k}\lp \vec{x},f(t) \rp\ee
where the function $\chi_{j, k}$ is defined by
\bqa
\chi_{j, k}(\vec{x},t) &=& 1 \qquad {\rm if \ } \        
|\vec{x}| \leq M^{j}\ M^{-k\lp \frac{1-2\al}{1+2\al}\rp}\ 
M^{ \frac{1-2\al}{1+2\al}},\
f(t) \leq M^{j+k}\no\\
                 &=& 0 \qquad {\rm otherwise \ }
\eqa

\item{} 
for $j\leq j_M$ and $k=0$ we have 
\be
\left |C^{j, 0}(\vec{x},t) \right | \ 
\leq \ M^{-2j} \ M^{\frac{4}{1+2\al}}\ 
2^{2\lp\frac{1-2\al}{1+2\al}\rp} \  
\chi_{j, 0}\lp \vec{x},f(t) \rp\ee
where the function $\chi_{j, 0}$ is defined by
\bqa
\chi_{j, 0}(\vec{x},t) &=& 1 \qquad {\rm if \ } \        
|\vec{x}| \leq M^{j}\ ,\
f(t) \leq M^{j}\no\\
                 &=& 0 \qquad {\rm otherwise \ }
\eqa

\item{} 
for $j= j_M+1$  we have 
\be
\left |C^{j_M+1 0}(\vec{x},t) \right | \ 
\leq \ M^{-2j_M} \ 2^{2\lp\frac{1-2\al}{1+2\al}\rp} \  
\chi_{j_M+1, 0}\lp f(t)\rp
\ \frac{K_p}{\lp 1+ M^{-j_M}|\vec{x}|\rp^p}
\ee
where the function $\chi_{j_M+1, 0}$ is defined by
\bqa
\chi_{j_M+1, 0}(t) &=& 1 \qquad {\rm if \ } \    
f(t) \leq M^{j_M}\no\\
                 &=& 0 \qquad {\rm otherwise \ }
\eqa
and the spatial decay for $|\vec{x}|$ comes from the
decay of the function $F$ in (\ref{cdecomp}).
\end{itemize}

\paragraph{Integration volume}
The region of spatial integration (for a scale propagator)
is now fixed by the $\chi_{j,k}$ domain. Therefore
\begin{itemize}

\item{} 
for $j< j_M$ and $k>0$ we have 
\be
V_{j,k} = |\vec{x}|^3\ f(t) \leq M^{4j} \ 
M^{-k\lp \frac{2-8\al}{1+2\al}\rp}\ 
M^{3\lp \frac{1-2\al}{1+2\al}\rp}\ 
\ee
\item{} 
for $j\leq j_M$ and $k=0$ we have 
\be
V_{j,k} = |\vec{x}|^3\ f(t) \leq M^{4j} \ 
\ee

\item{} 
for $j= j_M+1$  we have 
\be
V_{j,k} = |\vec{x}|^3\ f(t) \leq M^{4j_M} \ 
\ee

\end{itemize}

As we have seen, the tree propagator is used in two
cases, namely to bound the sum over cubes in the Hadamard
bound (see (\ref{hadb})) and to perform the sum over trees.
In the Hadamard bound we must have
\be
F_{jk} =: |C^{jk}|^2 \ M^{4j}\ M^k\ \leq K\ M^{-\vep k} 
\label{opt1}\ee
for some constants $K, \vep >0$ ($K$ is actually proportional to
some constant power of $M$). The decay $ M^{-\vep k}$ is 
necessary  to sum  over $k$.
Inserting the $\al$ depending bounds for $C^{jk}$ we have,
for $k>0$
\be
F_{jk} \leq  M^{-4j} \ 
M^{-k\lp \frac{8\al}{1+2\al}\rp}
M^{2 \frac{3+2\al}{1+2\al}}\ 
2^{2\frac{1-2\al}{1+2\al}} \  
 \ M^{4j}\ M^k = M^{k\left [ 1-\lp \frac{8\al}{1+2\al}\rp
\right ]}
\ M^{2 \frac{3+2\al}{1+2\al}}\ 
2^{2\frac{1-2\al}{1+2\al}} 
\ee
and (\ref{opt1}) is true for
\be
 1-\lp \frac{8\al}{1+2\al}\rp < 0\ \Rightarrow \ 
\al > \frac{1}{6}
\ee

On the other hand when summing over the tree structure we 
must ensure that
\be
F_{jk} =: |C^{jk}| \ V_{jk} \leq K\ M^{2j} \ M^{-\vep k}
\label{opt2}\ee
for some constants $K, \vep >0$ ($K$ is actually proportional to
some constant power of $M$). Again the decay $ M^{-\vep k}$ is 
necessary to sum over $k$.
Inserting the values for $ |C^{jk}|$ and  $V_{jk}$ we have
\bqa
F_{jk} & \leq &  \ M^{-2j} \ 
M^{-k\lp \frac{4\al}{1+2\al}\rp}
M^{ \frac{3+2\al}{1+2\al}}\ 
2^{\frac{1-2\al}{1+2\al}} \   M^{4j} \ 
M^{-k\lp \frac{2-8\al}{1+2\al}\rp}\ 
M^{3\lp \frac{1-2\al}{1+2\al}\rp}\no\\
& \leq &
 M^{2j} \ M^{-k\lp \frac{2-4\al}{1+2\al}\rp}\ 
M^{2\lp \frac{3-2\al}{1+2\al}\rp}\ 2^{\frac{1-2\al}{1+2\al}}
\eqa
and (\ref{opt2}) is true for
\be
2-8\al > 0 \  \Rightarrow \ 
\al < \frac{1}{2}
\ee
Therefore the parameter $\al$ then can take values only
in the open interval $( \frac{1}{6}, \frac{1}{2})$.
Actually we choose the value $\al=\frac{1}{4}$
which corresponds to
\be
V_{jk} = M^{4j}.
\ee
For this value the band volume
does not depend on $k$ which is 
consistent with the choice of $j$ as the 
real band slicing, while  $k$ is just an
auxiliary band slicing.

\medskip
\noindent{{\bf Acknowledgment}}
M. Disertori acknowledges partial support of
NSF  grant DMS 97-29992 for this work.

\medskip
\noindent{\large{\bf References}}
\medskip

%{\tenrm

\noindent [AR1] A. Abdesselam and V. Rivasseau, Rev. Math.
Phys. Vol. 9 No 2, 123 (1997) 
\vskip.1cm

\noindent [AR2] A. Abdesselam and V.  Rivasseau, Trees, forests and jungles: a
botanical garden for cluster expansions, in Constructive Physics, ed by
V. Rivasseau, Lecture Notes in Physics 446, Springer Verlag, 1995.

\vskip.1cm
\noindent [BF] G. A. Battle and P Federbush,
A note on cluster expansions, tree graph identities, extra
1/N! factors!!! Lett. Math. Phys. {\bf 8}, 55, (1984).

\vskip.1cm
\noindent [BG] G. Benfatto and G. Gallavotti,
Perturbation theory of the Fermi surface in a quantum liquid.
A general quasi particle formalism and one dimensional systems, 
Journ. Stat. Phys. {\bf 59} 541 (1990).

\vskip.1cm

\noindent [BGPS] G. Benfatto,  G. Gallavotti, A. Procacci and B. Scoppola,
Commun. Math. Phys. {\bf 160}, 93 (1994).
\vskip.1cm

\noindent [BM] F. Bonetto and V. Mastropietro,
Commun. Math. Phys. {\bf 172}, 57 (1995).

\noindent [DR1] M. Disertori and V. Rivasseau, Interacting Fermi liquid 
in two dimensions at finite temperature, Part I: Convergent Attributions.
\vskip.1cm

\noindent [DR2] M. Disertori and V. Rivasseau, Interacting Fermi liquid 
in two dimensions at finite temperature, Part II: Renormalization,
\vskip.1cm

\noindent [FMRT] J. Feldman, J. Magnen, V. Rivasseau and E. Trubowitz,
An infinite Volume Expansion for Many Fermion Green's Functions,
Helv. Phys. Acta {\bf 65}
(1992) 679.
\vskip.1cm

\noindent [FT1]  J. Feldman and E. Trubowitz, 
Perturbation theory for Many Fermion Systems, Helv. Phys. Acta {\bf 63}
(1991) 156.
\vskip.1cm

\noindent [FT2]  J. Feldman and E. Trubowitz, The flow of an Electron-Phonon
System to the Superconducting State, Helv. Phys. Acta {\bf 64}
(1991) 213.
\vskip.1cm

\noindent [GN]  G. Gallavotti and F. Nicol{\`o}, Renormalization theory
in four dimensional scalar fields, I and II, Commun. Math. Phys. {\bf 100},
545 and {\bf 101}, 247 (1985).
\vskip.1cm

\noindent [MR] J. Magnen and V. Rivasseau, A Single 
Scale Infinite Volume Expansion for
Three Dimensional Many Fermion Green's Functions,
Math.  Phys. Electronic  Journal, Volume 1,  1995.
\vskip.1cm

\noindent{[R]} V. Rivasseau,
{\it From perturbative to constructive Renormalization,}
Princeton University Press, 1991.

\noindent [S] M. Salmhofer,
Continuous renormalization for Fermions and Fermi liquid theory,
Commun. Math. Phys.{\bf 194}, 249 (1998).
%}

\end{document}